\documentclass[letterpaper,twocolumn]{aastex6}
\shorttitle{K2 Candidate Planetary Systems III: A Dense Neptune}
\shortauthors{Dressing et al.}
\usepackage{epsfig}
\usepackage{amsmath}
\usepackage{rotating}
\usepackage{natbib} 
\usepackage{enumerate}
\usepackage{hyperref}
\usepackage{url}
\usepackage{longtable}
\usepackage{color}

\begin{document}
    
\def\mearth{{\rm\,M_\oplus}}                                                    
\def\msun{{\rm\,M_\odot}}                                                       
\def\rsun{{\rm\,R_\odot}}                                                       
\def\rearth{{\rm\,R_\oplus}} ннн
\def\fearth{{\rm\,F_\oplus}}
\def\lsun{{\rm\,L_\odot}}                                                          
\def\kepler {{\emph{Kepler}\,}}                                              
\newcommand{\teff}{\ensuremath{T_{\mathrm{eff}}}}                               
\newcommand{\logg}{\ensuremath{\log g}}


\title{Characterizing K2 Candidate Planetary Systems Orbiting Low-Mass Stars III:\\A High Mass \& Low Envelope Fraction for the Warm Neptune K2-55b\altaffilmark{1}}
\altaffiltext{1}{Some of the data presented herein were obtained at the W.\ M.\ Keck Observatory, which is operated as a scientific partnership among the California Institute of Technology, the University of California and the National Aeronautics and Space Administration. The Observatory was made possible by the generous financial support of the W.M. Keck Foundation.}
\author{Courtney D. Dressing\altaffilmark{1,2}}
\author{Evan Sinukoff\altaffilmark{3,4}}
\author{Benjamin J. Fulton\altaffilmark{3,4,5,6}}
\author{Eric D. Lopez\altaffilmark{7}}
\author{Charles A. Beichman\altaffilmark{6}}
\author{Andrew W. Howard\altaffilmark{4}}
\author{Heather A. Knutson\altaffilmark{8}}
\author{Michael Werner\altaffilmark{9}}
\author{Bj\"{o}rn Benneke\altaffilmark{10}}
\author{Ian J. M. Crossfield\altaffilmark{11}}
\author{Howard Isaacson\altaffilmark{1}}
\author{Jessica Krick\altaffilmark{12}}
\author{Varoujan Gorjian\altaffilmark{9}}
\author{John Livingston\altaffilmark{13}}
\author{Erik A. Petigura\altaffilmark{8,14}}
\author{Joshua E. Schlieder\altaffilmark{7}}
\author{Rachel L. Akeson\altaffilmark{6}}
\author{Konstantin Batygin\altaffilmark{8}}
\author{Jessie L.~Christiansen\altaffilmark{6}}
\author{David R. Ciardi\altaffilmark{6}}
\author{Justin R. Crepp\altaffilmark{15}}
\author{Erica J. Gonzales\altaffilmark{5,15}}
\author{Kevin Hardegree-Ullman\altaffilmark{6,16}}
\author{Lea A. Hirsch\altaffilmark{1}}
\author{Molly Kosiarek\altaffilmark{5,15}}
\author{Lauren M. Weiss\altaffilmark{10,17}}
\altaffiltext{1}{Astronomy Department, University of California, Berkeley, CA 94720, USA}
\altaffiltext{2}{{\tt dressing@berkeley.edu}}
\altaffiltext{3}{Institute for Astronomy, University of Hawai`i at M\={a}noa, Honolulu, HI 96822, USA}
\altaffiltext{4}{Cahill Center for Astrophysics, California Institute of Technology, Pasadena, CA 91125, USA}
\altaffiltext{5}{National Science Foundation Graduate Research Fellow}
\altaffiltext{6}{IPAC-NExScI, Mail Code 100-22, Caltech, 1200 E. California Blvd., Pasadena, CA 91125, USA}
\altaffiltext{7}{NASA Goddard Space Flight Center, Greenbelt, MD 20771, USA}
\altaffiltext{8}{Division of Geological \& Planetary Sciences, California Institute of Technology, Pasadena, CA 91125, USA}
\altaffiltext{9}{Jet Propulsion Laboratory, California Institute of Technology, USA}
\altaffiltext{10}{Institut de Recherche sur les Exoplan\`{e}tes, Universit\'{e} de Montr\'{e}al, Canada}
\altaffiltext{11}{Department of Physics, and Kavli Institute for Astrophysics and  Space  Research,  Massachusetts Institute  of  Technology, Cambridge, MA 02139, USA}
\altaffiltext{12}{IPAC, Mail Code 314-6, Caltech, 1200 E. California Blvd., Pasadena, CA 91125}
\altaffiltext{13}{Department of Astronomy, The University of Tokyo, 7-3-1 Hongo, Bunkyo-ku, Tokyo 113-0033, Japan}
\altaffiltext{14}{NASA Hubble Fellow}
\altaffiltext{15}{Department of Astronomy and Astrophysics, University of California, Santa Cruz, CA 95064, USA}
\altaffiltext{16}{Department of Physics and Astronomy, University of Toledo, Toledo, OH, 43606, USA}
\altaffiltext{17}{Trottier Fellow}
\vspace{0.5\baselineskip} 
\date{\today}
\slugcomment{Accepted to the Astronomical Journal on June 25, 2018}

\begin{abstract}
K2-55b is a Neptune-sized planet orbiting a K7 dwarf with a radius of $0.715^{+0.043}_{-0.040}\rsun$, a mass of $0.688\pm0.069\msun$, and an effective temperature of $4300^{+107}_{-100}$~K. Having characterized the host star using near-infrared spectra obtained at IRTF/SpeX, we observed a transit of K2-55b with \emph{Spitzer/IRAC} and confirmed the accuracy of the original \emph{K2} ephemeris for future follow-up transit observations. Performing a joint fit to the \emph{Spitzer/IRAC} and \emph{K2} photometry, we found a planet radius of $4.41^{+0.32}_{-0.28}\rearth$, an orbital period of $2.84927265_{-6.42\times10^{-6}}^{+6.87\times10^{-6}}$~days, and an equilibrium temperature of roughly 900K. We then measured the planet mass by acquiring twelve radial velocity (RV) measurements of the system using HIRES on the 10~m Keck I Telescope. Our RV data set precisely constrains the mass of K2-55b to $43.13^{+5.98}_{-5.80}\mearth$, indicating that K2-55b has a bulk density of $2.8_{-0.6}^{+0.8}$~g~cm$^{-3}$ and can be modeled as a rocky planet capped by a modest H/He envelope ($M_{\rm envelope} = 12\pm3\% M_p$). K2-55b is denser than most similarly sized planets, raising the question of whether the high planetary bulk density of K2-55b could be attributed to the high metallicity of K2-55. The absence of a substantial volatile envelope despite the large mass of K2-55b poses a challenge to current theories of gas giant formation. We posit that K2-55b may have escaped runaway accretion by migration, late formation, or inefficient core accretion or that K2-55b was stripped of its envelope by a late giant impact.   \end{abstract}


\maketitle

\section{Introduction}
The NASA \emph{K2} mission is continuing the legacy of the original \emph{Kepler} mission by using the \emph{Kepler} spacecraft to search for transiting planets orbiting roughly $10,000 - 30,000$ stars in multiple fields along the ecliptic. Although restricted to the ecliptic plane by pointing requirements emplaced by the loss of a second reaction wheel in May~2013, \emph{K2} has the freedom to observe a wider variety of stars than the original \emph{Kepler} mission because the field of view changes every few months  \citep{putnam+wiemer2014, howell_et_al2014}. The \emph{K2} target lists are entirely community-driven and Guest Observer proposers have seized the opportunity to study planets and stars in diverse settings.  \emph{K2} has already probed multiple star clusters and is surveying stars with a diverse array of ages, metallicities, and masses. Low-mass stars are particularly well-represented among \emph{K2} targets: 41\% of selected Guest Observer targets are expected to be M and K~dwarfs \citep{huber_et_al2016}. 

The selection bias toward smaller stars is driven by the dual desires to probe stellar habitable zones and to detect small planets. Although the brief, roughly 80-day duration of each \emph{K2} Campaign window is too short to detect multiple transits of planets in the habitable zones of Sun-like stars, the window is just long enough to search for potentially habitable planets orbiting cool stars. Furthermore, the deeper transit depths of planets orbiting smaller stars increase the likelihood that \emph{K2} will be able to detect small planets using only short segments of data with relatively few transits. 

As of 2018 March~28, the \emph{K2} mission had already enabled the detection of 480~planet candidates and 262~confirmed planets \citep[NASA Exoplanet Archive K2 Candidates Table, ][]{akeson_et_al2013}. In this paper, we concentrate on the confirmed planet K2-55b, a Neptune-sized planet orbiting a moderately bright late K~dwarf ($V = 13.546$, $Ks = 10.471$). Compared to a typical \emph{K2} confirmed planet, K2-55b is larger ($4.38^{+0.29}_{-0.25}\rearth$ versus the median radius of $2.3\rearth$) and has a much shorter orbital period ($2.84927265_{-6.42\times10^{-6}}^{+6.87\times10^{-6}}$~d compared to the median value of 7.9~d). The host star \mbox{K2-55} (EPIC~205924614) is much cooler ($4300^{+107}_{-100}$~K versus 5476~K) and slightly smaller ($0.715^{+0.043}_{-0.040}\rsun$ versus $0.87\rsun$) than the average host star of a \emph{K2} confirmed planet. At [Fe/H]~$=0.376\pm0.095$, K2-55 is also one of the more metal-rich stars targeted by \emph{K2}. 

The high metallicity of K2-55 presents a convenient opportunity to test how stellar metallicity, which we assume to be a proxy for the initial metal content in the protoplanetary disk, influences the formation and evolution of planetary systems. Accordingly, the primary objective of this paper is to determine the bulk density of K2-55b and investigate possible compositional models. 

Adventageously, measuring the mass of Neptune-sized planets like K2-55b also provides a way to probe the critical core mass required to commence runaway accretion and form giant planets. For larger planets, degeneracies in interior structure models typically thwart attempts to approximate core masses unless they can be inferred indirectly \citep[e.g., via eccentricity measurements,][]{batygin_et_al2009, kramm_et_al2012, becker+batygin2013, buhler_et_al2016, hardy_et_al2017}. Our secondary goal for this paper is therefore to use K2-55b as a test case for investigating the formation of massive planets. 

We begin by reviewing the discovery, validation, and system characterization of K2-55b in Section~\ref{sec:discovery}. Next, we describe our new \emph{Spitzer} and \emph{Keck/HIRES} observations of K2-55 in Section~\ref{sec:observations} and analyze them in Section~\ref{sec:phot} and \ref{sec:rv}, respectively. We then discuss the implications of our bulk density estimate for the composition and formation of K2-55b in Section~\ref{sec:discussion} before concluding in Section~\ref{sec:conclusions}.

\section{The Discovery of K2-55b}
\label{sec:discovery}
\subsection{\emph{K2} Observations of K2-55}
K2-55 (EPIC~205924614) was observed by the NASA \emph{K2} mission during Campaign 3, which extended from 2014 November 14 until 2015 February 3. Like the majority of \emph{K2} targets, \mbox{K2-55} was observed in long-cadence mode using 30-minute integrations. The \emph{K2} photometry of \mbox{K2-55} is publicly available on MAST.\footnote{\url{https://archive.stsci.edu/canvas/k2hlsp_plot.html?k2=205924614&c=3}}

Although subsequent spectroscopic analyses have revealed that \mbox{K2-55} is a dwarf star, the target was initially proposed by Dennis Stello on behalf of the KASC Working Group~8, the astroSTEP and APOKASC collaborations, and the GALAH team. Interestingly, K2-55 was not included in guest observer proposals focused on dwarf stars. For more details about the inclusion or exclusion of K2-55 in various K2 guest observer proposals, see Appendix \ref{sec:appendix}.

\subsection{Detection and Validation of K2-55b}
The \emph{K2} mission does not provide official lists of planet candidates, but \mbox{K2-55b} was detected by multiple teams using independent pipelines. The candidate was initially reported by \citet{vanderburg_et_al2016} as a $4.4\rearth$ planet in a 2.8-day orbit around a 4237K star with a radius of roughly $0.65\rsun$. \citet{vanderburg_et_al2016} calculated the stellar properties using the V-K color-temperature relation from \citet{boyajian_et_al2013} and flagged the star as a possible giant.

\citet{schmitt_et_al2016} also reported the discovery of \mbox{K2-55b} as PHOI-3~b, a transiting planet with a planetary/stellar radius ratio of $0.0574^{+0.0032}_{-0.0010}$ and an orbital period of 2.8~days. \citet{schmitt_et_al2016} did not characterize the host star and therefore did not report a physical planet radius for PHOI-3~b. They did obtain Keck/NIRC2 imaging to search for nearby stellar companions and reported a lack of stellar companions between $0\farcs25$ and $2\farcs00$ from the target with sensitivities of $\Delta m = 4.00$ and $\Delta m = 6.07$, respectively.

\mbox{K2-55b} was also detected by \citet{barros_et_al2016}, who reported transit events with a depth of 0.372\% and a total duration of 2.093~hr, and by \citet{crossfield_et_al2016}. In addition to re-discovering the planet, \citet{crossfield_et_al2016} used the {\tt VESPA} framework \citep{morton2012, morton2015} to validate \mbox{K2-55b} as a bona fide planet with a radius of $3.82 \pm 0.32 \rearth$. The \citet{crossfield_et_al2016} false positive analysis incorporated $K$-band high contrast imaging acquired with Keck/NIRC2 and high-resolution spectra obtained with Keck/HIRES that restricted the possibility of stellar blends. Specifically, the AO imagery ruled out the presence of stars $\Delta m_{Ks} = 8$ fainter than \mbox{K2-55} at a separation of $0\farcs5$ and  $\Delta m_{Ks} = 9$ fainter at a separation of $1''$. Similarly, a spectroscopic search for secondary stellar lines in the Keck/HIRES spectra \citep{kolbl_et_al2015} placed a limit of 1\% on the brightness of any secondary stars within $0\farcs4$. Overall, \citet{crossfield_et_al2016} calculated a false positive probability (FPP) of $1.7\times 10^{-9}$, well below their adopted validation threshold of ${\rm FPP} < 1\%$. 

\subsection{Stellar Classification}
In their analysis, \citet{crossfield_et_al2016} assumed $R_\star = 0.630\pm 0.050\rsun$,  $M_\star = 0.696\pm 0.047\msun$, and $T_{\rm eff} = 4456 \pm 148$~K. These initial estimates were based on the optical and near-infrared photometry available in the Ecliptic Plane Input Catalog \citep[EPIC,][]{huber_et_al2016}.

\citet{martinez_et_al2017} and \citet{dressing_et_al2017a} later revised the classification of \mbox{K2-55} by acquiring near-infrared spectra at NTT/SOFI ($R \approx 1000$) and IRTF/SpeX ($R \approx 2000$), respectively. \citet{dressing_et_al2017a} classified the star as a K7 dwarf with \mbox{$R_\star = 0.715_{-0.040}^{+0.043}\rsun$},  \mbox{$M_\star =0.688 \pm 0.015 \msun$}, and $T_{\rm eff} = 4300_{-100}^{+107}$~K. \citet{martinez_et_al2017} reported consistent but less precise parameters of \mbox{$R_\star = 0.769 \pm 0.063\rsun$},  $M_\star =0.785 \pm 0.059$, and \mbox{$T_{\rm eff} = 4240 \pm 259$~K}. These temperature constraints are consistent with the estimate of 4422K from Gaia DR2 \citep{gaia_et_al2016, gaia_et_al2018}. For the remainder of this paper, we adopt the stellar classification from \citet{dressing_et_al2017a} with the larger mass error of $\pm0.069\msun$ reported by \citet{dressing_et_al2017b}. Note that this revised stellar radius is 13\% larger than the value used in \citet{crossfield_et_al2016}, suggesting that the planet is larger than previously reported by \citet{crossfield_et_al2016}.

\subsection{Improved Transit Parameters}
After classifying cool dwarfs hosting \emph{K2} candidate planetary systems in \citet{dressing_et_al2017a}, we combined our revised stellar classifications with new transit fits of the \emph{K2} photometry to produce a catalog of planet properties for \emph{K2} cool dwarf systems. As explained in \citet{dressing_et_al2017b}, we estimated the planet properties by using the {\tt BATMAN Python} package \citep{kreidberg2015} to generate a transit models based on the formalism presented in \citet{mandel+agol2002}. We then estimated the errors on planet properties by running a Markov Chain Monte Carlo analysis using the {\tt emcee Python} package \citep{foreman-mackey_et_al2013, goodman+weare2010}. 

During the transit analysis, we varied the orbital period ($P$), the time of transit ($T_C$), the planet-to-star radius ratio ($R_p/R_\star$), the scaled semi-major axis ($a/R_\star$), the inclination ($i$), the eccentricity ($e$), the longitude of periastron ($\omega$), and two quadratic limb darkening parameters ($u_1$ and $u_2$). We fit for $\sqrt{e}\cos \omega$ and $\sqrt{e}\sin \omega$ to increase the efficiency of sampling low-eccentricity orbits \citep[e.g.,][]{eastman_et_al2013} and projected the limb darkening parameters into the $q_1 - q_2$ coordinate-space proposed by \citet{kipping2013}. We also incorporated our knowledge of the stellar density by including a prior on the scaled semi-major axis \citep{seager+mallen-ornelas2003, sozzetti_et_al2007, torres_et_al2008}.

In order to reduce the likelihood of systematic biases in our planet properties, we fit the \emph{K2} photometry returned by three different data reduction pipelines. First, we analyzed the photometry returned by the K2SFF pipeline \citep{vanderburg+johnson2014, vanderburg_et_al2016} and found a planet/star radius ratio of $R_p/R_\star = 0.056_{-0.001}^{+0.002}$. Next, we re-fit the transit parameters using photometry reduced with the K2SC pipeline \citep{aigrain_et_al2016} and the k2phot pipeline \citep{petigura_et_al2015}. In both cases, we found consistent planet/star radius ratios of $R_p/R_\star =0.056_{-0.001}^{+0.002}$ and $R_p/R_\star =0.055_{-0.001}^{+0.002}$, respectively. 

All of these values are in agreement with the previous estimate of $R_p/R_\star = 0.0552 \pm 0.0013$ \citep{crossfield_et_al2016}, which was based on fits to the k2phot photometry. \citet{vanderburg_et_al2016} and \citet{schmitt_et_al2016} found larger (but also consistent) values of $R_p/R_\star = 0.05814$ and $R_p/R_\star = 0.0574^{+0.0032}_{-0.0019}$, respectively.

Combining the stellar radius of $R_\star = 0.715_{-0.040}^{+0.043}\rsun$ \citep{dressing_et_al2017a} with the planet-to-star radius ratio of $R_p/R_\star = 0.056_{-0.001}^{+0.002}$ yields a planet radius of $4.38^{+0.29}_{-0.25} \rearth$ \citep{dressing_et_al2017b}. Our estimate is consistent with the radius of $4.63\pm0.40\rearth$ estimated by \citet{martinez_et_al2017}, but significantly larger than the value of $3.82 \pm 0.32 \rearth$ found by \citet{crossfield_et_al2016}. We attribute the planet radius discrepancy to differences in the assumed stellar radius; the revised estimates determined by \citet{martinez_et_al2017} and \citet{dressing_et_al2017a,dressing_et_al2017b} were larger than the value assumed by \citet{crossfield_et_al2016}. 

\section{Observations}
\label{sec:observations}

\subsection{Spitzer/IRAC Photometry}
\label{ssec:spitzer}
In order to refine the transit ephemeris estimated from the \emph{K2} data, we observed an additional transit of K2-55b using the Infrared Array Camera (IRAC) on the \emph{Spitzer Space Telescope} (GO~11026, PI~Werner). We began monitoring K2-55 at BJD = 2457430.636 (February 12, 2016) and collected data points every 12~s until BJD=2457430.891 for a total observation period of 6.1~h. Based on our previous analysis of the \emph{K2} photometry, we expected that K2-55b would begin transiting 2~h into the requested observation window and finish egress 1.9~h later. Our planned observation therefore included 4.2~h of out-of-transit flux baseline to aid our analysis of the transit event. 

Prior to beginning our science observations, we obtained 30~min of ``pre-observation'' data to allow the telescope temperature to stabilize after slewing from the preceding target \citep{grillmair_et_al2012}. We conducted these pre-observations in peak-up mode using the Pointing Calibration and Reference Sensor to improve the positioning of \mbox{K2-55} during our science observations. For both sets of observations, we elected to conduct observations in Channel~2 (4.5~$\mu$m) rather than Channel~1 (3.6~$\mu$m) due to the lower amplitude of intra-pixel sensitivity variations visible in Channel~2 data \citep{ingalls_et_al2012}. 

\subsection{Keck/HIRES Radial Velocities}
Between 12~August~2016 and 25~December 2016, we obtained twelve~observations of K2-55 using the High Resolution Echelle Spectrometer \citep[HIRES,][]{vogt_et_al1994} on the 10~m Keck I Telescope on the summit of Maunakea. HIRES is a slit-fed spectrograph and a demonstrated single measurement precision of approximately 1.5~m~s$^{-1}$ for observations with SNR of 200 and 1~m~s$^{-1}$  for SNR of 500 \citep{fischer_et_al2016}. Although the spectrometer has a wavelength range of \mbox{364--797}~nm, we restricted our radial velocity analysis to the \mbox{510--620}~nm region covered by the iodine reference cell, which was mounted in front of the spectrometer entrance slit for all of our radial velocity observations. Following standard California Planet Search (CPS) procedures \citep{howard_et_al2010b}, we obtained our radial velocity observations using the ``C2'' decker ($0\farcs87 \times 14"$ slit) for a spectral resolution of 55,000. We terminated the exposures after 45~minutes giving \mbox{SNR pixel$^{-1}$ = 60--90} near 550 nm, depending on sky conditions.

On 22~September 2016, we also obtained a higher resolution ``template'' observation with the iodine cell removed to aid in the process of disentangling the stellar and iodine spectra. Our template observation was taken using the ``B3'' decker ($0\farcs57 \times 14"$ slit) to reach a higher resolution of roughly 70,000. As in previous CPS publications, we determined RVs by forward-modeling the iodine-free template spectra, a high-quality iodine transmission spectrum, and the instrumental response \citep{marcy+butler1992, valenti_et_al1995, butler_et_al1996, howard_et_al2009}. We present the measured RVs and uncertainties in Table~\ref{tab:rvs}.

\begin{deluxetable}{cccc}
\tablecolumns{4}
\tabletypesize{\normalsize}
\tablecaption{Relative Radial Velocities\label{tab:rvs}}
\tablehead{
\multicolumn{2}{c}{Observation Date} & 
\multicolumn{2}{c}{Radial Velocity (m~s$^{-1}$)}\\
\colhead{BJD-2450000} & 
\colhead{UT} &
\colhead{Value} &
\colhead{Error}
}
\startdata
7612.873042 & 08/12/2016 & 3.66 & 1.89 \\
7614.003359 & 08/13/2016 & -19.49 & 1.88 \\
7651.986215 & 09/20/2016 & а26.53 & 2.2 \\
7668.943278 & 10/07/2016 & а17.93 & 2.06 \\
7678.910917 & 10/17/2016 & -17.57 & 2.63 \\
7679.739888 & 10/18/2016 & -14.18 & 2.06 \\
7697.863996 & 11/05/2016 & а28.79 & 2.17 \\
7713.740959 & 11/21/2016 & -13.72 & 1.83 \\
7718.783696 & 11/26/2016 & -15.18 & 2.24 \\
7745.740906 & 12/23/2016 & а15.45 & 3.26 \\
7746.727362 & 12/24/2016 & 8.9 & 2.51 \\
7747.741953 & 12/25/2016 & -33.2 & 2.98\\
\enddata                  
\end{deluxetable}

\section{Analysis of the Photometry}
\label{sec:phot}
We refined the radius estimate and ephemeris of \mbox{K2-55~b} by fitting the transits observed by \emph{Spitzer} and \emph{K2}. Having already fit the \emph{K2} data separately in \citet{dressing_et_al2017b}, we began this analysis by considering the \emph{Spitzer} data alone. We then conducted a simultaneous fit of both the \emph{Spitzer} photometry and the \emph{K2} photometry to further constrain the properties of the planet. 

\subsection{Generating Light Curves from \emph{Spitzer} Data}
We considered a variety of fixed and variable apertures when extracting the photometric light curves from the \emph{Spitzer} observations. Our investigation was motivated by previous \emph{Spitzer} analyses demonstrating that a wise choice of extraction aperture can minimize the scatter and red-noise component of the resulting residuals \citep{knutson_et_al2012, lewis_et_al2013, todorov_et_al2013, kammer_et_al2015, benneke_et_al2017}. As in earlier studies, our full photometry extraction procedure included determining and removing the sky background, estimating the position of the star on the detector array using flux-weighted centroiding, and then summing the total flux within particular circular apertures.

In the fixed case, we tested 36~aperture radii spanning the range between 1.5~pixels and 5~pixels at 0.1~pixel spacing. When exploring time variable apertures, we began by determining the scaling of the noise pixel parameter $\beta = \left(\sum_n I_n\right)^2 / \left(\sum_n I_n^2 \right)$, where $I_n$ is the intensity measured in pixel $n$ \citep{mighell2005}. We then rescaled the noise pixel aperture radius as $r = a\sqrt{\beta} + c$, where we considered scaling factors $0.6 \leq a \leq 1.2$ and shifts $-0.8 \leq c \leq 0.4$. 

We also investigated whether binning the data before fitting would improve performance. For each choice of aperture, we generated eight binned versions of K2-55 photometry using between 2 and 9 points per bin for effective integration times of $24-108$~s.  We analyzed these data sets along with the unbinned data.  

Finally, we experimented with trimming the data. As noted by \citet{ge_et_al2018}, proper trimming of pre- and post-transit data can improve the quality of fits to \emph{Spitzer} data exhibiting curved systematics. We considered 21 possible trim durations ranging between 0 hours and 1 hour at either end of the light curve and allowed the ending trim duration and the starting trim duration to assume different values. 

We selected the ideal binning, aperture, and pre- and post-transit trimming by fitting the \emph{Spitzer} light curve using the full range of parameter choices and inspecting each fit. After extracting and trimming each light curve, we fit the systematics and transit signal as described in Section~\ref{ssec:pld}, re-binned the residuals in progressively larger bins, checked the scaling of the noise with increasing bin size, and assessed how well each fit reproduced the expected square-root noise scaling. We performed this initial parameter exploration using an eccentric model for the orbit of K2-55b ($e = 0.125$, $\omega = 196^\circ$) but the estimated transit properties are nearly identical for eccentric and circular orbits (see Section~\ref{ssec:jointfit} and Table~\ref{tab:joint}).

\begin{figure}[tbp]
\centering
\includegraphics[width=0.5\textwidth]{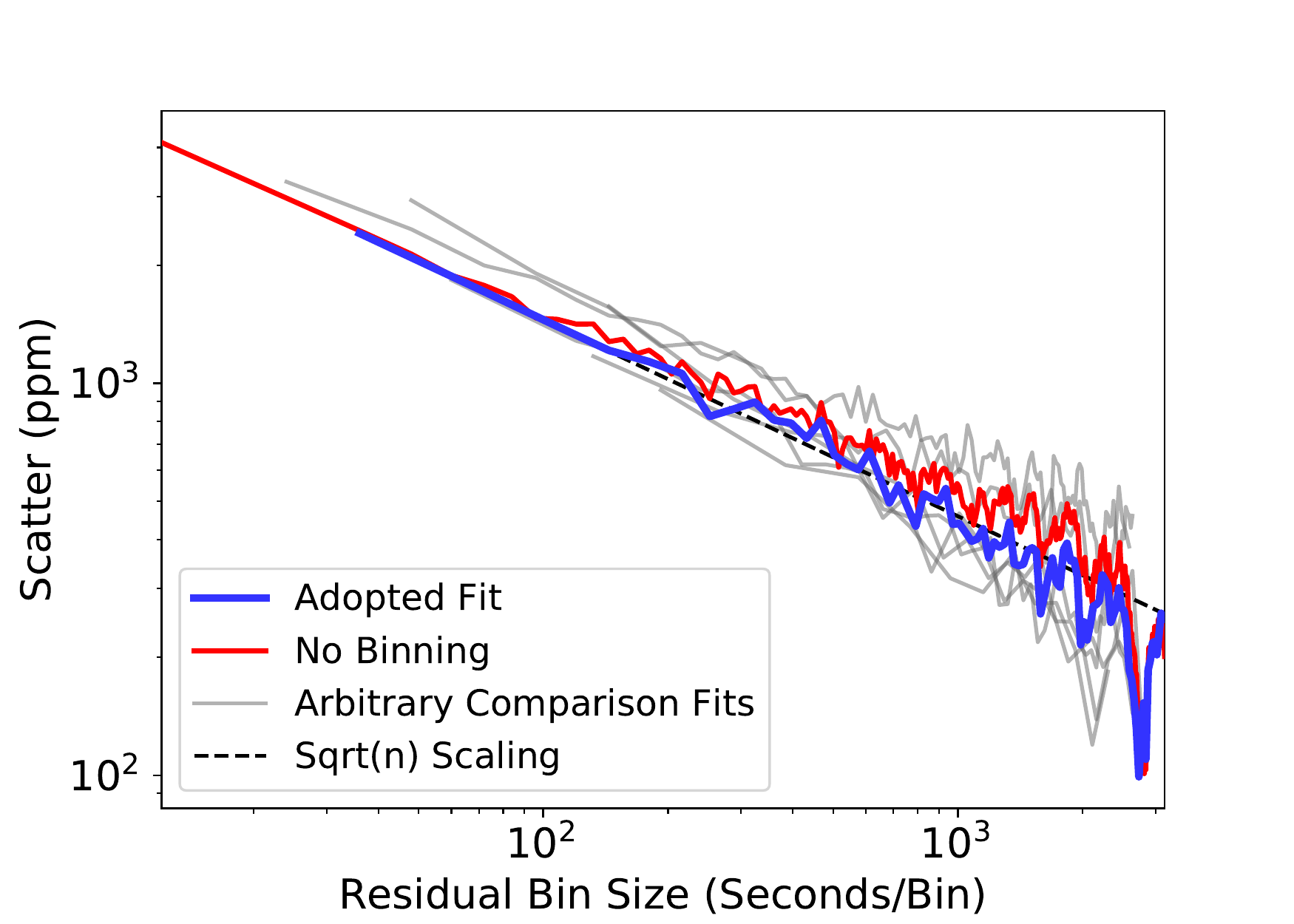}
\caption{Performance of various fits to the \emph{Spitzer} photometry compared to the expected square-root noise scaling (black dashed line). Each line displays the RMS of the residuals after fitting the light curve and rebinning the residuals to bins spanning a certain number of seconds. The adopted fit (thick blue line) has both the lowest red noise and the lowest scatter. For comparison, the red line shows the performance of a fit using the same aperture and trimming but no pre-fit binning and the gray lines show the performance of twenty other reductions of the photometry. The fits shown in light gray use different apertures, pre-fit binning, and trim durations.}
\label{fig:pick_aperture}
\end{figure}

For the remainder of the paper, we investigate the light curve produced using the best combination of fit parameters: a fixed aperture radius of 2.7 pixels, a binning of 3 points per bin (36 seconds per bin), a starting trim of 0.3 hours, and no trimming at the end of the light curve. As shown in Figure~\ref{fig:pick_aperture}, this light curve has the lowest red-noise component and the lowest scatter of all of the light curves we considered. 

\subsection{Fitting the {Spitzer} Data}
\label{ssec:pld}
We analyzed our \emph{Spitzer} data using the Pixel-Level Decorrelation (PLD) technique first introduced by \citet{deming_et_al2015} and later modified by \citet{benneke_et_al2017}. Specifically, we modeled the observed flux $D(t_i)$ at each timestamp $t_i$ as the multiplicative combination of a sensitivity function $S(t_i)$ and a transit model $f(t_i)$. We then maximized the likelihood
\begin{equation}
\mathcal{L}  = \sum_{i=1}^N \frac{1}{\sqrt{2 \pi \sigma^2}}\exp\left(-\frac{(D(t_i) - S(t_i) \cdot f(t_i))^2}{2 \sigma^2}\right) \,,
\end{equation}
where $\sigma$ is a photometric scatter parameter fit simultaneously with $S(t_i)$ and $f(t_i)$. We allowed $\sigma$ to vary between 0.00001 and 0.3. For the instrument model $S(t_i)$, we assumed that the sensitivity can be described by the linear combination of the raw counts $D_k (t_i)$ of each pixel $k$ within a $5\times5$ pixel region centered on the star and a linear ramp with slope $m$:
\begin{equation}
S(t_i)  = \frac{ \sum_{k=1}^{25} w_k D_k(t_i)}{\sum_{k=1}^{25} D_k(t_i)} + m \cdot t_i \, ,
\label{eq:sensitivity}
\end{equation}
where the $w_k$ are the time-independent PLD weights given to each pixel.

We generated the transit model $f(t_i)$ by using the {\tt BATMAN} python package \citep{kreidberg2015} to solve the equations of \citet{mandel+agol2002}. Unlike the \emph{K2} photometry, our \emph{Spitzer} time series contains only a single transit event. We therefore fixed the orbital period to that found by \citet{dressing_et_al2017b} and fit for the transit midpoint $T_0$, planet/star radius ratio $R_p/R_\star$, scaled semi-major axis ratio $a/R_\star$, and orbital inclination $i$. For our adopted model, we assumed that K2-55b had a circular orbit based on our analysis of the RV data (see Section~\ref{sec:rv}), but we note that this choice does not significantly alter the transit profile. We estimated quadratic limb darkening coefficients in the \emph{Spitzer} bandpass by interpolating the values tabulated by \citet{claret+bloemen2011}. Accordingly, we set the coefficients to $u_1 = 0.0824$ and $u_2 = 0.1531$. We restricted the orbital inclination to $70^\circ < i < 90^\circ$ and required that the transit midpoint fall within the \emph{Spitzer} data set. 

In addition to verifying the orbital ephemeris predicted from the \emph{K2} data, our \emph{Spitzer} data also provide an opportunity confirm the depth of the transit event. In Figure~\ref{fig:rrstar}, we compare the planet/star radius ratios estimated from our independent fits to the \emph{K2} and \emph{Spitzer} data. Although we find tighter radius ratio constraints from the \emph{K2} data ($R_p/R_\star = 0.056_{-0.001}^{+0.002}$) than from the \emph{Spitzer} data ($0.0562_{-0.0025}^{+0.0030}$), our results are nearly identical. Table~\ref{tab:joint} contains all of the model parameters from the \emph{Spitzer}-only fit.

\begin{deluxetable*}{cccccc}
\tablecolumns{6}
\tabletypesize{\normalsize}
\tablecaption{Transit and Systematic Parameters from the Photometric Analysis\label{tab:joint}}
\tablehead{
\colhead{} & 
\colhead{} & 
\multicolumn{4}{c}{Model}\\
\cline{3-6}
\colhead{Parameter} & 
\colhead{Units} & 
\colhead{\emph{Spitzer} circular} &
\colhead{\emph{Spitzer}+\emph{K2} circular} &
\colhead{\emph{Spitzer}+\emph{K2} fixed $e$} &
\colhead{\emph{Spitzer}+\emph{K2} variable $e$}
}
\startdata
$T_0$\tablenotemark{a} & d &  ${-6.13\times10^{-5}}_{-0.0012}^{+0.0013}$ & $7.96\times{10^{-5}} \pm 0.00019$ & $-1.83\times{10^{-5}}_{-0.00024}^{+0.00022}$ & $2.27\times {10^{-13}}_{-0.00024}^{+0.00021}$\\
$P$ & d & 2.849274 (fixed) & $2.84927265_{-6.42\times10^{-6}}^{+6.87\times10^{-6}}$ & $2.84927261_{-6.38\times10^{-6}}^{+6.94\times10^{-6}}$ & $2.84927252_{-6.60\times10^{-6}}^{+7.01\times10^{-6}}$     \\
$R_p/R_{\star, K2}$ & $\cdots$ & $\cdots$ & $0.0559_{-0.0012}^{+0.0030}$ &  $0.559_{-0.0011}^{+0.0029}$  & $0.0561_{-0.0013}^{+0.0031}$\\
$R_p/R_{\star, S}$ & $\cdots$  & $0.0562_{-0.0025}^{+0.0030}$ & $0.0557_{-0.0023}^{+0.0022}$ & $0.0557_{-0.0023}^{+0.0022}$ & $0.0557_{-0.0022}^{+0.0023}$\\
$a/R_{\star}$ & $\cdots$ & $9.53_{-3.06}^{+1.54}$ & $10.55_{-1.38}^{+0.64}$  & $10.86_{-1.37}^{+0.64}$  & $10.50_{-1.37}^{+1.14}$\\
$i$ & deg   & $86.82_{-4.07}^{+2.26}$ & $88.05_{-1.75}^{+1.36}$ & $88.17_{-1.62}^{+1.27}$ & $87.98_{-1.70}^{+1.33}$\\
$\sigma_{K2}$ & $\cdots$   & $\cdots$ & $0.000167_{-3.7\times10^{-6}}^{+3.9\times10^{-6}}$  & $0.000167_{-3.7\times10^{-6}}^{+3.9\times10^{-6}}$& $0.000167_{-3.8\times10^{-6}}^{+3.9\times10^{-6}}$\\
$\sigma_S$  & $\cdots$    & $0.0024_{-7.3\times10^{-5}}^{+7.7\times10^{-5}}$ & $0.0024_{-7.3\times10^{-5}}^{+7.8\times10^{-5}}$ & $0.0024_{-7.2\times10^{-5}}^{+7.6\times10^{-5}}$ & $0.0024_{-7.1\times10^{-5}}^{+7.5\times10^{-5}}$\\
$\sqrt e \sin \omega$ & $\cdots$ & $\cdots$ & $\cdots$   & -0.10 (fixed) & $-0.08_{-0.18}^{+0.21}$\\
$\sqrt e \cos \omega$ & $\cdots$ & $\cdots$ & $\cdots$ & -0.34 (fixed) & $-0.29_{-0.09}^{+0.12}$ \\
$e$ & $\cdots$ & $\cdots$ & $\cdots$  & 0.125 (fixed) & $0.127_{-0.055}^{+0.057}$\\
$\omega$ & rad & $\cdots$ & $\cdots$ & -2.86 (fixed)   & $-2.88_{-0.64}^{+0.61}$\\
\enddata      
\tablenotetext{a}{For ease of comparison, we display the time of transit center minus BKJD = 2150.42286667.}            
\end{deluxetable*}

\begin{figure}[tbp]
\centering
\includegraphics[width=0.48\textwidth]{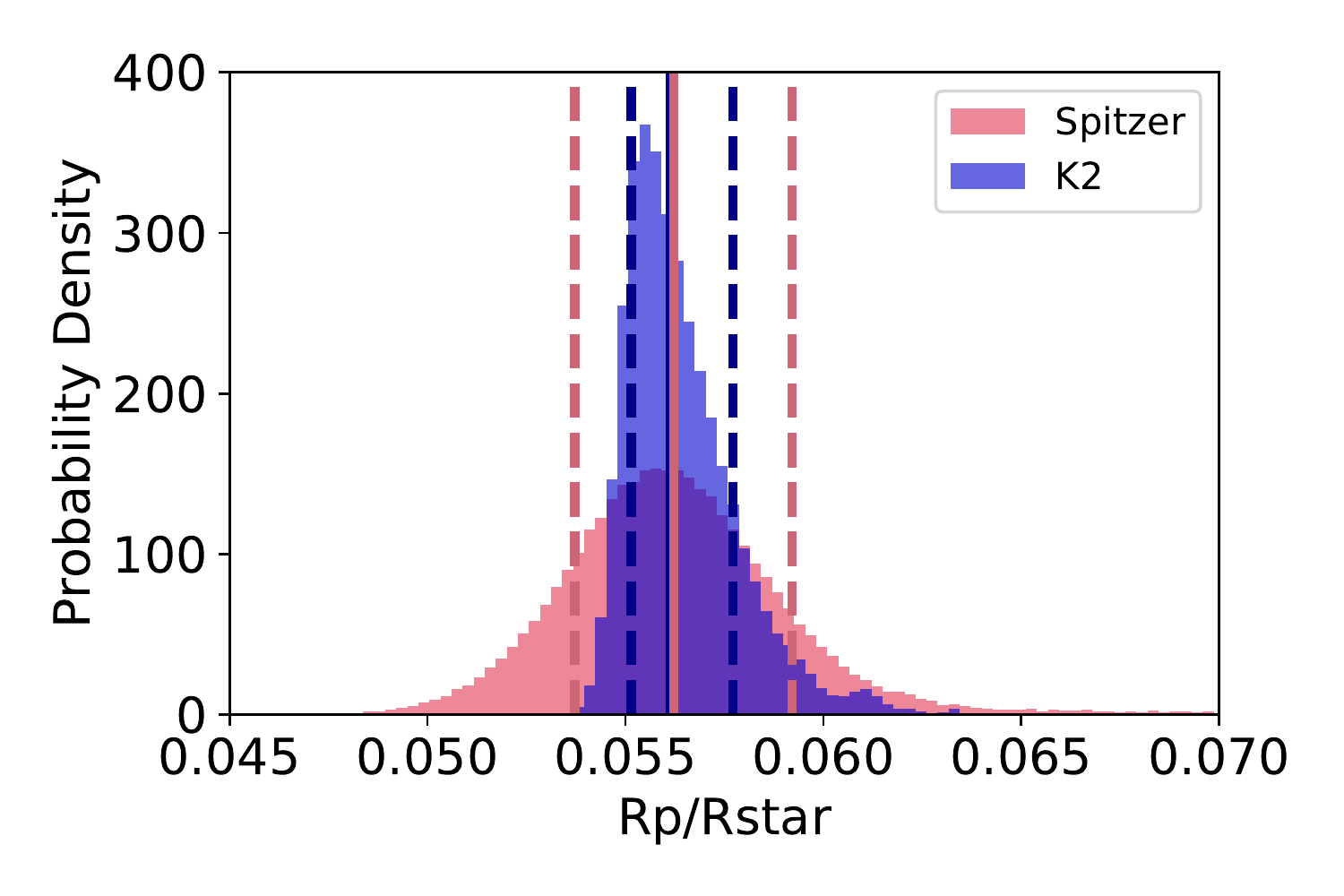}
\caption{Comparison of planet/star radius ratios estimated by fitting the \emph{K2} (blue) and \emph{Spitzer} (coral) data independently. The solid and dashed lines mark the median value and $1\sigma$ errors, respectively. \label{fig:rrstar}}
\label{fig:rrstar}
\end{figure}

\subsection{Fitting the \emph{Spitzer} and \emph{K2} Data Simultaneously}
\label{ssec:jointfit}
After fitting the \emph{Spitzer} photometry separately, we conducted a joint fit of the \emph{Spitzer} and \emph{K2} photometry to further contrain the planet parameters. For our joint fit, we used fixed quadratic limb darkening parameters set by consulting the limb darkening tables in \citet{claret+bloemen2011}. Specifically, we adopted $u_1 = 0.7306$ and $u_2 = 0.0338$ for the \emph{Kepler} bandpass and $u_1 = 0.0824$ and $u_2 = 0.1531$ for the \emph{Spitzer} bandpass. These values are the parameters estimated by \citet{claret+bloemen2011} for a 4250K star with $\log g = 4.5$ and [Fe/H] = 0.3. 

The free parameters in our joint fit were the orbital period $P$, the transit midpoint $T_0$, the planet/star radius ratio in both the \emph{Spitzer} and \emph{K2} bandpasses ($R_p/R_{\star,{Spitzer}}$, $R_p/R_{\star,{\rm K2}}$), the scaled semi-major axis ratio $a/R_\star$, the orbital inclination $i$, and two photometric scatter terms ($\sigma_{Spitzer}$, $\sigma_{\rm K2}$). As for the \emph{Spitzer}-only fit, we assumed a circular orbit for K2-55b based on our analysis of the RV data. For comparison, we repeated the analysis using an eccentric orbit ($e = 0.125$, $\omega = 196^\circ$) and found little variation in the resulting parameters. We also ran a third analysis in which we used the results of our RV analysis to impose Gaussian priors on $e$ and $\omega$ and allowed the parameters to vary. All three fits yield consistent planet properties and $R_p/R_{\star,{Spitzer}}=0.056\pm0.002$ in all cases.

\begin{figure*}[tbp]
\centering
\includegraphics[width=0.49\textwidth]{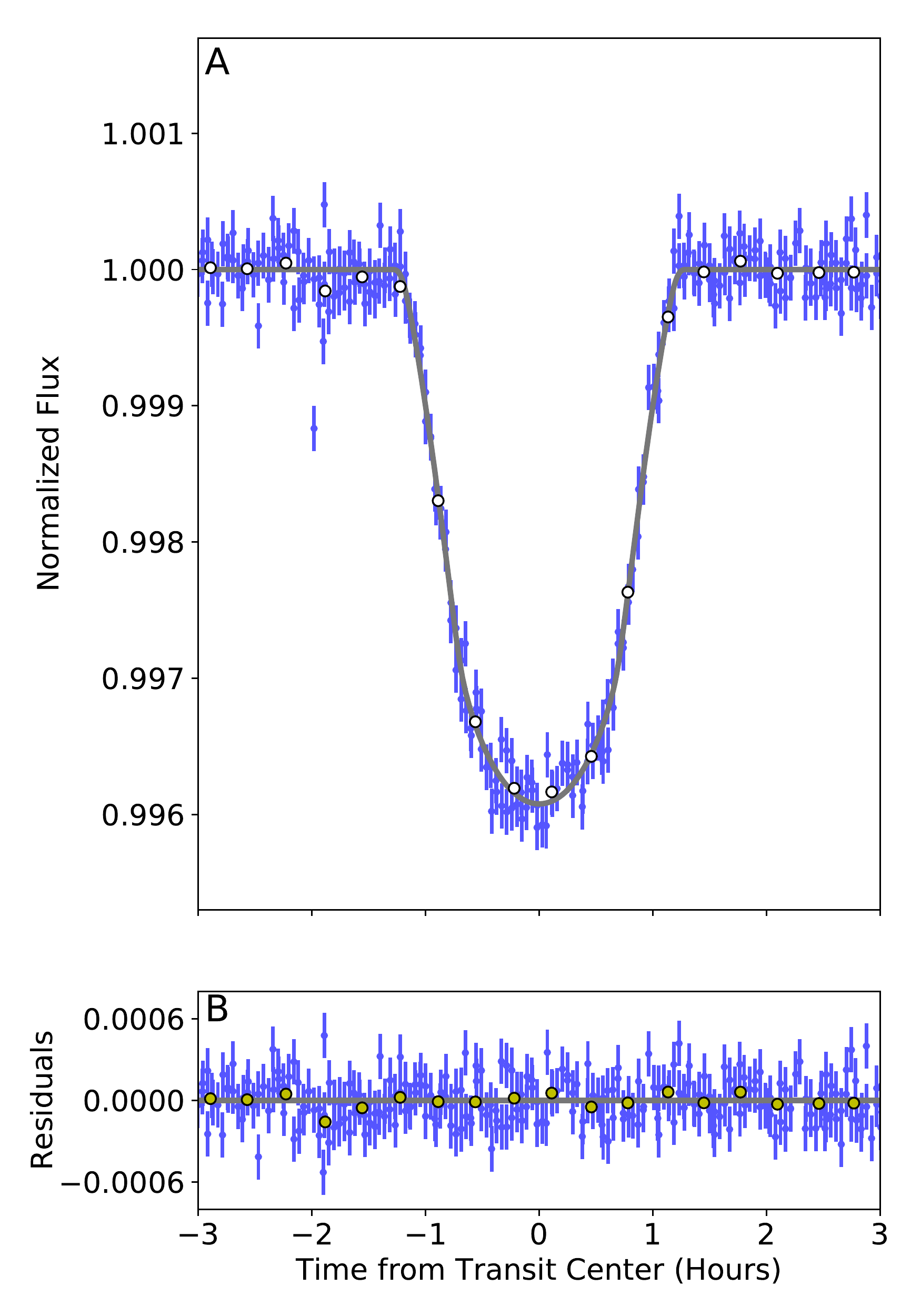}
\includegraphics[width=0.49\textwidth]{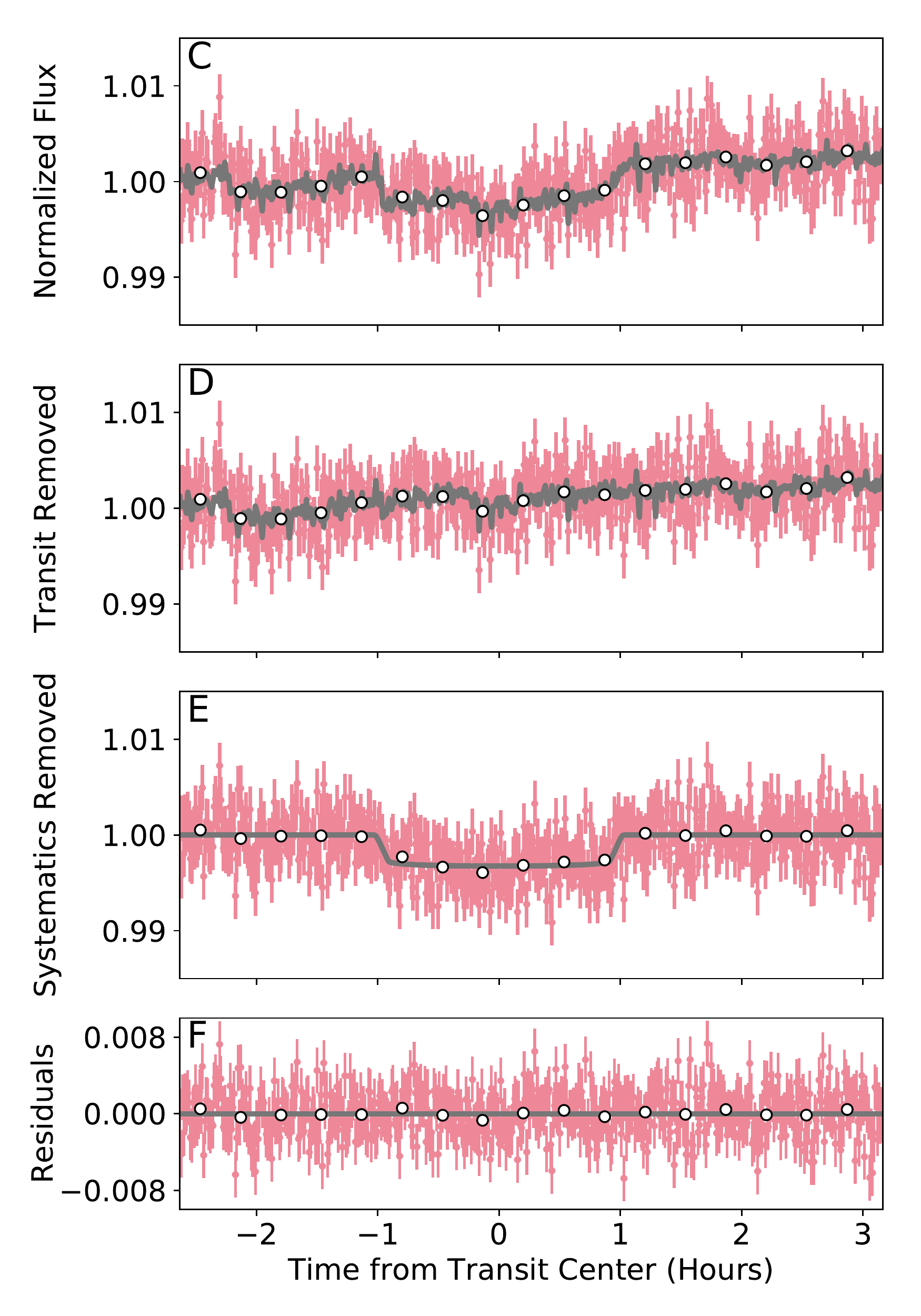}
\caption{Joint fit to the \emph{K2} and \emph{Spitzer} photometry. In all panels, the white points show the data binned to 20-minute increments. The errors on the binned data are smaller than the data points. \emph{Panel A:} Light curve model (gray line) and phase-folded \emph{K2} photometry (blue points) versus time. Note that the transit appears slightly v-shaped due to the relatively long 30-minute integration times used by \emph{K2}. \emph{Panel B: } Residuals to the \emph{K2} fit. \emph{Panel C: } Full light curve model (gray line) versus raw \emph{Spitzer} photometry (red points). \emph{Panel~D:}~Systematics model (gray) versus \emph{Spitzer} photometry after removing the best-fit transit model. \emph{Panel E:} Transit model (gray) versus systematics-corrected \emph{Spitzer} photometry. \emph{Panel F:} Residuals to the full fit. }
\label{fig:tfits}
\end{figure*}

We adopt the circular fit as our chosen model and display the results in Figure~\ref{fig:tfits}. We also summarize the results in Table~\ref{tab:joint}. The residuals to the full fit follow Gaussian distributions with a median value of \mbox{$-1.1 \times 10^{-5}$} and a standard deviation of $0.00017$ for the \emph{K2} data and $0.0001$ and $0.0024$, respectively, for the \emph{Spitzer} data. The primary benefit to analyzing the \emph{Spitzer} data along with the \emph{K2} data is that the errors on the transit mid-point and period decreased by factors of 1.9 and 4.0 compared to analyzing the \emph{K2} data alone. Accordingly, the uncertainty on the transit midpoint for an observation in late 2020 (perhaps by JWST) has decreased from 30 minutes to 7 minutes, significantly reducing the amount of telescope time needed to ensure that the full transit is observed.

We tested the influence of our choice of limb darkening parameters by repeating the variable eccentricity analysis using two different sets of limb darkening parameters. In particular, we considered one set of alternative parameters corresponding to a 4000K star with $\log g = 4.0$ and [Fe/H] = 0.2 ($u_{1,Kepler}  = 0.7858$, $u_{2, Kepler} = -0.0163$, $u_{1,Spitzer}  = 0.0827$, $u_{2, Spitzer} = 0.1443$) and a second set corresponding to a 4500K star with $\log g = 5.0$ and [Fe/H] = 0.5 ($u_{1,Kepler}  = 0.6895$, $u_{2, Kepler} = 0.067$, $u_{1,Spitzer}  = 0.0791$, $u_{2, Spitzer} = 0.1594$). Regardless of our specific choice of limb darkening parameters, we found consistent results for the planet properties. 

\begin{figure*}[tbp]
\centering
\includegraphics[width=0.48\textwidth]{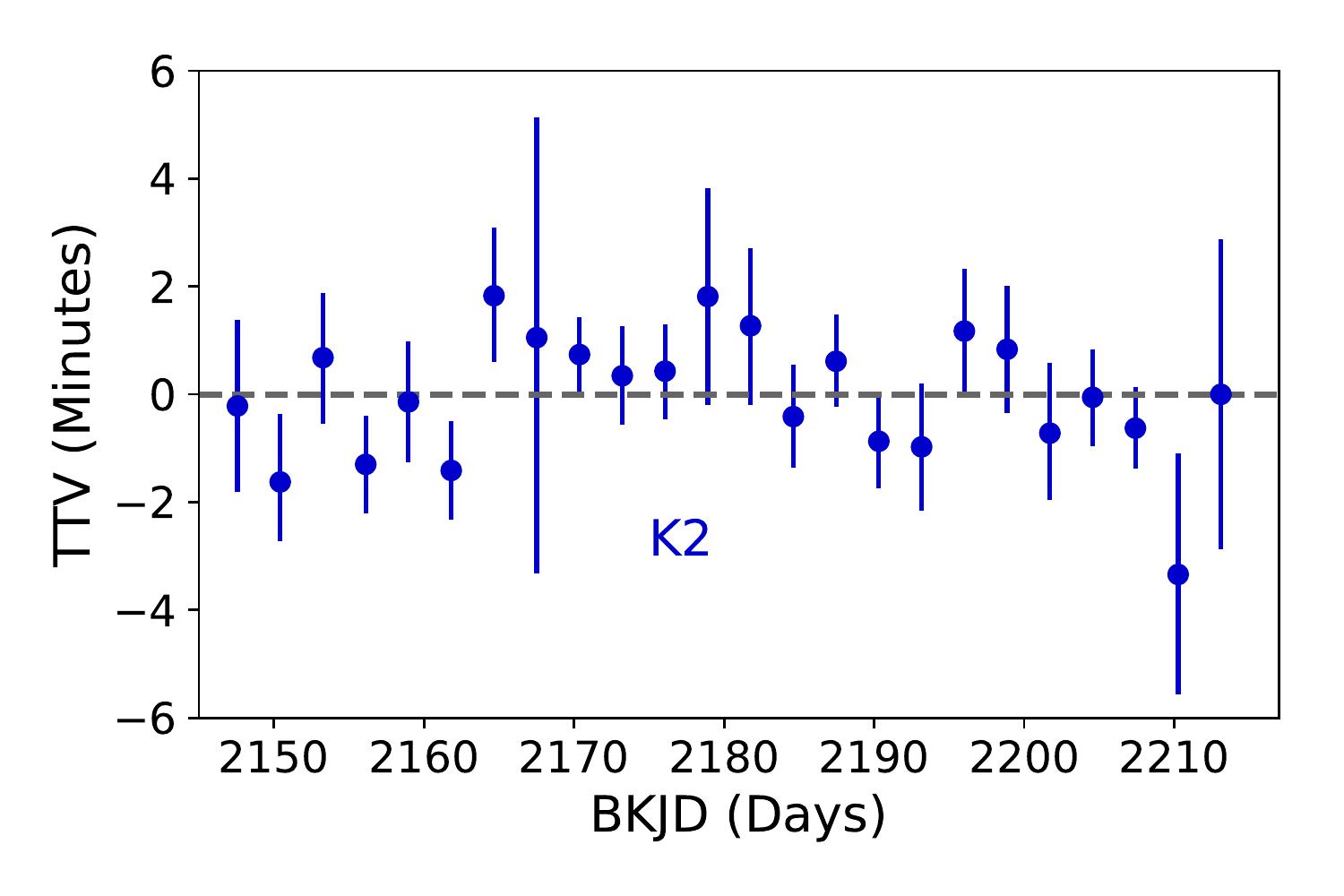}
\includegraphics[width=0.48\textwidth]{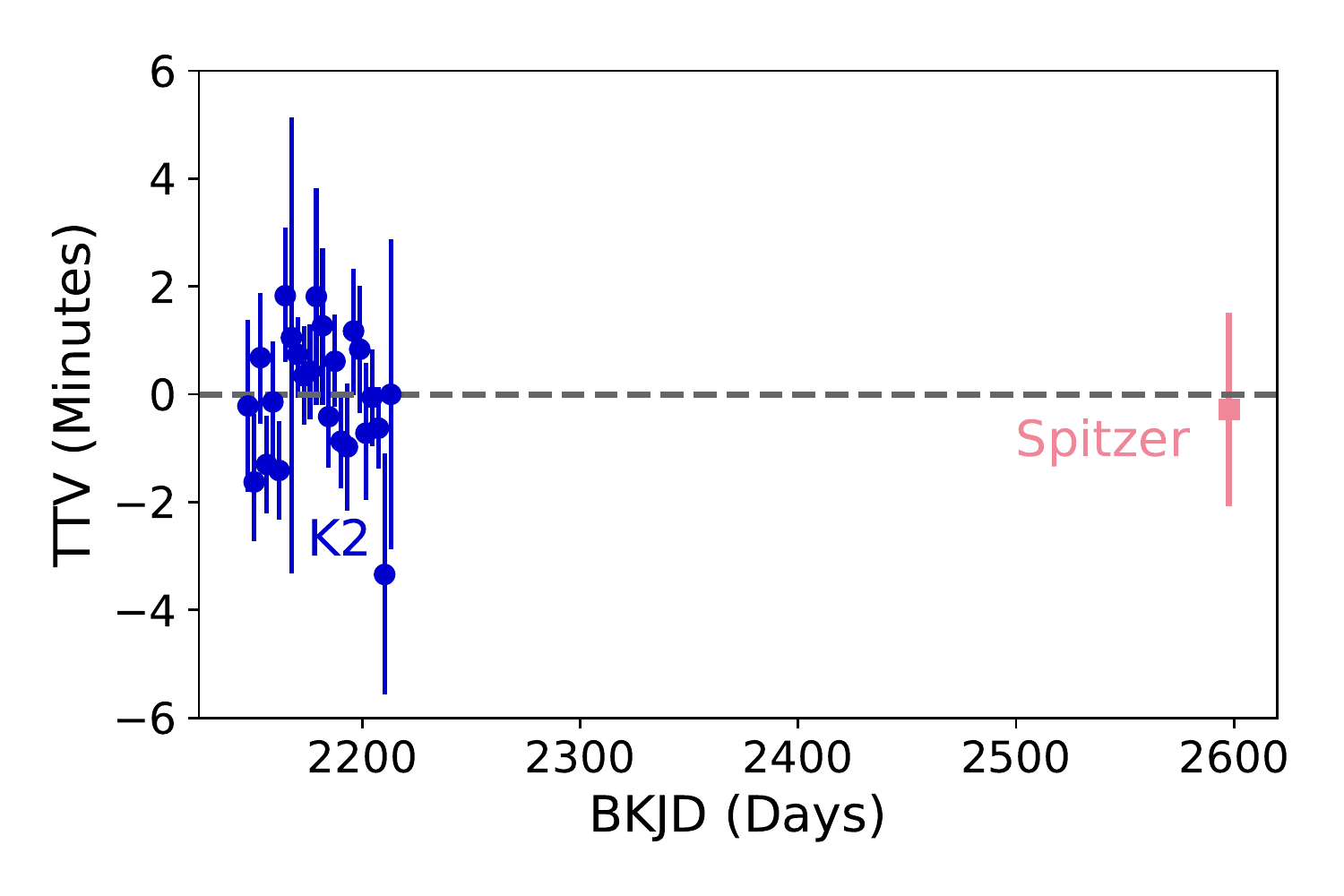}
\caption{Observed transit times of K2-55b relative to the best-fit linear ephemeris provided in Table~\ref{tab:system}. The transit times measured from both the \emph{K2} data (blue circles in both panels) and the \emph{Spitzer} data (coral square, right panel only) are consistent with a linear ephemeris. \emph{Left:} Zoomed-in view of transit times measured from \emph{K2} data. \emph{Right: } All measured transit times.}
\label{fig:ttv}
\vspace{1em}
\end{figure*}

\subsection{Searching for Transit Timing Variations}
Once we had determined the best-fit system parameters, we checked for transit timing variations (TTVs) by inspecting each transit event individually. Specifically, we found the transit midpoints that minimized the difference between the observed data points and the best-fit transit model. We then rescaled the errors so that the reduced $\chi^2$ was equal to unity and slid the transit model along until the $\chi^2$ increased by 1. As shown in Figure~\ref{fig:ttv}, the transit midpoints we measured for the 24~transits visible in the \emph{K2} data are consistent with a linear ephemeris. Although there is a hint of curvature, fitting the transit times with a quadratic ephemeris does not improve the fit enough to justify the introduction of additional free parameters ($\Delta BIC = 10$). Accordingly, we expected that our prediction of the \emph{Spitzer} transit midpoint would be accurate to within a few hours even in the worst case scenario. Indeed, our \emph{Spitzer}-only fit yielded a transit midpoint of \mbox{BJD = 2457430.75882} within one minute ($<1\sigma$) of our predicted value of \mbox{BJD = 2457430.75902.} 

\section{Analysis of the Radial Velocity Data}
\label{sec:rv}
As in other recent CPS publications \citep[e.g.,][]{christiansen_et_al2016, sinukoff_et_al2017a, sinukoff_et_al2017b}, we analyzed the radial velocities using the publicly-available {\tt RadVel} {\tt Python} package\footnote{\url{https://github.com/California-Planet-Search/radvel}} \citep{fulton_et_al2018}. We first performed a maximum-likelihood fit to the RVs and then determined errors by running a Markov-Chain Monte Carlo (MCMC) analysis around the maximum-likelihood solution. When assessing various solutions, we incorporated stellar jitter into the likelihood $\mathcal{L}$ by adopting the same likelihood function as \citet{howard_et_al2014} and \cite{dumusque_et_al2014}:
\begin{equation}
\ln \mathcal{L} = - \sum_i \left[ \frac{\left(v_i - v_m \left(t_i \right)\right)^2}{2 \left(\sigma^2_i + \sigma^2_{sj}\right)} + \ln \sqrt{2 \pi \left (\sigma^2_i + \sigma^2_{sj}\right) }\right] \,
\end{equation}
where the subscript $i$ denotes the individual data points at times $t_i$, $v_i$ are the measured RVs, $v_m (t_i)$ are the modeled RVs, $\sigma_i$ are the instrumental errors on the measured RVs, and $\sigma_{sj}$ is the stellar jitter.

{\tt RadVel} conducts MCMC analyses using the affine-invariant {\tt emcee} sampler \citep{foreman-mackey_et_al2013} and includes built-in tests for convergence. Specifically, we initialized eight ensembles of RadVel runs each containing 100~parallel MCMC chains clustered near the maximum-likelihood solution. To ensure that the chains were well-mixed and properly converged, we discarded the initial segment of each chain as ``burn-in'' and ran the MCMC analysis for at least 1000 additional steps. We then compared the chains across ensembles of RadVel runs and confirmed that they arrived at consistent parameter values. More formally, we tested for converge by computing the Gelman-Rubin potential scale reduction factor $\hat{R}$ \citep{gelman+rubin1992} and requiring that $\hat{R} < 1.01$. In order to compensate for the effects of autocorrelation on parameter estimates, we also required that our chains contained at least 1000 effective independent draws for each parameter as suggested by \citet{ford2006}. 

The \emph{K2} photometry of K2-55 revealed a single transiting planet at an orbital period of $2.85$~days and no evidence for additional transiting planets. Accordingly, we began our RV fits by considering only a single planet on a Keplerian orbit. We then restricted our fits to circular orbits to test whether the additional model complexity of varying $e$ and $\omega$ was warranted by the data. Finally, we experimented with fitting linear and quadratic trends to the data to check for the presence of additional, non-transiting planets in the system. In all cases, we fixed the stellar jitter to the value of $\sigma_j =  5.34$~m~s$^{-1}$ found when fitting the data using a single, eccentric planet. 

\begin{figure*}[tbhp]
\centering
\includegraphics[width=1\textwidth]{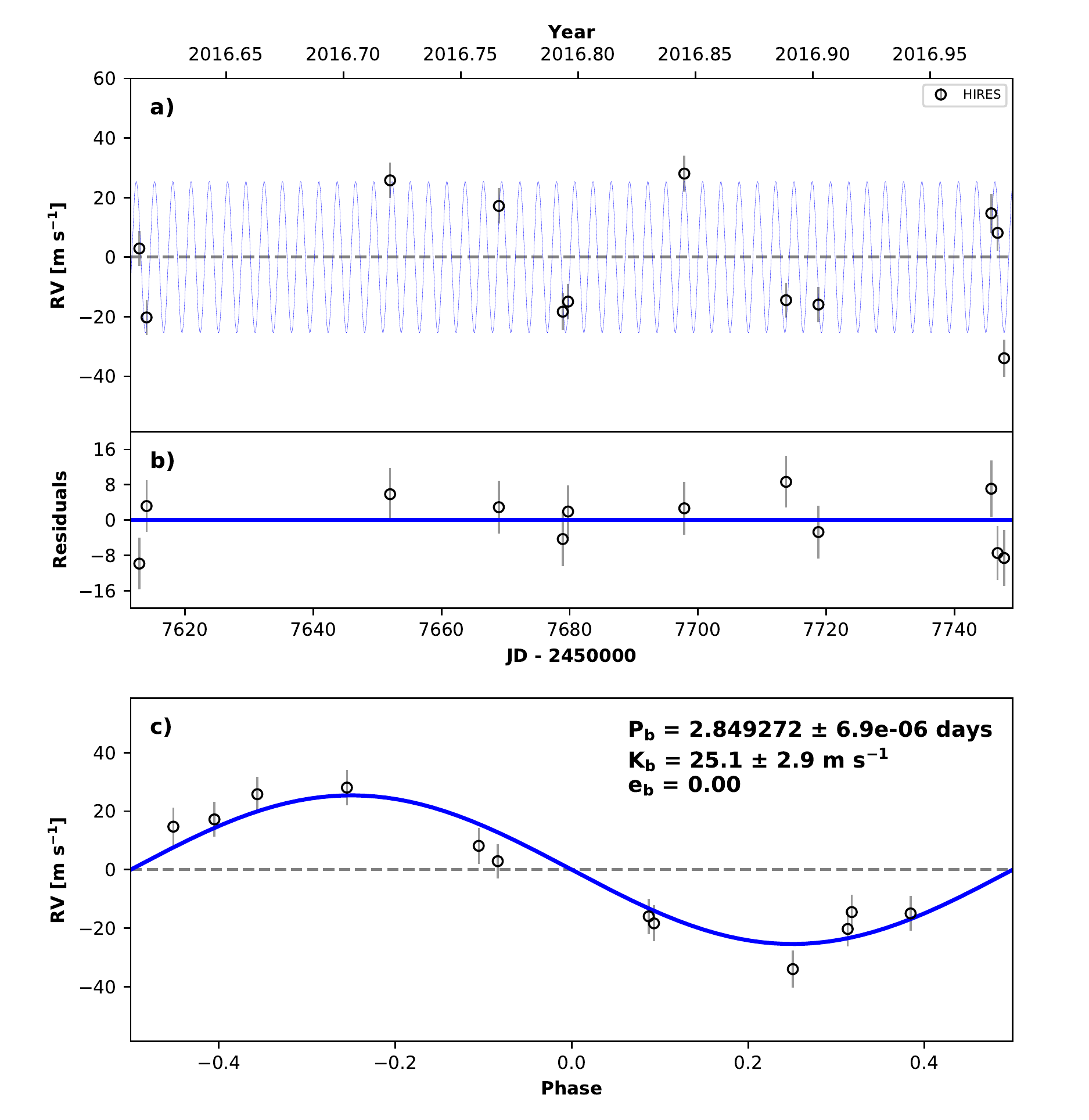}
\caption{\emph{Top: } Best-fit 1-planet circular orbital model (blue line) for K2-55 overlaid on our Keck/HIRES data (circles with errors). Note that the plotted model is the maximum likelihood model while the orbital parameters listed in Table \ref{tab:system} are the median values of the posterior distributions. We add in quadrature the RV jitter term(s) listed in Table \ref{tab:system} with the measurement uncertainties for all RVs. \emph{Middle: } Residuals to the best fit 1-planet model. \emph{Bottom: } RVs phase-folded to the ephemeris of K2-55b compared to the phase-folded model.}
\label{fig:rv}
\end{figure*}

As shown in Table~\ref{tab:fitcomp}, we found consistent masses for K2-55b regardless of whether the model included eccentricity or a long-term trend. All of these models appear to produce reasonable fits to the RV data, but they vary in the number of free parameters. In order to determine the appropriate level of complexity for our 12-point RV data set, we calculated the Bayesian Information Criterion \citep[BIC,][]{schwarz1978} and report the results in Table~\ref{tab:fitcomp}. Our BIC analysis revealed that the model containing a single planet on an eccentric orbit and no long-term trend fit the data better than a model containing a single planet on a circular orbit and no long-term trend, but that the additional parameters required to fit eccentric orbits were not justified by the performance of the fit ($\Delta BIC = 1.75$).  We saw no compelling evidence for a long-term variation in the data: adding a linear or quadratic trend to the eccentric planet model increased the BIC by $\Delta BIC = 2.48$ or $\Delta BIC = 3.96$, respectively, which indicates that the trend-free model is preferred. We display our adopted model and the Keck/HIRES data in Figure~\ref{fig:rv}. 

\begin{deluxetable*}{cccccccc}
\tablecolumns{8}
\tabletypesize{\normalsize}
\tablecaption{RV Model Comparison\tablenotemark{a}\label{tab:fitcomp}}
\tablehead{
\colhead{} & 
\colhead{} & 
\multicolumn{6}{c}{Model}\\
\cline{3-8}
\colhead{Parameter} & 
\colhead{Units} & 
\colhead{circ} &
\colhead{circ + linear} &
\colhead{circ + quad} &
\colhead{ecc} &
\colhead{ecc + linear} &
\colhead{ecc + quad}
}
\startdata
$e$ & $\cdots$ &  $\cdots$ & $\cdots$  & $\cdots$  & $0.124^{+0.054}_{-0.055}$ & $0.125^{+0.062}_{-0.060}$  & $0.119^{+0.064}_{-0.061}$ \\
$\omega$ & rad & $\cdots$  & $\cdots$  & $\cdots$  & $-2.87^{+0.57}_{-0.65}$ & $-2.83^{+0.64}_{-0.72}$  &   $-3.13^{+0.80}_{-0.82}$\\
$\gamma$  & m s$^{-1}$ & $0.7\pm2.1$ & $0.7\pm2.3$ & $3.5\pm3.2$ & $0.6^{+1.7}_{-1.8}$  & $0.6\pm1.9$  &  $2.9\pm2.9$ \\
$\dot{\gamma}$ & m s$^{-1}$ d$^{-1}$ & $\cdots$  & $-0.004^{+0.049}_{-0.050}$ & $-0.027\pm 0.052$  & $\cdots$ & $0.0003^{+0.042}_{-0.044}$  & $-0.018^{+0.045}_{-0.047}$ \\
$\ddot{\gamma}$ &  m s$^{-1}$ d$^{-2}$ & $\cdots$ & $\cdots$  & $-0.0014 \pm 0.0011$ & $\cdots$ & $\cdots$  &  $-0.0012\pm0.0011$\\
$\sigma$ &  m s$^{-1}$ & $6.8^{+2.3}_{-1.6}$ & $7.4^{+2.7}_{-1.8}$  & $7.0^{+2.9}_{-1.8}$ & $5.3^{+2.2}_{-1.4}$ & $5.9^{+2.7}_{-1.6}$  & $5.8^{+3.1}_{-1.8}$ \\
$K$ & m s$^{-1}$ & $25.1^{+2.9}_{-3.0}$ & $25.0 \pm 3.2$  & $24.7^{+3.0}_{-3.2}$ & $25.8^{+2.5}_{-2.6}$  & $25.7^{+2.8}_{-3.0}$  & $25.5^{+2.7}_{-3.1}$ \\
$M_p$ & $\mearth$ & $43.13^{+5.98}_{-5.80}$  & $43.00^{+6.36}_{-6.18}$  & $42.54^{+6.16}_{-6.11}$  & $43.99^{+5.33}_{-5.30}$ & $43.74^{+5.72}_{-5.87}$  & $43.41^{+5.73}_{-6.11}$  \\
$BIC\tablenotemark{b}$ & $\cdots$ & 87.21 & 89.69  & 89.72 & 85.46 & 87.94  & 89.42  \\
$\Delta BIC$ & $\cdots$ & 1.75  & 4.23 & 4.26 & $\cdots$ & 2.48  & 3.96 \\
\enddata      
\tablenotetext{a}{Reference epoch for $\gamma$, $\dot{\gamma}$, and $\ddot{\gamma}$ is BJD~2457689.754631. }            
\tablenotetext{b}{In order to compute the BIC used for the model comparison, we fixed the jitter to $\sigma_j =  5.34$~m~s$^{-1}$.}            
\end{deluxetable*}

The orbital period of K2-55b is short enough that we might have expected the orbit to be tidally circularized. According to \citet{goldreich+soter1966}, the circularization timescale for a planet with mass $M_p$ and radius $R_p$ on a modestly eccentric orbit around a star of mass $M_\star$ is: 
\begin{equation}
t_{\rm circ}  = \frac{4}{63}\frac{1}{\sqrt{G M_\star^3}}\frac{M_p a^{13/2}Q'}{R_p^5} \,
\end{equation}
where $G$ is the gravitational constant, $a$ is the semimajor axis of the planet, and the factor $Q'$ scales inversely with dissipation efficiency. As noted by \citet{mardling2007}, $Q'$ is a modified $Q$-value and related to the tidal quality factor $Q$ by the Love number $k_p$ such that $Q' = 3Q/2k_p$. 

 We do not know the tidal quality factor or Love number of K2-55b, but adopting Neptune-like values of $9000 < Q < 36000$ \citep{zhang+hamilton2008} and $k_2 = 0.41$ \citep{bursa1992} yields circularization timescales of 110-450~Myr. These timescales are much shorter than the expected age of the system, indicating that K2-55b may actually have a higher $Q$ if the planet really does have nonzero eccentricity. For instance, a tidal quality factor of $Q = 10^5$ would yield a circularization time of 6~Gyr. 
 
 Building on the work of \citet{agundez_et_al2014}, \citet{morley_et_al2017} reported a similarly high dissipation factor for GJ~436b ($Q' \approx 10^5 - 10^6$) and hypothesized that the interior structures of close-in Neptune-sized planets may differ from those of the more distant ice giants in our solar system. A high $Q$ for K2-55b would be consistent with this theory. In the future, occasional monitoring of K2-55 over a timescale longer than our original 120-day baseline will help constrain the eccentricity and interior structure of K2-55b. For now, we adopt the circular solution and infer that K2-55b has a mass of $43.13^{+5.98}_{-5.80}\mearth$. Although our model comparison test revealed that the current RV data set is better fit by an eccentric orbit than by a circular orbit, the difference is small ($\Delta BIC =  1.75$) and the choice of a circular orbit does not significantly affect the resulting planet mass estimates (\mbox{$\Delta m_p =  0.86\mearth = 0.15\sigma$}). 

\section{Discussion}
\label{sec:discussion}
Now that we have constrained the radius (Section~\ref{sec:phot}) and mass of K2-55b (Section~\ref{sec:rv}), we devote the remainder of the paper to discussing the implications of our results. We begin in Section~\ref{ssec:context} by determining the bulk density of K2-55b and comparing the planet to other similarly sized planets both within and beyond the Solar System. We then consider possible compositions for K2-55b in Section~\ref{ssec:composition}. When compared to other planets with similar masses or radii, we find that K2-55b has a surprisingly high density and low inferred envelope fraction. 

In order to understand whether K2-55b is truly an odd planet or simply one example drawn from a class of planets with a diverse array of properties, we examine the overall frequency of intermediate-sized planets and the possible connections between planet occurrence and system properties (Section~\ref{ssec:freqnep}). We then review the compositional diversity of intermediate-sized planets in Section~\ref{ssec:compnep} and propose several scenarios explaining the formation of K2-55b in Section~\ref{ssec:formation}. Finally, we consider possible atmospheric models for K2-55b in Section~\ref{ssec:atmosphere} and discuss the prospects for follow-up atmospheric characterization studies.

\subsection{Placing K2-55b in Context}
\label{ssec:context}
Combining our photometrically-derived planet radius estimate of $4.41^{+0.32}_{-0.28}\rearth$ with our radial velocity mass constraint of $43.13^{+5.98}_{-5.80}\mearth$, we find that K2-55b has a bulk density of $2.8_{-0.6}^{+0.8}$~g~cm$^{-3}$. Although \mbox{K2-55} is only 14\% larger than Neptune ($3.87\rearth$) and 11\% larger than Uranus ($3.98\rearth$) it is significantly more massive than either ice giant: K2-55b ($43.13^{+5.98}_{-5.80}\mearth$) is 2.5 times as massive as Neptune ($17.15\mearth$), three times as massive as Uranus ($14.54\mearth$), and nearly half the mass of Saturn ($95.16\mearth$). As a result, the bulk density of K2-55b ($2.8_{-0.6}^{+0.8}$~g~cm$^{-3}$) is 120\% and 71\% higher than those of Uranus (1.271~g~cm$^{-3}$) and Neptune (1.638~g~cm$^{-3}$), respectively. The interior structure of K2-55b is therefore quite distinct from that of the ice giants in our Solar System. Despite the similar sizes of all three planets, K2-55b must have a lower fraction of volatiles or ices than either Uranus or Neptune.

In order to better compare K2-55b to other exoplanets, we queried the Confirmed Planets Table from the NASA Exoplanet Archive\footnote{We note that the NASA Exoplanet Archive was missing the stellar effective temperature and metallicity of GJ~436. We adopt $T_{\rm eff} = 3416\pm54$~K \citep{vonbraun_et_al2012} and [Fe/H] = $+0.02 \pm 0.20$ \citep{lanotte_et_al2014}.} \citep{akeson_et_al2013} and selected all planets orbiting single stars\footnote{We omitted the circumbinary Kepler-413b from Figures~\ref{fig:massradius} and \ref{fig:planets}. Although Kepler-413b may resemble K2-55b in terms of mass, radius, and bulk density ($M_p = 67^{+22}_{-21}\mearth$, $R_p = 4.35\pm0.10\rearth$, $\rho_p = 3.2\pm 1.0$~g~cm$^{-3}$), the two planets likely followed different formation pathways. Furthermore, Kepler-413b has only coarse mass constraints based on photometric-dynamical modeling \citep{kostov_et_al2014}.}  with densities measured  to better than 50\% as of 2018~March~28. In Figure~\ref{fig:massradius}, we place K2-55b and the other well-constrained planets on the mass-radius diagram. K2-55b resides near several other planets with masses $>30\mearth$ and radii $<6\rearth$: K2-27~b \citep{vaneylen_et_al2016, petigura_et_al2017}, K2-39~b \citep{vaneylen_et_al2016, petigura_et_al2017}, K2-98~b \citep{barragan_et_al2016}, K2-108~b \citep{petigura_et_al2017},  Kepler-101~b \citep{bonomo_et_al2014}, and WASP-156~b \citep{demangeon_et_al2018}. All of these planets orbit stars that are hotter and more massive than K2-55. The coolest host stars are \mbox{K2-39} ($T_{\rm eff} = 4912$~K), an evolved star with a radius of $2.93\rsun$, and WASP-156 ($T_{\rm eff} = 4910$~K), a metal-rich K3 star with [Fe/H]$=0.24\pm0.12$. K2-55 stands out as the smallest, lowest mass host star harboring a massive transiting planet ($M_p > 30\mearth$). 

\subsection{The Composition of K2-55b}
\label{ssec:composition}
The density of K2-55b ($2.8_{-0.6}^{+0.8}$~g~cm$^{-3}$) is intermediate between the values expected for terrestrial planets and gas giants, suggesting that K2-55b has a heterogeneous composition containing both heavy elements and low-density volatiles. Accordingly, we model K2-55b as a two-layer planet consisting of a rocky core capped by a low-density H/He envelope. We note that that \mbox{K2-55b} might also contain ices \citep{rogers_et_al2011}, but variations in the core water abundance of Neptune-sized planets have a negligible influence on the radius-composition relation compared to changes in the H/He envelope fraction. \citep{lopez+fortney2014}. Furthermore, the degeneracies between icy interiors and rocky interiors are impossible to break with mass and radius measurements alone \citep{adams_et_al2008, figueria_et_al2009}.

For our two-layer model, we use the internal structure and thermal evolution models developed by \citet{lopez+fortney2014}, who generated an ensemble of model planets spanning a variety of planet masses ($M_p$), envelope fractions ($M_{env}/M_p$), and planet insolation flux ($F_p$). \citet{lopez+fortney2014} then evolved the planets forward in time and tracked the evolution of the planet radii. The resulting grid of planet masses, radii, envelope fractions, insolation fluxes, and ages has been used to infer the compositions of a multitude of planets \citep[e.g.,][]{wolfgang+lopez2015}. The studies most germane to our analysis of K2-55b are those of \citet{petigura_et_al2016} and \citet{petigura_et_al2017}, who employed the models of \citet{lopez+fortney2014} to analyze a set of sub-Saturns. As defined by \citet{petigura_et_al2016,petigura_et_al2017}, ``Sub-Saturns'' are planets with radii of $4-8\rearth$. At $4.41^{+0.32}_{-0.28}\rearth$, K2-55b could therefore be described as a ``small sub-Saturn.''  

The \citet{petigura_et_al2017} planet sample included 19 sub-Saturns with densities measured to precisions of 50\% or better. Although tightly restricted in radius to \mbox{$4.0\rearth<R_p<7.8\rearth$}, the \citet{petigura_et_al2017} sub-Saturn sample spans a broad mass range of $4.8-69.9\mearth$ and a correspondingly large density range of $0.09-2.40\,{\rm g}\,{\rm cm}^{-3}$. The observed masses and radii of the planets in their sample could be explained by envelope fractions of $7-60\%$ H/He by mass. 

Interpolating the same \citet{lopez+fortney2014} models to investigate the composition of K2-55b, we find that our estimated mass of $43.13^{+5.98}_{-5.80}$ and radius of $4.41^{+0.32}_{-0.28}\rearth$ are consistent with an envelope fraction of $12\pm3\%$.  This inferred envelope fraction is on the low end of the range observed by \citet{petigura_et_al2017}, underscoring the point that K2-55 has an exceptionally low gas fraction for its mass. K2-55b is denser than any of the planets in the \citet{petigura_et_al2017} sample and more massive than all but four of the 19~planets they considered.

\begin{figure*}[tbhp]
\centering
\includegraphics[width=0.49\textwidth]{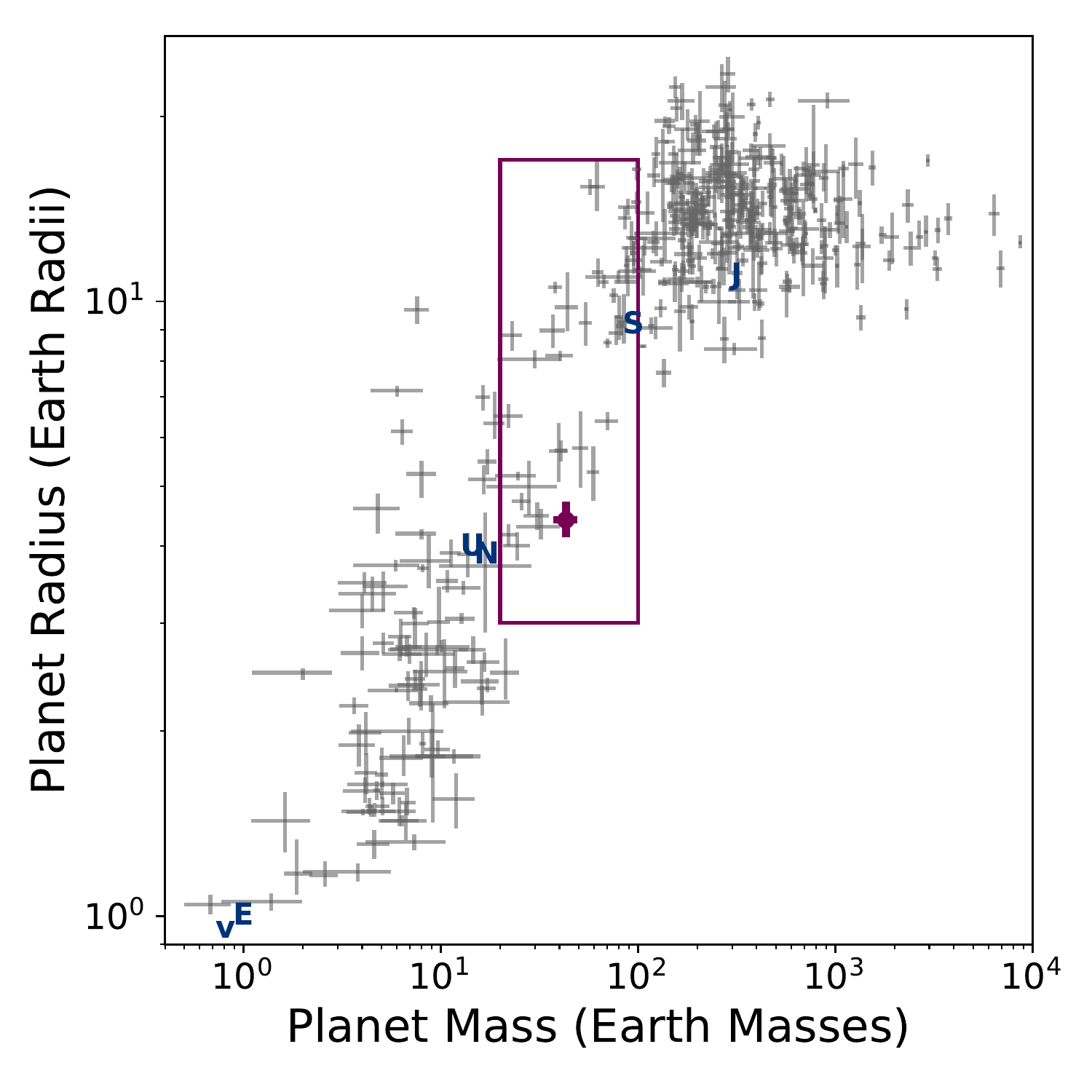}
\includegraphics[width=0.49\textwidth]{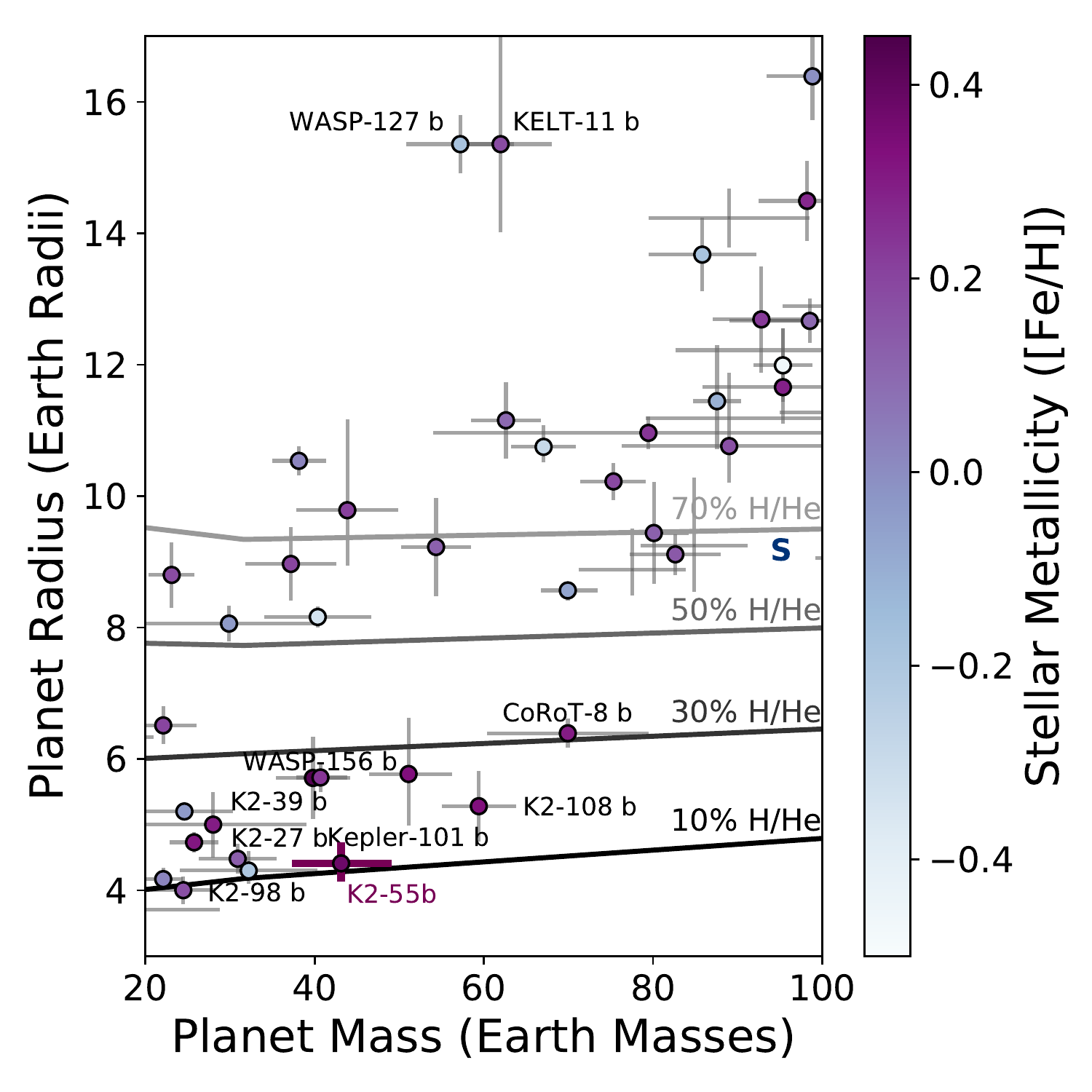}
\caption{Mass and radius of K2-55b (point with thick purple error bars) compared to those of other small planets (points with thin gray error bars). \emph{Left: } K2-55b compared to all confirmed planets from the NASA Exoplanet Archive \citep{akeson_et_al2013} with densities measured to better than 50\% as of 2018 March~28. \emph{Right: } Zoomed-in view comparing K2-55b to the subset of confirmed planets with masses between $20\mearth$ and $100\mearth$ (i.e., planets with masses within roughly a factor of two of the mass of K2-55b) and to the two-layer models from \citet[][thick gray lines]{lopez+fortney2014}. All points (including the point for K2-55b) are color-coded by the metallicity of the host star as indicated by the color-bar and the points closest to K2-55b are labeled. We also mark 
KELT-11b \citep{pepper_et_al2017} and WASP-127b \citep{lam_et_al2017} because they are far from the main population of planets. For reference, the purple rectangle in the left panel indicates the boundaries of the smaller region displayed in the right panel and the navy letters in both panels mark the locations of Solar System planets.}
\label{fig:massradius}
\end{figure*}

\begin{figure*}[tbhp]
\centering
\includegraphics[width=0.49\textwidth]{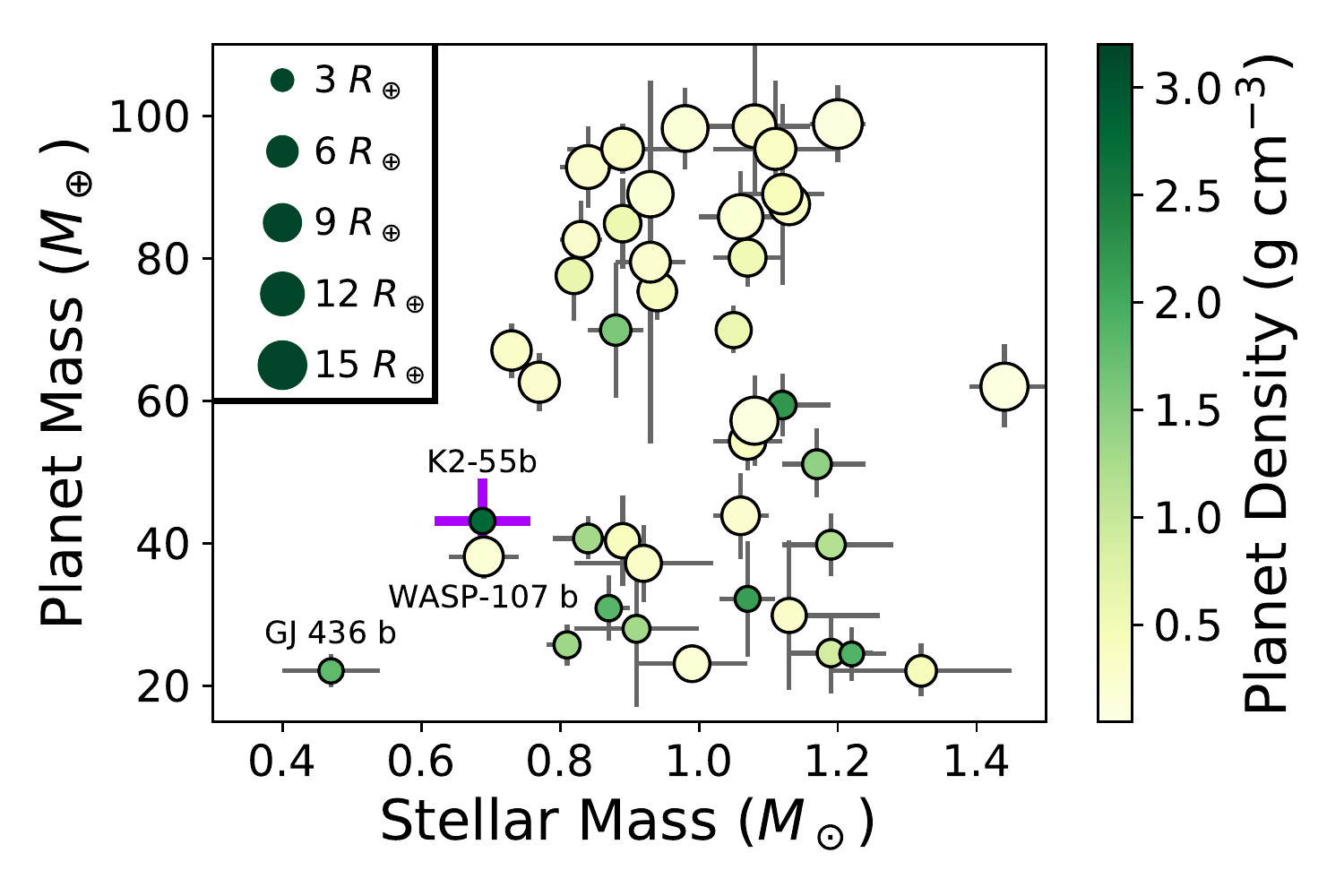}
\includegraphics[width=0.49\textwidth]{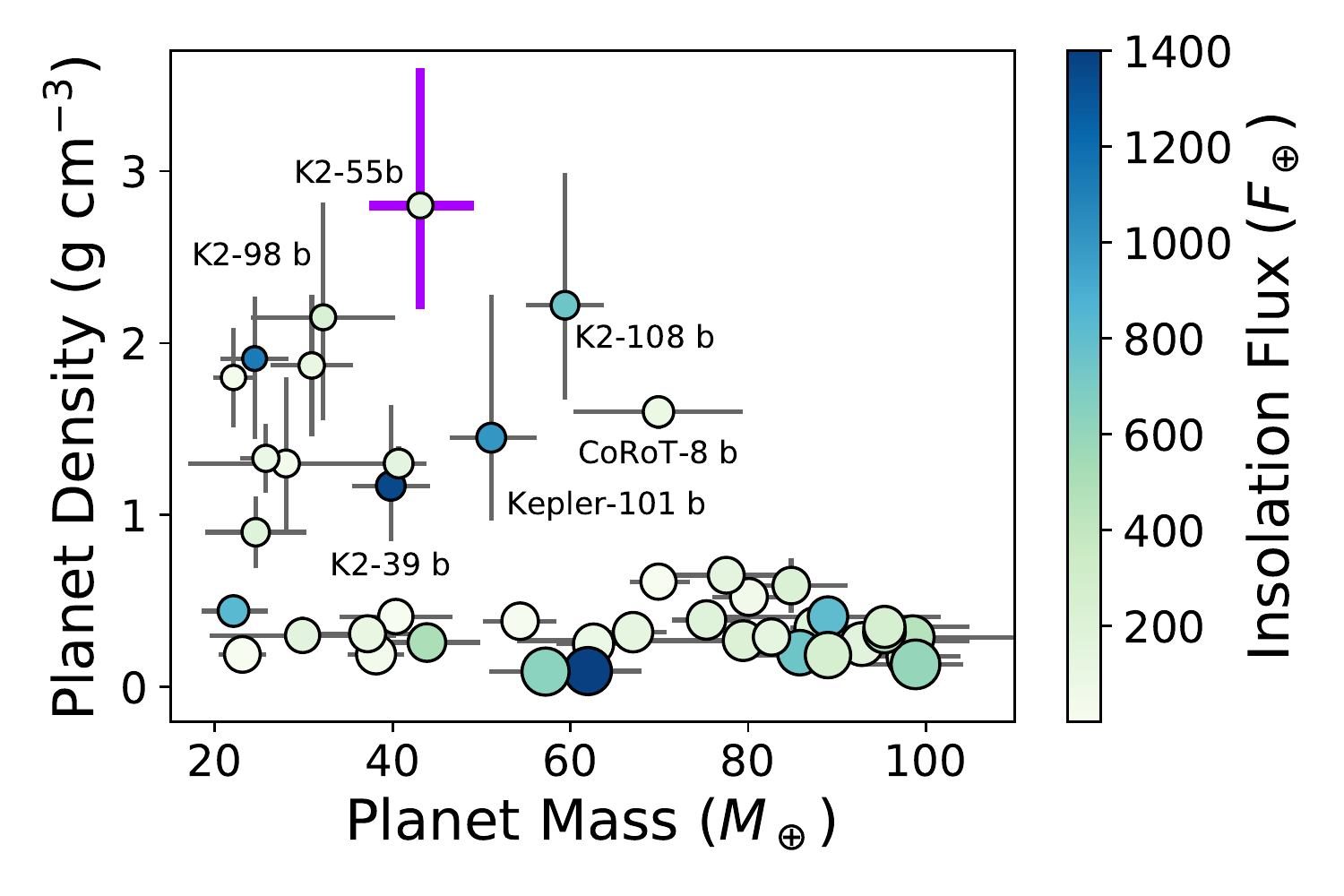}
\includegraphics[width=0.49\textwidth]{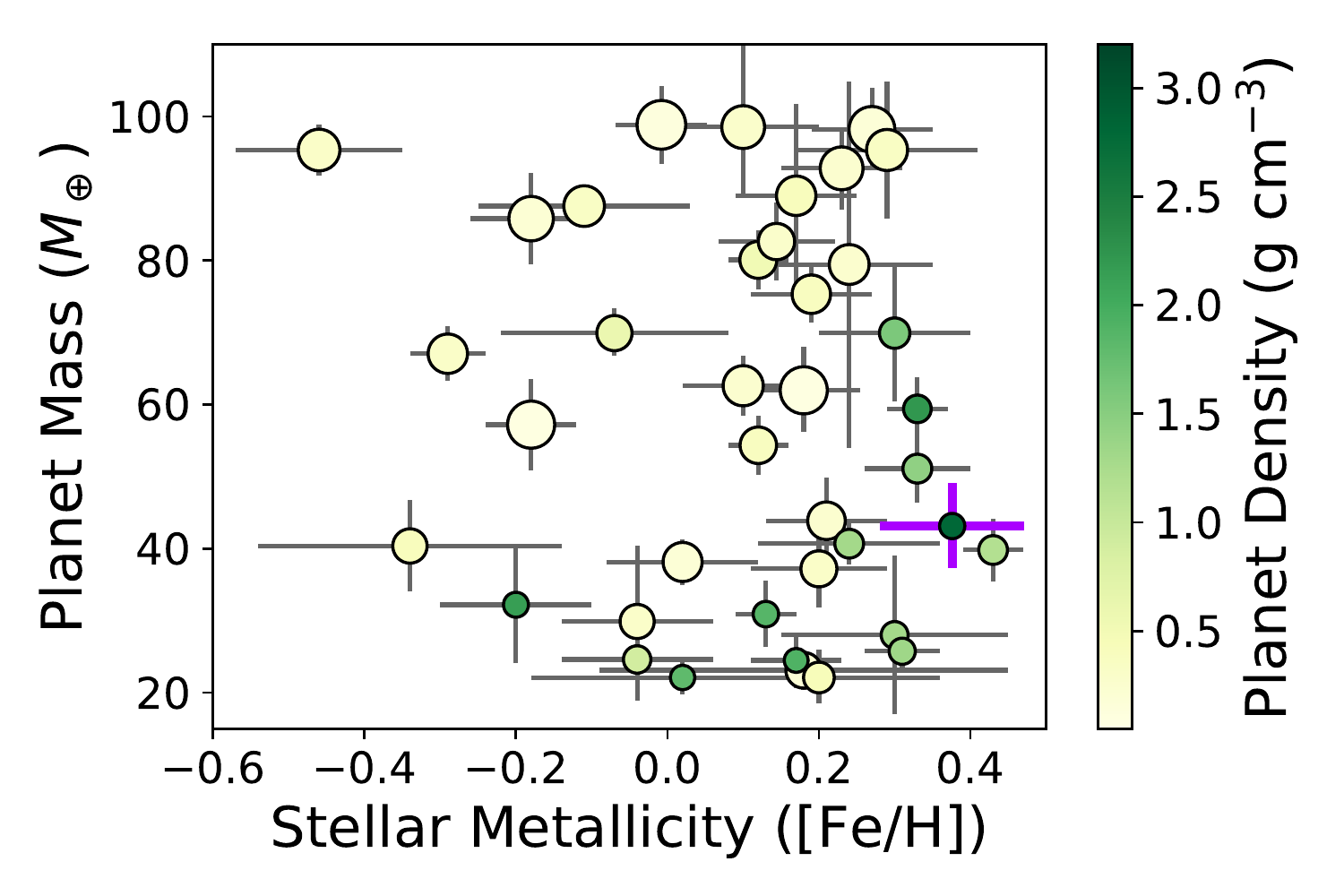}
\includegraphics[width=0.49\textwidth]{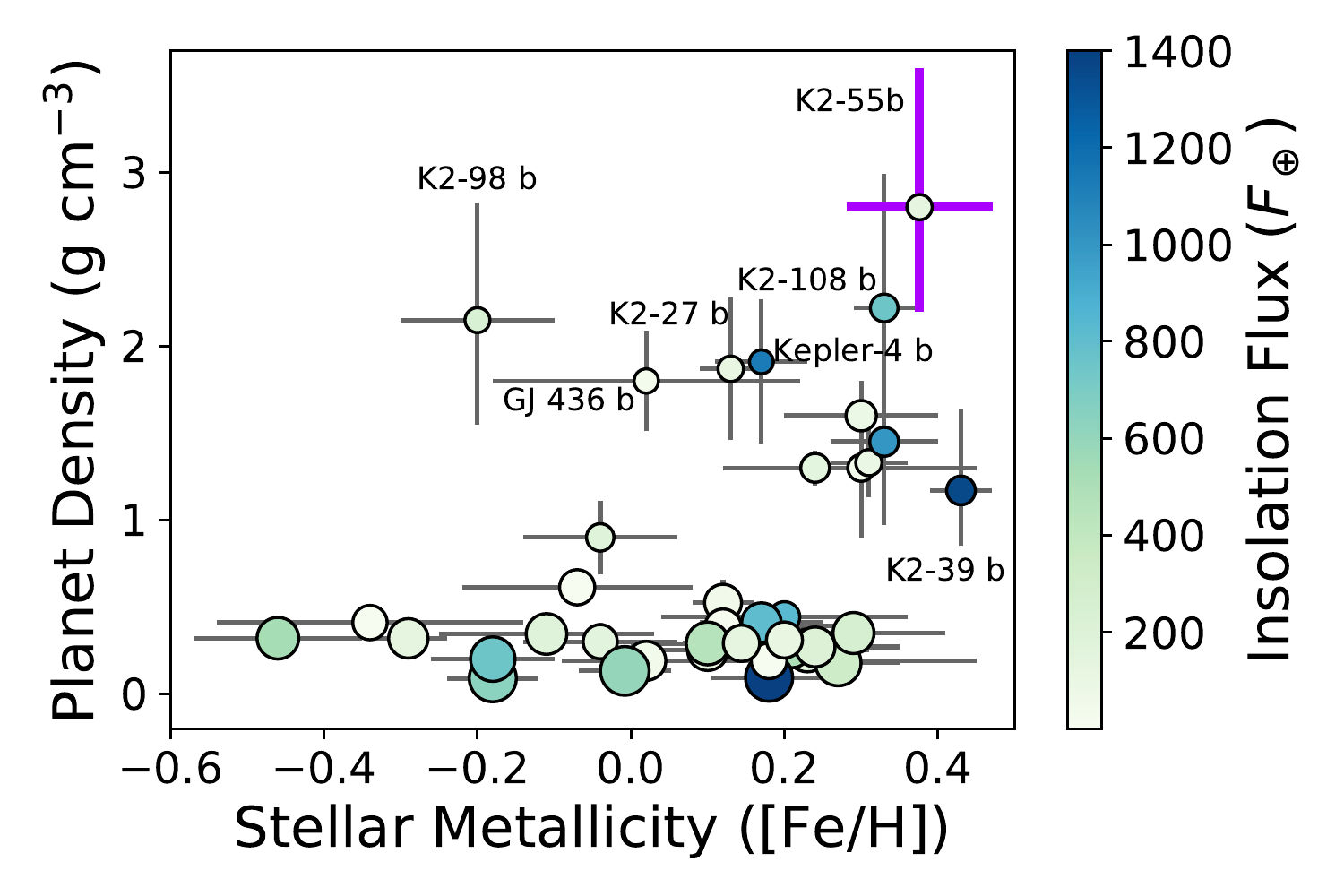}
\caption{Comparison of the planets in the right panel of Figure~\ref{fig:massradius} (circles with thin gray errorbars) to K2-55b (circle with thick purple error bars). The data points are scaled by planet radius and colored by planet density (left panel) or insolation flux (right panel) as indicated by the legends. Key planets are labeled for reference. \emph{Top Left: } Planet mass versus stellar mass.  \emph{Top Right: } Planet density versus planet mass. Note that K2-55b is the densest planet in the sample. \emph{Bottom Left: } Planet mass versus stellar metallicity. \emph{Bottom Right: } Planet density versus stellar metallicity.}
\label{fig:planets}
\end{figure*}

Considering all planets with $20\mearth < M_p < 100\mearth$ and radii of $3\rearth < R_p < 17\rearth$ (i.e., all of the planets in the right panel of Figure~\ref{fig:massradius}), we find that the median host star has an effective temperature of $5449$~K and a mass of $0.99\msun$. The full range spans $3416 - 6270$~K and $0.47 - 1.44\msun$. As shown in Figure~\ref{fig:planets}, the only host star less massive than K2-55 is GJ~436, further emphasizing that K2-55b may be a curiously massive planet given the mass of its host star. Figure~\ref{fig:planets} also reveals that K2-55b is denser than all of the planets in the right panel of Figure~\ref{fig:massradius}. The combination of our high bulk density estimate for K2-55b ($2.8_{-0.6}^{+0.8}$~g~cm$^{-3}$) and the high metallicity of K2-55 might suggest that K2-55b formed from a protoplanetary disk with an unusually deep reservoir of solid material.

\subsection{The Frequency of Planets with Intermediate Radii}
\label{ssec:freqnep}
In general, Neptune-sized planets are more common than Jupiter-sized planets, but much rarer than smaller planets \citep[e.g.,][]{youdin2011, howard_et_al2012, dressing+charbonneau2013, fressin_et_al2013, petigura_et_al2013b, fulton_et_al2017}. Using the full \emph{Kepler} data set and sub-dividing the stellar sample by spectral type, \citet{mulders_et_al2015} estimated that planets with radii of $4-5.7\rearth$ and periods of $2.0-3.4$~d occur at a rate of \mbox{$0.00022\pm0.00018$} planets per F~star, \mbox{$0.0011\pm0.0004$} planets per G~star, \mbox{$0.0016\pm0.0008$} planets per K~dwarf, and $<0.0069$~planets per M~dwarf. The detection of \mbox{K2-55b} is therefore less remarkable for the low mass of the host star than for the intermediate size of the planet: close-in Neptunes seldom occur regardless of host star spectral type.

The dependence of the hot Neptune occurrence rate on stellar metallicity is more complicated. The increased prevalence of gas giants orbiting metal-rich stars is well-established \citep[e.g.,][]{gonzalez1997, santos_et_al2004, fischer+valenti2005}, but the role of metallicity on the occurrence rates of smaller planets is less understood. Examining the \emph{Kepler} planet sample, \citet{buchhave_et_al2014} found that planets larger than $3.9\rearth$ orbit stars that are significantly more metal-rich than the hosts of smaller planets. \citet{buchhave_et_al2014} also noted that the host stars of $1.7-3.9\rearth$ planets are more metal-rich than the host stars of smaller planets, but \citet{schlaufman2015} countered that the data are better described by a continuous gradient of increasing metallicity with increasing planet radius from $1\rearth$ to $4\rearth$ rather than a sharp metallicity jump at $1.7\rearth$. 

In a related study, \citet{wang+fischer2015} observed that planet occurrence is positively correlated with stellar metallicity independent of planet size. In particular, they found that metal-rich stars ([Fe/H]~$>0.05$) were $9.30^{+5.62}_{-3.04}$ times more likely than metal-poor stars ([Fe/H]~$<-0.05$) to harbor planets with radii of $3.9-22\rearth$. The metallicity bias appears less pronounced for smaller planets ($2.03^{+0.29}_{-0.26}$ for $1.7\rearth<R_p<3.9\rearth$ and $1.72^{+0.19}_{-0.17}$ for $R_p < 1.7\rearth$), but the metallicity preference might be underestimated due to the observational bias against detecting transiting planets orbiting metal-rich stars due to the shallower transit depths caused by their larger radii. 

Considering the possible interplay between planet occurrence, stellar metallicity, and orbital period, \citet{mulders_et_al2016} found that short-period planets \mbox{($P < 10$~d)} are biased toward metal-rich host stars \mbox{([Fe/H]~$\simeq 0.15 \pm 0.05$~dex)} while longer period planets orbit stars with solar-like metallicities. While this trend toward higher stellar metallicities at shorter planet orbital periods is quite pronounced for the smallest planets ($<1.7\rearth$), the trend disappears for larger planets: host stars of $3.9 - 14\rearth$ planets typically have super-solar metallicities of $0.14\pm0.04$~dex regardless of planet orbital period. Accordingly, the realization that the host star of Neptune-sized \mbox{K2-55b} is metal-rich \mbox{([Fe/H]~=~$0.376\pm0.095$)} would be unsurprising even if the planet had an orbital period significantly longer than the observed value of 2.8~d.

\subsection{The Compositional Diversity of Planets with Intermediate Radii}
\label{ssec:compnep}
Concentrating on sub-Saturns, \citet{petigura_et_al2017} tested several different theories to explain the large dispersion in planet mass, density, and envelope fraction. \citet{petigura_et_al2017} noted that the envelope fractions of the hottest planets in their sample ($T_{\rm eq} > 1250$~K) were restricted to a smaller range of $10\% < M_{\rm env}/M_p < 30\%$ while the cooler planets spanned the full estimated range from $10\%-60\%$. The lack of hot planets with larger envelope fractions might indicate that photoevaporation prevents close-in sub-Saturns from retaining large quantities of volatiles. However, photoevaporation could not be the only explanation for the observed diversity of sub-Saturn compositions because \citet{petigura_et_al2017} did not observe a strong correlation between present-day planet equilibrium temperature and envelope fraction. In agreement with \citet{petigura_et_al2017}, the right panels of Figure~\ref{fig:planets} do not display a strong relationship between planet density and insolation flux. The most highly irradiated planet \citep[KELT-11b,][]{pepper_et_al2017} has a bulk density of $0.93$~g~cm$^{-3}$, but less irradiated planets like K2-55b ($F_p < 200\fearth$) span a wide range of densities from $0.09 - 2.2$~g~cm$^{-3}$. 

Similarly, \citet{petigura_et_al2017} failed to detect a correlation between host star metallicity and envelope fraction, demonstrating that disk metallicity changes alone cannot explain the observed densities of sub-Saturns. The lack of a correlation between stellar metallicity and envelope fraction was slightly surprising because \citet{thorngren_et_al2016} had previously noted an anti-correlation between planet metal abundance (approximated as \mbox{$Z_p = M_{\rm core}/M_p$}) and planet mass for planets with masses of $30-3000\mearth$. The \citet{petigura_et_al2017} planet sample included more lower mass planets than the original \citet{thorngren_et_al2016} sample, which allowed \citet{petigura_et_al2017} to learn that the previously detected anti-correlation does not appear to extend to planets with masses below $30\mearth$. \citet{petigura_et_al2017} suggested that perhaps the extinction of the trend at lower masses is a manifestation of different formation pathways for gas giants and lower mass planets. K2-55b is a more massive sub-Saturn and falls nicely on the 
relation found by \citet{thorngren_et_al2016} between planet metal enrichment relative to stellar metallicity ($Z_{\rm planet}/Z_{*}$) and planet mass. Specifically, the \citet{thorngren_et_al2016} relation predicts a planet metal enrichment ratio of $Z_{\rm planet}/Z_{*} = 24$ for a $44\mearth$ planet and the ratio for K2-55b is $Z_{\rm planet}/Z_{*} = 26$.

The high planet mass of K2-55b and the super-solar metallicity of K2-55 are also consistent with the finding by \citet{petigura_et_al2017} that stars with higher metallicities tend to host more massive sub-Saturns. The positive correlation between stellar metallicity and sub-Saturn mass may suggest that more massive planetary cores formed in more metal-rich protoplanetary disks \citep{petigura_et_al2017}. As shown in the bottom panels of Figure~\ref{fig:planets}, the densest sub-Saturns tend to orbit the most metal-rich host stars. This trend is particularly pronounced in the bottom right panel, which displays a clear separation between the denser sub-Saturns and the low density larger planets. 

Intriguingly, \citet{petigura_et_al2017} also noted that more massive sub-Saturns tend to have moderately eccentric orbits and orbit stars without other detected planets while less massive sub-Saturns tend to follow more circular orbits and reside in systems with multiple transiting planets. As a $43.13^{+5.98}_{-5.80}\mearth$ planet in a system with no other detected planets, K2-55b might therefore be expected to have an eccentric orbit. Additional observations are required to tighten the constraints on the orbital eccentricity of K2-55b and better discriminate between eccentric and circular models.

\subsection{Possible Formation Scenarios for K2-55b}
\label{ssec:formation}
Under the core accretion model of planet formation, planetesimals collide to form protoplanetary cores, which then acquire gaseous envelopes \citep{perri+cameron1974, mizuno_et_al1978, mizuno1980, stevenson1982, bodenheimer+pollack1986, pollack_et_al1996}. If the planet core is able to become sufficiently massive before the gaseous disk dissipates \citep[at roughly a few Myr,][]{williams+cieza2011} then the growing planet can enter a phase of runaway accretion in which the envelope grows rapidly. The onset of the ``core-accretion instability'' occurs when the mass of the planetary core exceeds the ``critical core mass,'' $M_{\rm crit}$. While numerous studies have estimated $M_{\rm crit}$ as roughly $10\mearth$ \citep[][and references therein]{ikoma_et_al2000}, \citet{rafikov2006, rafikov2011} demonstrated that variations in the assumed disk properties and planetesimal accretion rate can alter $M_{crit}$ by orders of magnitude, resulting in a wide range of $0.1-100\mearth$. In general, $M_{\rm crit}$ decreases with increasing distance from the star due to the cooler disk temperatures found in the outer disk. $M_{\rm crit}$ also decreases with increasing mean molecular weight, but this effect can be outbalanced by the stronger trend of increasing $M_{\rm crit}$ with increasing dust opacity \citep{hori+ikoma2011, nettelmann_et_al2011, piso+youdin2014}.
 
Although the mass of K2-55b is below the upper end of the $0.1\mearth < M_{\rm crit} < 100\mearth$ range found by \citet{rafikov2006, rafikov2011}, the absence of a large volatile envelope for a $43.13^{+5.98}_{-5.80}\mearth$ planet is noteworthy and at odds with general expectations from core accretion models. Na\"ively assuming that K2-55b formed in situ at 0.0347~au in a minimum mass solar nebula \citep[MMSN,][]{hayashi1981} with a $\Sigma_p = 33 F Z_{\rm rel} r^{-3/2}\,{\rm g}\,{\rm cm}^{-2}$ solid surface density profile, a total mass ratio $F = 1$, and a metal richness $Z_{\rm rel} = 0.33$ \citep{chiang+youdin2010}, there would have been only $0.01\mearth$ of solids available for building K2-55b. Adopting a more massive minimum mass extrasolar nebula (MMEN) solid surface density profile \citep{chiang+laughlin2013, gaidos2017} would yield roughly $0.16\mearth$ of solids. Although significantly higher than the estimate based on the MMSN, the solid mass locally available in the MMEN model is less than 0.4\% of the present-day mass of K2-55b, indicating that either K2-55b itself or the planetary building blocks that would become K2-55b \citep[e.g.,][]{hansen+murray2012, chatterjee+tan2014} must have migrated inward from farther out in the disk. 

Acknowledging the puzzling existence of a massive close-in planet with only a modest H/He envelope, we propose four possible formation scenarios for K2-55b:
\begin{enumerate}
\item{Classic type I migration into the inner disk cavity}
\item{Collisions of multiple planets}
\item{Post-formation atmospheric loss}
\item{Formation via less efficient core accretion}
\end{enumerate}  

Under the first scenario, uneven torques from the disk on K2-55b would have caused the planet to drift inward toward the host star \citep{ward1997, tanaka_et_al2002}. The Type I migration\footnote{For a recent review of Type I migration and disk-planet interactions in general, see \citet{kley+nelson2012}.} would have been halted after \mbox{K2-55b} entered the inner cavity between the disk and the star. K2-55b would have therefore escaped runaway accretion because it was trapped at the 2:1 resonance with the disk inner edge \citep[e.g.,][]{kuchner+lecar2002} rather than embedded within the disk. Although feasible, this argument is unsatisfying due to the fine-tuning required to have K2-55b cross the disk edge after reaching a large overall mass but before accumulating a substantial envelope.

In the second scenario, K2-55b might have been formed via collisions of smaller planets. For instance, \citet{boley_et_al2016} found that collisions of smaller planets in systems of tightly packed inner planets (STIPs) can produce gas-poor giant planets if the progenitor planets collide after the gas disk has dissipated. Another possible explanation is that the protoplanetary disk orbiting K2-55 might have been slightly misaligned with respect to the host star \citep[e.g.,][]{bate_et_al2010}, which could have been orbited by several less massive planets. Once the gas in the disk had dissipated, the continued contraction of the star along the Hayashi track could have driven a resonance through the system \citep{spalding+batygin2016}. The resonance would have perturbed the orbits of the smaller planets, causing them to collide with each other and form a more massive planet. 

The primary challenge facing the second explanation is that collisional velocities close to the star at the present-day orbital location of K2-55b are high enough that collisions are more likely to result  in fragmentation than growth \citep[][but see \citealt{wallace_et_al2017}]{leinhardt+stewart2012}. Unless the smaller planets collided farther out in the disk where collisional velocities were lower and the newly formed K2-55b subsequently migrated inward to 0.0347~au via planetesimal scattering, this scenario is unlikely to explain the formation of \mbox{K2-55b}. Alternatively, the presence of a gaseous envelope before the collision might have made the collision less destructive \citep[e.g.,][]{liu_et_al2015}. The logical observational test for this scenario is to measure the spin-orbit alignment of the system via the Rossiter-McLaughlin effect \citep{rossiter1924, mclaughlin1924}, but the host star is too faint to permit such a precise measurement with current facilities. 

A third possibility is that K2-55b formed as a ``regular'' sub-Saturn with a typical envelope fraction but then lost most of its envelope to a single late giant impact \citep[e.g.,][]{inamdar+schlichting2015, schlichting_et_al2015, liu_et_al2015, inamdar+schlichting2016}. More massive planets are less vulnerable to envelope loss via either photoevaporation or impacts \citep{lopez+fortney2013, inamdar+schlichting2015}, suggesting that a late giant impact could have had a more catastrophic effect for K2-55b than for a Saturn-mass planet. 

Our fourth formation scenario for K2-55b is that the planet formed via ``conventional'' core accretion, but that our incomplete understanding of core accretion causes us to overestimate the efficiency of planet formation. We note that the relatively small envelopes of Uranus and Neptune mandate that the gas disk dissipated just after the planets reached their final masses \citep[e.g.,][]{pollack_et_al1996, dodson-robinson+bodenheimer2010} and that producing super-Earths rather than mini-Neptunes requires delaying planet formation until most of the gas is depleted \citep{lee_et_al2014, lee+chiang2016}. Alternatively, super-Earths might form in a gas-rich disk but with dust-rich atmospheres that delay cooling and prevent them from acquiring enough gas to trigger runaway accretion \citep{lee_et_al2014, lee+chiang2015}. 

Instead of requiring that the gas in the K2-55 protoplanetary disk dissipated just as K2-55b was beginning to accrete an envelope, an alternative formation scenario is that K2-55b grew via pebble accretion \citep{lambrechts+johansen2012}. As the pebbles accreted, they would have heated the growing planet and consequently turned to dust due to the high temperature of the atmosphere. The dusty atmosphere would have inhibited cooling and prevented K2-55b from accreting an envelope \citep{lega+lambrechts2016}. 

Although the pebble heating explanation is appealing, \citet{lee+chiang2015} note that pebble accretion can block runaway accretion only for planets with low-mass cores ($M_{\rm core} < 5\mearth$); a youthful version of \mbox{K2-55b} would be too massive to escape runaway gas accretion. Nevertheless, the modern high density of K2-55b might be attributed to gas-stealing late giant impacts \citep{inamdar+schlichting2015,inamdar+schlichting2016}. If K2-55b actually has an eccentric orbit, tidal heating may have also warmed the planet and helped block runaway accretion \citep{ginzburg+sari2017}. While the specific formation pathway for K2-55b is uncertain, the sheer variety of possible explanations demonstrates that further theoretical and observational work is required to better understand core accretion and planet formation in general. Studying additional planets in the same size range as K2-55b will help determine which scenario (or combination of scenarios) best explains the formation of dense Neptune-sized planets.

\subsection{Prospects for Atmospheric Investigations}
\label{ssec:atmosphere}

Although K2-55b alone cannot solve all of the mysteries of planet formation, determining the composition of the envelope may help constrain where and how \mbox{K2-55b} formed. At the most basic level, determining the mean molecular weight of the atmosphere would reveal whether our simplistic two-layer model of a rocky core surrounded by a H/He envelope is sufficient or whether K2-55b is better explained by a lower-density core containing a large admixture of ices and a higher-density water-rich envelope. More sophisticated measurements of the relative abundances of particular molecules would enable tests of the various formation scenarios outlined in Section~\ref{ssec:formation} and perhaps spur the genesis of new formation scenarios. For instance, measuring a superstellar C/O ratio would provide further evidence that K2-55b formed beyond the snow line and subsequently migrated inward \citep{oberg_et_al2011}. On the other hand, measuring a substellar C/O ratio could indicate that K2-55b formed inside the ice line \citep{mordasini_et_al2016}.

Transmission spectra would also reveal whether the atmosphere of K2-55b is clear or shrouded by clouds or hazes. \citet{morley_et_al2015} predicted a transition at equilibrium temperatures near 1000K between predominantly hazy atmospheres for cooler planets and predominantly clear atmospheres for hotter planets. \citet{crossfield+kreidberg2017} note that observations of warm Neptunes ($2\rearth < R_p < 6\rearth$, $500{\rm K} < T_{\rm eff} < 1000{\rm K}$) are consistent with this theory, but that the observations cannot yet differentiate between high mean molecular weight atmospheres and high-altitude clouds or hazes for the majority of planets with apparently featureless spectra. Furthermore, the \citet{crossfield+kreidberg2017} sample contains only six warm Neptunes. K2-55b has an equilibrium temperature of roughly 900K and would be an interesting addition to this small sample.

In order to test whether such observations might be feasible, we used the publicly-available  ExoTransmit package \citep{kempton_et_al2017} to generate model atmospheres for K2-55b. We considered a wide variety of atmospheric compositions with a range of C/O ratios. In all cases, the high surface gravity of \mbox{K2-55b} ($22\,{\rm m}\,{\rm s}^{-2} = 2 g_{\rm Neptune}$) muted the dynamic range of atmospheric features, rendering detailed atmospheric characterization challenging. 

Overall, the full range of transit depths is expected to span approximately 150~ppm if the atmosphere has roughly solar composition. Increasing the C/O ratio of a solar metallicity model atmosphere from \mbox{C/O = 0.2} to \mbox{C/O = 1.2} would increase the transit depth by \mbox{50-100~ppm} in the most informative regions ($2-2.5\mu$m and $3-4\mu$m) and produce negligible effects elsewhere in the spectrum. Distinguishing between a water-dominated atmosphere and a carbon dioxide-dominated atmosphere would require detecting differences of roughly 20~ppm. Accordingly, the first-order investigation of whether the atmosphere of K2-55b has a low or high mean molecular weight would be relatively straightforward (assuming that the investigation is not foiled by clouds), but determining detailed molecular abundances would require a more significant investment of telescope time. 

The atmosphere of K2-55b could also be probed during secondary eclipse. Assuming an albedo of 0.15 and an equilibrium temperature of 900K, the estimated secondary eclipse depth is 140~ppm. This modest signal would be challenging to detect with \emph{Spitzer} (\mbox{SNR = 0.8}), but would be detectable with \emph{JWST/MIRI} (\mbox{SNR = 7-8}). For reference, GJ~436b has a secondary eclipse depth of \mbox{$155\pm22$~ppm} at $3.6\mu m$ \citep{morley_et_al2017}, but GJ~436 (\mbox{$V = 10.613$}, \mbox{$Ks = 6.073$}) is significantly brighter than K2-55 (\mbox{$V = 13.55$}, \mbox{$Ks = 10.471$}).

\begin{deluxetable*}{lcl}
\tablecolumns{3}
\tabletypesize{\normalsize}
\tablecaption{K2-55 System Parameters\label{tab:system}}
\tablehead{
\colhead{Parameter} & 
\colhead{Value and $1\sigma$ Errors} &
\colhead{Ref.\tablenotemark{a}}
}
\startdata
\multicolumn{3}{c}{K2-55 (star) = EPIC 205924614} \\
\hline
Right ascension & $22^{\rm h}15^{\rm m}00.462^{\rm s}$ & 1\\
Declination & $-17^{\rm d}15^{\rm m}02.55^{\rm s}$ & 1\\ 
$V$ magnitude& 13.546 & 1\\
Kepler magnitude & 13.087 & 1\\
2MASS K magnitude & 10.471 & 1 \\
$T_{\rm eff}$ (K) & $4300_{-100}^{+107}$ & 2\\
$R_\star (\rsun)$ & $0.715_{-0.040}^{+0.043}$ & 2\\
$M_\star (\msun)$ & $0.688 \pm 0.069$ & 2\\
$[{\rm Fe}/{\rm H}]$ & $0.376 \pm 0.095$ & 2\\
$\log g$ & $4.566 \pm 0.036$ & 2\\
Systemic Velocity\tablenotemark{b}(m s$^{-1}$) & $0.7 \pm 2.1$ & 6\\
RV Jitter (m s$^{-1}$) & $6.8^{+2.3}_{-1.6}$ & 6 \\
Parallax (mas) & $6.240 \pm 0.028$ & 4 \\
Distance (pc) & $159.52^{+0.73}_{-0.72}$  & 5 \\
\hline
\multicolumn{3}{c}{K2-55b (planet) = EPIC 205924614.01} \\
\hline
\emph{Transit and orbital parameters} & & \\
Orbital period $P$ (days) & $2.84927265_{-6.42\times10^{-6}}^{+6.87\times10^{-6}}$ & 6\\
Transit epoch $T_C$ (BJD) & $2456983.4229 \pm 0.00019$& 6\\
$a$ (AU) & $0.0347 \pm 0.001$ & 3\\ 
$R_p/R_\star$ & $0.056_{-0.001}^{+0.003}$ & 6\\
$a/R_\star$ & $10.55_{-1.38}^{+0.64}$ & 6\\
Inc (deg) & $88.05_{-1.75}^{+1.36}$& 6\\
Impact parameter & $0.36_{-0.24}^{+0.23}$ & 6\\
Longitude of periastron $\omega$ (rad)  & fixed to $\pi/2$ & 6\\
Orbital eccentricity $e$  & fixed to 0 & 6\\
RV semi-amplitude $K$ (m s$^{-1}$) & $25.1^{+2.9}_{-3.0}$ & 6\\
\hline
\emph{Planetary parameters} & & \\
$R_p$ ($\rearth$) & $4.41^{+0.32}_{-0.28} $ & 6\\
$M_p$ ($\mearth$) & $43.13^{+5.98}_{-5.80}$ & 6\\
$\rho_p$ (g cm$^{-3}$) &  $2.8_{-0.6}^{+0.8}$ & 6\\
$F_p$ ($F_\oplus$) & $141.3_{-23.5}^{+28.8}$ & 3\\
$T_{\rm eq}$ (K)\tablenotemark{c} & $900$  & 6\\
H/He envelope fraction & $12\pm3$\%  & 6\\
\enddata
\tablenotetext{a}{\textbf{References.} (1)~\citet{huber_et_al2016}, (2)~\citet{dressing_et_al2017a}, (3)~\citet{dressing_et_al2017b}, (4)~\citet{gaia_et_al2018}, (5)~\citet{bailer-jones_et_al2018},  (6)~This Paper}
\tablenotetext{b}{Systemic velocity at BJD~2457689.754631.} 
\tablenotetext{c}{Assuming a Bond albedo of 0.15.}
\end{deluxetable*}

\section{Conclusions}
\label{sec:conclusions}
By adding new \emph{Spitzer}/IRAC and Keck/HIRES observations to extant \emph{K2} and IRTF/SpeX data, we have investigated the composition and formation of K2-55b, a Neptune-sized planet orbiting a metal-rich K7~dwarf. Our \emph{Spitzer}/IRAC data confirmed that K2-55b does not exhibit transit timing variations and verified the accuracy of the $K2$ ephemeris for future transit observations. Our Keck/HIRES data revealed a high mass of $43.13^{+5.98}_{-5.80}\mearth$, which resulted in a bulk density estimate of $2.8_{-0.6}^{+0.8}$~g~cm$^{-3}$ when combined with the radius estimate of $4.41^{+0.32}_{-0.28}\rearth$ from our joint fit to the \emph{K2} and \emph{Spitzer} photometry. By comparing our mass and radius estimates to theoretical models \citep{lopez+fortney2014}, we found that K2-55b can be described by a rocky core surrounded by a modest H/He envelope comprising $12\pm3\%$ of the total planet mass. The full system parameters are displayed in Table~\ref{tab:system}.

Although the envelopes of many similar sized planets contain up to 60\% of the total planet mass \citep{petigura_et_al2017}, only 10\% of the mass of K2-55b is expected to reside in the envelope. The relatively low envelope fraction was surprising because the estimated core mass of K2-55b is significantly larger than the typically quoted value of $10\mearth$ required to spur runaway accretion \citep{ikoma_et_al2000}. We proposed four possible explanations for the absence of a massive envelope: (1) K2-55b drifted into the inner cavity of the disk via Type~I migration just as the envelope was starting to accumulate; (2) K2-55b formed via the collisions of multiple smaller planets after the gas disk dissipated; (3) K2-55b formed with a substantial envelope that was later removed by a giant impact; (4) K2-55b appears unusual only because our understanding of core accretion is incomplete. 

Distinguishing among these scenarios (and others not listed here) will require expanding the sample of Neptune-sized planets with well-constrained densities. Fortunately, there are multiple pathways to find those planets. The NASA \emph{K2} mission is currently searching for transiting planets orbiting tens of thousands of stars in the ecliptic plane, including some cool dwarfs with high metallicities, and more ground-based surveys are beginning operations each year. Although many RV-detected planets will not transit and are therefore poor targets for compositional analyses, knowledge of the orbital periods and approximate masses of non-transiting planets still informs models of planet formation and evolution. 

Beginning later this year, the NASA Transiting Exoplanet Survey Satellite \citep[\emph{TESS},][]{ricker_et_al2014} will conduct a nearly all-sky survey for transiting planets orbiting nearby bright stars. Due to the wide-field nature of the survey, TESS will naturally survey stars with a wide range of metallicities and masses. In the late 2020s, the ESA PLATO mission \citep{rauer_et_al2014} will uncover even more transiting planets orbiting bright stars and precisely constrain host star properties using asteroseimology. Future follow-up observations with extremely precise radial velocity spectrographs will constrain the masses of transiting planets and permit further investigations of the correlations of the compositions of Neptune-sized planets and the minimum mass required to instigate runaway accretion. Atmospheric investigations with \emph{JWST}, \emph{HST}, and \emph{Spitzer} will be particularly useful for tracing present-day planet properties backward to formation scenarios.

\begin{acknowledgments}
This work was performed in part under contract with the Jet Propulsion Laboratory (JPL) funded by NASA through the Sagan Fellowship Program executed by the NASA Exoplanet Science Institute. C.D.D., A.W.H., and I.J.M.C. acknowledge support from the \emph{K2} Guest Observer Program. A.W.H. acknowledges support for our \emph{K2} team through a NASA Astrophysics Data Analysis Program grant and observing support from NASA at Keck Observatory. E.A.P. acknowledges support from Hubble Fellowship grant. We thank the anonymous referee for providing helpful comments that improved the quality of the paper. 

This paper includes data collected by the \emph{K2} mission, which is funded by the NASA Science Mission directorate. The W.M. Keck Observatory is operated as a scientific partnership among the California Institute of Technology, the University of California and the National Aeronautics and Space Administration. The Observatory was made possible by the generous financial support of the W.M. Keck Foundation.  This work is based in part on observations made with the
Spitzer Space Telescope, which is operated by the Jet Propulsion Laboratory, California Institute of Technology under a contract with NASA. This research has made use of the NASA Exoplanet Archive, which is operated by the California Institute of Technology, under contract with the National Aeronautics and Space Administration under the Exoplanet Exploration Program.

 This work has made use of data from the European Space Agency (ESA) mission
{\it Gaia} (\url{https://www.cosmos.esa.int/gaia}), processed by the {\it Gaia}
Data Processing and Analysis Consortium (DPAC,
\url{https://www.cosmos.esa.int/web/gaia/dpac/consortium}). Funding for the DPAC
has been provided by national institutions, in particular the institutions
participating in the {\it Gaia} Multilateral Agreement.

The authors wish to recognize and acknowledge the very significant cultural role and reverence that the summit of Maunakea has always had within the indigenous Hawaiian community.  We are most fortunate to have the opportunity to conduct observations from this mountain. 
\end{acknowledgments}

\facilities{IRTF (SpeX), Keck:I (HIRES), Spitzer (IRAC)}
\software{emcee \citep{foreman-mackey_et_al2013}, ExoTransmit \citep{kempton_et_al2017}, RadVel \citep{fulton_et_al2018}}

\bibliography{../../../mdwarf_biblio}

\begin{thebibliography}{}
\expandafter\ifx\csname natexlab\endcsname\relax\def\natexlab#1{#1}\fi

\bibitem[{{Adams} {et~al.}(2008){Adams}, {Seager}, \&
  {Elkins-Tanton}}]{adams_et_al2008}
{Adams}, E.~R., {Seager}, S., \& {Elkins-Tanton}, L. 2008, \apj, 673, 1160

\bibitem[{{Ag{\'u}ndez} {et~al.}(2014){Ag{\'u}ndez}, {Venot}, {Selsis}, \&
  {Iro}}]{agundez_et_al2014}
{Ag{\'u}ndez}, M., {Venot}, O., {Selsis}, F., \& {Iro}, N. 2014, \apj, 781, 68

\bibitem[{{Aigrain} {et~al.}(2016){Aigrain}, {Parviainen}, \&
  {Pope}}]{aigrain_et_al2016}
{Aigrain}, S., {Parviainen}, H., \& {Pope}, B.~J.~S. 2016, \mnras, 459, 2408

\bibitem[{{Akeson} {et~al.}(2013){Akeson}, {Chen}, {Ciardi}, {Crane}, {Good},
  {Harbut}, {Jackson}, {Kane}, {Laity}, {Leifer}, {Lynn}, {McElroy}, {Papin},
  {Plavchan}, {Ram{\'{\i}}rez}, {Rey}, {von Braun}, {Wittman}, {Abajian},
  {Ali}, {Beichman}, {Beekley}, {Berriman}, {Berukoff}, {Bryden}, {Chan},
  {Groom}, {Lau}, {Payne}, {Regelson}, {Saucedo}, {Schmitz}, {Stauffer},
  {Wyatt}, \& {Zhang}}]{akeson_et_al2013}
{Akeson}, R.~L., {Chen}, X., {Ciardi}, D., {et~al.} 2013, \pasp, 125, 989

\bibitem[{{Bailer-Jones} {et~al.}(2018){Bailer-Jones}, {Rybizki}, {Fouesneau},
  {Mantelet}, \& {Andrae}}]{bailer-jones_et_al2018}
{Bailer-Jones}, C.~A.~L., {Rybizki}, J., {Fouesneau}, M., {Mantelet}, G., \&
  {Andrae}, R. 2018, ArXiv e-prints, arXiv:1804.10121

\bibitem[{{Barrag{\'a}n} {et~al.}(2016){Barrag{\'a}n}, {Grziwa}, {Gandolfi},
  {Fridlund}, {Endl}, {Deeg}, {Cagigal}, {Lanza}, {Prada Moroni}, {Smith},
  {Korth}, {Bedell}, {Cabrera}, {Cochran}, {Cusano}, {Csizmadia},
  {Eigm{\"u}ller}, {Erikson}, {Guenther}, {Hatzes}, {Nespral}, {P{\"a}tzold},
  {Prieto-Arranz}, \& {Rauer}}]{barragan_et_al2016}
{Barrag{\'a}n}, O., {Grziwa}, S., {Gandolfi}, D., {et~al.} 2016, \aj, 152, 193

\bibitem[{{Barros} {et~al.}(2016){Barros}, {Demangeon}, \&
  {Deleuil}}]{barros_et_al2016}
{Barros}, S.~C.~C., {Demangeon}, O., \& {Deleuil}, M. 2016, \aap, 594, A100

\bibitem[{{Bate} {et~al.}(2010){Bate}, {Lodato}, \& {Pringle}}]{bate_et_al2010}
{Bate}, M.~R., {Lodato}, G., \& {Pringle}, J.~E. 2010, \mnras, 401, 1505

\bibitem[{{Batygin} {et~al.}(2009){Batygin}, {Bodenheimer}, \&
  {Laughlin}}]{batygin_et_al2009}
{Batygin}, K., {Bodenheimer}, P., \& {Laughlin}, G. 2009, \apjl, 704, L49

\bibitem[{{Becker} \& {Batygin}(2013)}]{becker+batygin2013}
{Becker}, J.~C., \& {Batygin}, K. 2013, \apj, 778, 100

\bibitem[{{Benneke} {et~al.}(2017){Benneke}, {Werner}, {Petigura}, {Knutson},
  {Dressing}, {Crossfield}, {Schlieder}, {Livingston}, {Beichman},
  {Christiansen}, {Krick}, {Gorjian}, {Howard}, {Sinukoff}, {Ciardi}, \&
  {Akeson}}]{benneke_et_al2017}
{Benneke}, B., {Werner}, M., {Petigura}, E., {et~al.} 2017, \apj, 834, 187

\bibitem[{{Bodenheimer} \& {Pollack}(1986)}]{bodenheimer+pollack1986}
{Bodenheimer}, P., \& {Pollack}, J.~B. 1986, \icarus, 67, 391

\bibitem[{{Boley} {et~al.}(2016){Boley}, {Granados Contreras}, \&
  {Gladman}}]{boley_et_al2016}
{Boley}, A.~C., {Granados Contreras}, A.~P., \& {Gladman}, B. 2016, \apjl, 817,
  L17

\bibitem[{{Bonomo} {et~al.}(2014){Bonomo}, {Sozzetti}, {Lovis}, {Malavolta},
  {Rice}, {Buchhave}, {Sasselov}, {Cameron}, {Latham}, {Molinari}, {Pepe},
  {Udry}, {Affer}, {Charbonneau}, {Cosentino}, {Dressing}, {Dumusque},
  {Figueira}, {Fiorenzano}, {Gettel}, {Harutyunyan}, {Haywood}, {Horne},
  {Lopez-Morales}, {Mayor}, {Micela}, {Motalebi}, {Nascimbeni}, {Phillips},
  {Piotto}, {Pollacco}, {Queloz}, {S{\'e}gransan}, {Szentgyorgyi}, \&
  {Watson}}]{bonomo_et_al2014}
{Bonomo}, A.~S., {Sozzetti}, A., {Lovis}, C., {et~al.} 2014, \aap, 572, A2

\bibitem[{{Boyajian} {et~al.}(2013){Boyajian}, {von Braun}, {van Belle},
  {Farrington}, {Schaefer}, {Jones}, {White}, {McAlister}, {ten Brummelaar},
  {Ridgway}, {Gies}, {Sturmann}, {Sturmann}, {Turner}, {Goldfinger}, \&
  {Vargas}}]{boyajian_et_al2013}
{Boyajian}, T.~S., {von Braun}, K., {van Belle}, G., {et~al.} 2013, \apj, 771,
  40

\bibitem[{{Buchhave} {et~al.}(2014){Buchhave}, {Bizzarro}, {Latham},
  {Sasselov}, {Cochran}, {Endl}, {Isaacson}, {Juncher}, \&
  {Marcy}}]{buchhave_et_al2014}
{Buchhave}, L.~A., {Bizzarro}, M., {Latham}, D.~W., {et~al.} 2014, \nat, 509,
  593

\bibitem[{{Buhler} {et~al.}(2016){Buhler}, {Knutson}, {Batygin}, {Fulton},
  {Fortney}, {Burrows}, \& {Wong}}]{buhler_et_al2016}
{Buhler}, P.~B., {Knutson}, H.~A., {Batygin}, K., {et~al.} 2016, \apj, 821, 26

\bibitem[{{Bur\v{s}a}(1992)}]{bursa1992}
{Bur\v{s}a}, M. 1992, Earth Moon and Planets, 59, 239

\bibitem[{{Butler} {et~al.}(1996){Butler}, {Marcy}, {Williams}, {McCarthy},
  {Dosanjh}, \& {Vogt}}]{butler_et_al1996}
{Butler}, R.~P., {Marcy}, G.~W., {Williams}, E., {et~al.} 1996, \pasp, 108, 500

\bibitem[{{Chatterjee} \& {Tan}(2014)}]{chatterjee+tan2014}
{Chatterjee}, S., \& {Tan}, J.~C. 2014, \apj, 780, 53

\bibitem[{{Chen} {et~al.}(2018){Chen}, {Knutson}, {Dressing}, {Morley},
  {Werner}, {Gorjian}, {Beichman}, {Benneke}, {Christiansen}, {Ciardi},
  {Crossfield}, {Howell}, {Krick}, {Livingston}, {Morales}, \&
  {Schlieder}}]{ge_et_al2018}
{Chen}, G., {Knutson}, H.~A., {Dressing}, C.~D., {et~al.} 2018, ArXiv e-prints,
  arXiv:1801.10177

\bibitem[{{Chiang} \& {Laughlin}(2013)}]{chiang+laughlin2013}
{Chiang}, E., \& {Laughlin}, G. 2013, \mnras, 431, 3444

\bibitem[{{Chiang} \& {Youdin}(2010)}]{chiang+youdin2010}
{Chiang}, E., \& {Youdin}, A.~N. 2010, Annual Review of Earth and Planetary
  Sciences, 38, 493

\bibitem[{{Christiansen} {et~al.}(2016){Christiansen}, {Clarke}, {Burke},
  {Jenkins}, {Bryson}, {Coughlin}, {Mullally}, {Thompson}, {Twicken},
  {Batalha}, {Haas}, {Catanzarite}, {Campbell}, {Kamal Uddin}, {Zamudio},
  {Smith}, \& {Henze}}]{christiansen_et_al2016}
{Christiansen}, J.~L., {Clarke}, B.~D., {Burke}, C.~J., {et~al.} 2016, \apj,
  828, 99

\bibitem[{{Claret} \& {Bloemen}(2011)}]{claret+bloemen2011}
{Claret}, A., \& {Bloemen}, S. 2011, \aap, 529, A75

\bibitem[{{Crossfield} \& {Kreidberg}(2017)}]{crossfield+kreidberg2017}
{Crossfield}, I.~J.~M., \& {Kreidberg}, L. 2017, \aj, 154, 261

\bibitem[{{Crossfield} {et~al.}(2016){Crossfield}, {Ciardi}, {Petigura},
  {Sinukoff}, {Schlieder}, {Howard}, {Beichman}, {Isaacson}, {Dressing},
  {Christiansen}, {Fulton}, {L{\'e}pine}, {Weiss}, {Hirsch}, {Livingston},
  {Baranec}, {Law}, {Riddle}, {Ziegler}, {Howell}, {Horch}, {Everett}, {Teske},
  {Martinez}, {Obermeier}, {Benneke}, {Scott}, {Deacon}, {Aller}, {Hansen},
  {Mancini}, {Ciceri}, {Brahm}, {Jord{\'a}n}, {Knutson}, {Henning}, {Bonnefoy},
  {Liu}, {Crepp}, {Lothringer}, {Hinz}, {Bailey}, {Skemer}, \&
  {Defrere}}]{crossfield_et_al2016}
{Crossfield}, I.~J.~M., {Ciardi}, D.~R., {Petigura}, E.~A., {et~al.} 2016,
  \apjs, 226, 7

\bibitem[{{Demangeon} {et~al.}(2018){Demangeon}, {Faedi}, {H{\'e}brard},
  {Brown}, {Barros}, {Doyle}, {Maxted}, {Cameron}, {Hay}, {Alikakos},
  {Anderson}, {Armstrong}, {Boumis}, {Bonomo}, {Bouchy}, {Delrez}, {Gillon},
  {Haswell}, {Hellier}, {Jehin}, {Kiefer}, {Lam}, {Lendl}, {Mancini},
  {McCormac}, {Norton}, {Osborn}, {Palle}, {Pepe}, {Pollacco}, {Prieto-Arranz},
  {Queloz}, {S{\'e}gransan}, {Smalley}, {Triaud}, {Udry}, {West}, \&
  {Wheatley}}]{demangeon_et_al2018}
{Demangeon}, O.~D.~S., {Faedi}, F., {H{\'e}brard}, G., {et~al.} 2018, \aap,
  610, A63

\bibitem[{{Deming} {et~al.}(2015){Deming}, {Knutson}, {Kammer}, {Fulton},
  {Ingalls}, {Carey}, {Burrows}, {Fortney}, {Todorov}, {Agol}, {Cowan},
  {Desert}, {Fraine}, {Langton}, {Morley}, \& {Showman}}]{deming_et_al2015}
{Deming}, D., {Knutson}, H., {Kammer}, J., {et~al.} 2015, \apj, 805, 132

\bibitem[{{Dodson-Robinson} \&
  {Bodenheimer}(2010)}]{dodson-robinson+bodenheimer2010}
{Dodson-Robinson}, S.~E., \& {Bodenheimer}, P. 2010, \icarus, 207, 491

\bibitem[{{Dressing} \& {Charbonneau}(2013)}]{dressing+charbonneau2013}
{Dressing}, C.~D., \& {Charbonneau}, D. 2013, \apj, 767, 95

\bibitem[{{Dressing} {et~al.}(2017{\natexlab{a}}){Dressing}, {Newton},
  {Schlieder}, {Charbonneau}, {Knutson}, {Vanderburg}, \&
  {Sinukoff}}]{dressing_et_al2017a}
{Dressing}, C.~D., {Newton}, E.~R., {Schlieder}, J.~E., {et~al.}
  2017{\natexlab{a}}, \apj, 836, 167

\bibitem[{{Dressing} {et~al.}(2017{\natexlab{b}}){Dressing}, {Vanderburg},
  {Schlieder}, {Crossfield}, {Knutson}, {Newton}, {Ciardi}, {Fulton},
  {Gonzales}, {Howard}, {Isaacson}, {Livingston}, {Petigura}, {Sinukoff},
  {Everett}, {Horch}, \& {Howell}}]{dressing_et_al2017b}
{Dressing}, C.~D., {Vanderburg}, A., {Schlieder}, J.~E., {et~al.}
  2017{\natexlab{b}}, \aj, 154, 207

\bibitem[{{Dumusque} {et~al.}(2014){Dumusque}, {Bonomo}, {Haywood},
  {Malavolta}, {S{\'e}gransan}, {Buchhave}, {Collier Cameron}, {Latham},
  {Molinari}, {Pepe}, {Udry}, {Charbonneau}, {Cosentino}, {Dressing},
  {Figueira}, {Fiorenzano}, {Gettel}, {Harutyunyan}, {Horne}, {Lopez-Morales},
  {Lovis}, {Mayor}, {Micela}, {Motalebi}, {Nascimbeni}, {Phillips}, {Piotto},
  {Pollacco}, {Queloz}, {Rice}, {Sasselov}, {Sozzetti}, {Szentgyorgyi}, \&
  {Watson}}]{dumusque_et_al2014}
{Dumusque}, X., {Bonomo}, A.~S., {Haywood}, R.~D., {et~al.} 2014, \apj, 789,
  154

\bibitem[{{Eastman} {et~al.}(2013){Eastman}, {Gaudi}, \&
  {Agol}}]{eastman_et_al2013}
{Eastman}, J., {Gaudi}, B.~S., \& {Agol}, E. 2013, \pasp, 125, 83

\bibitem[{{Figueira} {et~al.}(2009){Figueira}, {Pont}, {Mordasini}, {Alibert},
  {Georgy}, \& {Benz}}]{figueria_et_al2009}
{Figueira}, P., {Pont}, F., {Mordasini}, C., {et~al.} 2009, \aap, 493, 671

\bibitem[{{Fischer} \& {Valenti}(2005)}]{fischer+valenti2005}
{Fischer}, D.~A., \& {Valenti}, J. 2005, \apj, 622, 1102

\bibitem[{{Fischer} {et~al.}(2016){Fischer}, {Anglada-Escude}, {Arriagada},
  {Baluev}, {Bean}, {Bouchy}, {Buchhave}, {Carroll}, {Chakraborty}, {Crepp},
  {Dawson}, {Diddams}, {Dumusque}, {Eastman}, {Endl}, {Figueira}, {Ford},
  {Foreman-Mackey}, {Fournier}, {F{\H u}r{\'e}sz}, {Gaudi}, {Gregory},
  {Grundahl}, {Hatzes}, {H{\'e}brard}, {Herrero}, {Hogg}, {Howard}, {Johnson},
  {Jorden}, {Jurgenson}, {Latham}, {Laughlin}, {Loredo}, {Lovis}, {Mahadevan},
  {McCracken}, {Pepe}, {Perez}, {Phillips}, {Plavchan}, {Prato}, {Quirrenbach},
  {Reiners}, {Robertson}, {Santos}, {Sawyer}, {Segransan}, {Sozzetti},
  {Steinmetz}, {Szentgyorgyi}, {Udry}, {Valenti}, {Wang}, {Wittenmyer}, \&
  {Wright}}]{fischer_et_al2016}
{Fischer}, D.~A., {Anglada-Escude}, G., {Arriagada}, P., {et~al.} 2016, \pasp,
  128, 066001

\bibitem[{{Ford}(2006)}]{ford2006}
{Ford}, E.~B. 2006, \apj, 642, 505

\bibitem[{{Foreman-Mackey} {et~al.}(2013){Foreman-Mackey}, {Hogg}, {Lang}, \&
  {Goodman}}]{foreman-mackey_et_al2013}
{Foreman-Mackey}, D., {Hogg}, D.~W., {Lang}, D., \& {Goodman}, J. 2013, \pasp,
  125, 306

\bibitem[{{Fressin} {et~al.}(2013){Fressin}, {Torres}, {Charbonneau}, {Bryson},
  {Christiansen}, {Dressing}, {Jenkins}, {Walkowicz}, \&
  {Batalha}}]{fressin_et_al2013}
{Fressin}, F., {Torres}, G., {Charbonneau}, D., {et~al.} 2013, \apj, 766, 81

\bibitem[{{Fulton} {et~al.}(2018){Fulton}, {Petigura}, {Blunt}, \&
  {Sinukoff}}]{fulton_et_al2018}
{Fulton}, B.~J., {Petigura}, E.~A., {Blunt}, S., \& {Sinukoff}, E. 2018, \pasp,
  130, 044504

\bibitem[{{Fulton} {et~al.}(2017){Fulton}, {Petigura}, {Howard}, {Isaacson},
  {Marcy}, {Cargile}, {Hebb}, {Weiss}, {Johnson}, {Morton}, {Sinukoff},
  {Crossfield}, \& {Hirsch}}]{fulton_et_al2017}
{Fulton}, B.~J., {Petigura}, E.~A., {Howard}, A.~W., {et~al.} 2017, \aj, 154,
  109

\bibitem[{{Gaia Collaboration} {et~al.}(2018){Gaia Collaboration}, {Brown},
  {Vallenari}, {Prusti}, {de Bruijne}, {Babusiaux}, \&
  {Bailer-Jones}}]{gaia_et_al2018}
{Gaia Collaboration}, {Brown}, A.~G.~A., {Vallenari}, A., {et~al.} 2018, ArXiv
  e-prints, arXiv:1804.09365

\bibitem[{{Gaia Collaboration} {et~al.}(2016){Gaia Collaboration}, {Prusti},
  {de Bruijne}, {Brown}, {Vallenari}, {Babusiaux}, {Bailer-Jones}, {Bastian},
  {Biermann}, {Evans}, {Eyer}, {Jansen}, {Jordi}, {Klioner}, {Lammers},
  {Lindegren}, {Luri}, {Mignard}, {Milligan}, {Panem}, {Poinsignon},
  {Pourbaix}, {Randich}, {Sarri}, {Sartoretti}, {Siddiqui}, {Soubiran},
  {Valette}, {van Leeuwen}, {Walton}, {Aerts}, {Arenou}, {Cropper}, {Drimmel},
  {H{\o}g}, {Katz}, {Lattanzi}, {O'Mullane}, {Grebel}, {Holland}, {Huc},
  {Passot}, {Bramante}, {Cacciari}, {Casta{\~n}eda}, {Chaoul}, {Cheek}, {De
  Angeli}, {Fabricius}, {Guerra}, {Hern{\'a}ndez}, {Jean-Antoine-Piccolo},
  {Masana}, {Messineo}, {Mowlavi}, {Nienartowicz}, {Ord{\'o}{\~n}ez- Blanco},
  {Panuzzo}, {Portell}, {Richards}, {Riello}, {Seabroke}, {Tanga},
  {Th{\'e}venin}, {Torra}, {Els}, {Gracia- Abril}, {Comoretto},
  {Garcia-Reinaldos}, {Lock}, {Mercier}, {Altmann}, {Andrae}, {Astraatmadja},
  {Bellas-Velidis}, {Benson}, {Berthier}, {Blomme}, {Busso}, {Carry},
  {Cellino}, {Clementini}, {Cowell}, {Creevey}, {Cuypers}, {Davidson}, {De
  Ridder}, {de Torres}, {Delchambre}, {Dell'Oro}, {Ducourant}, {Fr{\'e}mat},
  {Garc{\'\i}a-Torres}, {Gosset}, {Halbwachs}, {Hambly}, {Harrison}, {Hauser},
  {Hestroffer}, {Hodgkin}, {Huckle}, {Hutton}, {Jasniewicz}, {Jordan},
  {Kontizas}, {Korn}, {Lanzafame}, {Manteiga}, {Moitinho}, {Muinonen},
  {Osinde}, {Pancino}, {Pauwels}, {Petit}, {Recio-Blanco}, {Robin}, {Sarro},
  {Siopis}, {Smith}, {Smith}, {Sozzetti}, {Thuillot}, {van Reeven}, {Viala},
  {Abbas}, {Abreu Aramburu}, {Accart}, {Aguado}, {Allan}, {Allasia},
  {Altavilla}, {{\'A}lvarez}, {Alves}, {Anderson}, {Andrei}, {Anglada Varela},
  {Antiche}, {Antoja}, {Ant{\'o}n}, {Arcay}, {Atzei}, {Ayache}, {Bach},
  {Baker}, {Balaguer-N{\'u}{\~n}ez}, {Barache}, {Barata}, {Barbier}, {Barblan},
  {Baroni}, {Barrado y Navascu{\'e}s}, {Barros}, {Barstow}, {Becciani},
  {Bellazzini}, {Bellei}, {Bello Garc{\'\i}a}, {Belokurov}, {Bendjoya},
  {Berihuete}, {Bianchi}, {Bienaym{\'e}}, {Billebaud}, {Blagorodnova},
  {Blanco-Cuaresma}, {Boch}, {Bombrun}, {Borrachero}, {Bouquillon}, {Bourda},
  {Bouy}, {Bragaglia}, {Breddels}, {Brouillet}, {Br{\"u}semeister},
  {Bucciarelli}, {Budnik}, {Burgess}, {Burgon}, {Burlacu}, {Busonero}, {Buzzi},
  {Caffau}, {Cambras}, {Campbell}, {Cancelliere}, {Cantat-Gaudin}, {Carlucci},
  {Carrasco}, {Castellani}, {Charlot}, {Charnas}, {Charvet}, {Chassat},
  {Chiavassa}, {Clotet}, {Cocozza}, {Collins}, {Collins}, {Costigan}, {Crifo},
  {Cross}, {Crosta}, {Crowley}, {Dafonte}, {Damerdji}, {Dapergolas}, {David},
  {David}, {De Cat}, {de Felice}, {de Laverny}, {De Luise}, {De March}, {de
  Martino}, {de Souza}, {Debosscher}, {del Pozo}, {Delbo}, {Delgado},
  {Delgado}, {di Marco}, {Di Matteo}, {Diakite}, {Distefano}, {Dolding}, {Dos
  Anjos}, {Drazinos}, {Dur{\'a}n}, {Dzigan}, {Ecale}, {Edvardsson}, {Enke},
  {Erdmann}, {Escolar}, {Espina}, {Evans}, {Eynard Bontemps}, {Fabre},
  {Fabrizio}, {Faigler}, {Falc{\~a}o}, {Farr{\`a}s Casas}, {Faye}, {Federici},
  {Fedorets}, {Fern{\'a}ndez-Hern{\'a}ndez}, {Fernique}, {Fienga}, {Figueras},
  {Filippi}, {Findeisen}, {Fonti}, {Fouesneau}, {Fraile}, {Fraser}, {Fuchs},
  {Furnell}, {Gai}, {Galleti}, {Galluccio}, {Garabato}, {Garc{\'\i}a-Sedano},
  {Gar{\'e}}, {Garofalo}, {Garralda}, {Gavras}, {Gerssen}, {Geyer}, {Gilmore},
  {Girona}, {Giuffrida}, {Gomes}, {Gonz{\'a}lez-Marcos},
  {Gonz{\'a}lez-N{\'u}{\~n}ez}, {Gonz{\'a}lez-Vidal}, {Granvik}, {Guerrier},
  {Guillout}, {Guiraud}, {G{\'u}rpide}, {Guti{\'e}rrez-S{\'a}nchez}, {Guy},
  {Haigron}, {Hatzidimitriou}, {Haywood}, {Heiter}, {Helmi}, {Hobbs},
  {Hofmann}, {Holl}, {Holland}, {Hunt}, {Hypki}, {Icardi}, {Irwin}, {Jevardat
  de Fombelle}, {Jofr{\'e}}, {Jonker}, {Jorissen}, {Julbe}, {Karampelas},
  {Kochoska}, {Kohley}, {Kolenberg}, {Kontizas}, {Koposov}, {Kordopatis},
  {Koubsky}, {Kowalczyk}, {Krone-Martins}, {Kudryashova}, {Kull}, {Bachchan},
  {Lacoste-Seris}, {Lanza}, {Lavigne}, {Le Poncin-Lafitte}, {Lebreton},
  {Lebzelter}, {Leccia}, {Leclerc}, {Lecoeur-Taibi}, {Lemaitre}, {Lenhardt},
  {Leroux}, {Liao}, {Licata}, {Lindstr{\o}m}, {Lister}, {Livanou}, {Lobel},
  {L{\"o}ffler}, {L{\'o}pez}, {Lopez-Lozano}, {Lorenz}, {Loureiro},
  {MacDonald}, {Magalh{\~a}es Fernandes}, {Managau}, {Mann}, {Mantelet},
  {Marchal}, {Marchant}, {Marconi}, {Marie}, {Marinoni}, {Marrese},
  {Marschalk{\'o}}, {Marshall}, {Mart{\'\i}n-Fleitas}, {Martino}, {Mary},
  {Matijevi{\v{c}}}, {Mazeh}, {McMillan}, {Messina}, {Mestre}, {Michalik},
  {Millar}, {Miranda}, {Molina}, {Molinaro}, {Molinaro}, {Moln{\'a}r},
  {Moniez}, {Montegriffo}, {Monteiro}, {Mor}, {Mora}, {Morbidelli}, {Morel},
  {Morgenthaler}, {Morley}, {Morris}, {Mulone}, {Muraveva}, {Musella},
  {Narbonne}, {Nelemans}, {Nicastro}, {Noval}, {Ord{\'e}novic},
  {Ordieres-Mer{\'e}}, {Osborne}, {Pagani}, {Pagano}, {Pailler}, {Palacin},
  {Palaversa}, {Parsons}, {Paulsen}, {Pecoraro}, {Pedrosa}, {Pentik{\"a}inen},
  {Pereira}, {Pichon}, {Piersimoni}, {Pineau}, {Plachy}, {Plum}, {Poujoulet},
  {Pr{\v{s}}a}, {Pulone}, {Ragaini}, {Rago}, {Rambaux}, {Ramos-Lerate},
  {Ranalli}, {Rauw}, {Read}, {Regibo}, {Renk}, {Reyl{\'e}}, {Ribeiro},
  {Rimoldini}, {Ripepi}, {Riva}, {Rixon}, {Roelens}, {Romero-G{\'o}mez},
  {Rowell}, {Royer}, {Rudolph}, {Ruiz-Dern}, {Sadowski}, {Sagrist{\`a}
  Sell{\'e}s}, {Sahlmann}, {Salgado}, {Salguero}, {Sarasso}, {Savietto},
  {Schnorhk}, {Schultheis}, {Sciacca}, {Segol}, {Segovia}, {Segransan},
  {Serpell}, {Shih}, {Smareglia}, {Smart}, {Smith}, {Solano}, {Solitro},
  {Sordo}, {Soria Nieto}, {Souchay}, {Spagna}, {Spoto}, {Stampa}, {Steele},
  {Steidelm{\"u}ller}, {Stephenson}, {Stoev}, {Suess}, {S{\"u}veges}, {Surdej},
  {Szabados}, {Szegedi-Elek}, {Tapiador}, {Taris}, {Tauran}, {Taylor},
  {Teixeira}, {Terrett}, {Tingley}, {Trager}, {Turon}, {Ulla}, {Utrilla},
  {Valentini}, {van Elteren}, {Van Hemelryck}, {van Leeuwen}, {Varadi},
  {Vecchiato}, {Veljanoski}, {Via}, {Vicente}, {Vogt}, {Voss}, {Votruba},
  {Voutsinas}, {Walmsley}, {Weiler}, {Weingrill}, {Werner}, {Wevers},
  {Whitehead}, {Wyrzykowski}, {Yoldas}, {{\v{Z}}erjal}, {Zucker}, {Zurbach},
  {Zwitter}, {Alecu}, {Allen}, {Allende Prieto}, {Amorim},
  {Anglada-Escud{\'e}}, {Arsenijevic}, {Azaz}, {Balm}, {Beck}, {Bernstein},
  {Bigot}, {Bijaoui}, {Blasco}, {Bonfigli}, {Bono}, {Boudreault}, {Bressan},
  {Brown}, {Brunet}, {Bunclark}, {Buonanno}, {Butkevich}, {Carret}, {Carrion},
  {Chemin}, {Ch{\'e}reau}, {Corcione}, {Darmigny}, {de Boer}, {de Teodoro}, {de
  Zeeuw}, {Delle Luche}, {Domingues}, {Dubath}, {Fodor}, {Fr{\'e}zouls},
  {Fries}, {Fustes}, {Fyfe}, {Gallardo}, {Gallegos}, {Gardiol}, {Gebran},
  {Gomboc}, {G{\'o}mez}, {Grux}, {Gueguen}, {Heyrovsky}, {Hoar}, {Iannicola},
  {Isasi Parache}, {Janotto}, {Joliet}, {Jonckheere}, {Keil}, {Kim},
  {Klagyivik}, {Klar}, {Knude}, {Kochukhov}, {Kolka}, {Kos}, {Kutka}, {Lainey},
  {LeBouquin}, {Liu}, {Loreggia}, {Makarov}, {Marseille}, {Martayan},
  {Martinez-Rubi}, {Massart}, {Meynadier}, {Mignot}, {Munari}, {Nguyen},
  {Nordlander}, {Ocvirk}, {O'Flaherty}, {Olias Sanz}, {Ortiz}, {Osorio},
  {Oszkiewicz}, {Ouzounis}, {Palmer}, {Park}, {Pasquato}, {Peltzer}, {Peralta},
  {P{\'e}turaud}, {Pieniluoma}, {Pigozzi}, {Poels}, {Prat}, {Prod'homme},
  {Raison}, {Rebordao}, {Risquez}, {Rocca-Volmerange}, {Rosen}, {Ruiz-Fuertes},
  {Russo}, {Sembay}, {Serraller Vizcaino}, {Short}, {Siebert}, {Silva},
  {Sinachopoulos}, {Slezak}, {Soffel}, {Sosnowska}, {Strai{\v{z}}ys}, {ter
  Linden}, {Terrell}, {Theil}, {Tiede}, {Troisi}, {Tsalmantza}, {Tur},
  {Vaccari}, {Vachier}, {Valles}, {Van Hamme}, {Veltz}, {Virtanen}, {Wallut},
  {Wichmann}, {Wilkinson}, {Ziaeepour}, \& {Zschocke}}]{gaia_et_al2016}
{Gaia Collaboration}, {Prusti}, T., {de Bruijne}, J.~H.~J., {et~al.} 2016,
  \aap, 595, A1

\bibitem[{{Gaidos}(2017)}]{gaidos2017}
{Gaidos}, E. 2017, \mnras, 470, L1

\bibitem[{{Gelman} \& {Rubin}(1992)}]{gelman+rubin1992}
{Gelman}, A., \& {Rubin}, D.~B. 1992, Stat. Sci., 7, 457

\bibitem[{{Ginzburg} \& {Sari}(2017)}]{ginzburg+sari2017}
{Ginzburg}, S., \& {Sari}, R. 2017, \mnras, 464, 3937

\bibitem[{{Goldreich} \& {Soter}(1966)}]{goldreich+soter1966}
{Goldreich}, P., \& {Soter}, S. 1966, \icarus, 5, 375

\bibitem[{{Gonzalez}(1997)}]{gonzalez1997}
{Gonzalez}, G. 1997, \mnras, 285, 403

\bibitem[{{Goodman} \& {Weare}(2010)}]{goodman+weare2010}
{Goodman}, J., \& {Weare}, J. 2010, Communications in Applied Mathematics and
  Computational Science, Vol.~5, No.~1, p.~65-80, 2010, 5, 65

\bibitem[{{Grillmair} {et~al.}(2012){Grillmair}, {Carey}, {Stauffer}, {Fisher},
  {Olds}, {Ingalls}, {Krick}, {Glaccum}, {Laine}, {Lowrance}, \&
  {Surace}}]{grillmair_et_al2012}
{Grillmair}, C.~J., {Carey}, S.~J., {Stauffer}, J.~R., {et~al.} 2012, in
  \procspie, Vol. 8448, Observatory Operations: Strategies, Processes, and
  Systems IV, 84481I

\bibitem[{{Hansen} \& {Murray}(2012)}]{hansen+murray2012}
{Hansen}, B.~M.~S., \& {Murray}, N. 2012, \apj, 751, 158

\bibitem[{{Hardy} {et~al.}(2017){Hardy}, {Harrington}, {Hardin}, {Madhusudhan},
  {Loredo}, {Challener}, {Foster}, {Cubillos}, \& {Blecic}}]{hardy_et_al2017}
{Hardy}, R.~A., {Harrington}, J., {Hardin}, M.~R., {et~al.} 2017, \apj, 836,
  143

\bibitem[{{Hayashi}(1981)}]{hayashi1981}
{Hayashi}, C. 1981, Progress of Theoretical Physics Supplement, 70, 35

\bibitem[{{Hori} \& {Ikoma}(2011)}]{hori+ikoma2011}
{Hori}, Y., \& {Ikoma}, M. 2011, \mnras, 416, 1419

\bibitem[{{Howard} {et~al.}(2009){Howard}, {Johnson}, {Marcy}, {Fischer},
  {Wright}, {Henry}, {Giguere}, {Isaacson}, {Valenti}, {Anderson}, \&
  {Piskunov}}]{howard_et_al2009}
{Howard}, A.~W., {Johnson}, J.~A., {Marcy}, G.~W., {et~al.} 2009, \apj, 696, 75

\bibitem[{{Howard} {et~al.}(2010){Howard}, {Johnson}, {Marcy}, {Fischer},
  {Wright}, {Bernat}, {Henry}, {Peek}, {Isaacson}, {Apps}, {Endl}, {Cochran},
  {Valenti}, {Anderson}, \& {Piskunov}}]{howard_et_al2010b}
---. 2010, \apj, 721, 1467

\bibitem[{{Howard} {et~al.}(2012){Howard}, {Marcy}, {Bryson}, {Jenkins},
  {Rowe}, {Batalha}, {Borucki}, {Koch}, {Dunham}, {Gautier}, {Van Cleve},
  {Cochran}, {Latham}, {Lissauer}, {Torres}, {Brown}, {Gilliland}, {Buchhave},
  {Caldwell}, {Christensen-Dalsgaard}, {Ciardi}, {Fressin}, {Haas}, {Howell},
  {Kjeldsen}, {Seager}, {Rogers}, {Sasselov}, {Steffen}, {Basri},
  {Charbonneau}, {Christiansen}, {Clarke}, {Dupree}, {Fabrycky}, {Fischer},
  {Ford}, {Fortney}, {Tarter}, {Girouard}, {Holman}, {Johnson}, {Klaus},
  {Machalek}, {Moorhead}, {Morehead}, {Ragozzine}, {Tenenbaum}, {Twicken},
  {Quinn}, {Isaacson}, {Shporer}, {Lucas}, {Walkowicz}, {Welsh}, {Boss},
  {Devore}, {Gould}, {Smith}, {Morris}, {Prsa}, {Morton}, {Still}, {Thompson},
  {Mullally}, {Endl}, \& {MacQueen}}]{howard_et_al2012}
{Howard}, A.~W., {Marcy}, G.~W., {Bryson}, S.~T., {et~al.} 2012, \apjs, 201, 15

\bibitem[{{Howard} {et~al.}(2014){Howard}, {Marcy}, {Fischer}, {Isaacson},
  {Muirhead}, {Henry}, {Boyajian}, {von Braun}, {Becker}, {Wright}, \&
  {Johnson}}]{howard_et_al2014}
{Howard}, A.~W., {Marcy}, G.~W., {Fischer}, D.~A., {et~al.} 2014, \apj, 794, 51

\bibitem[{{Howell} {et~al.}(2014){Howell}, {Sobeck}, {Haas}, {Still},
  {Barclay}, {Mullally}, {Troeltzsch}, {Aigrain}, {Bryson}, {Caldwell},
  {Chaplin}, {Cochran}, {Huber}, {Marcy}, {Miglio}, {Najita}, {Smith},
  {Twicken}, \& {Fortney}}]{howell_et_al2014}
{Howell}, S.~B., {Sobeck}, C., {Haas}, M., {et~al.} 2014, PASP, 126, 398

\bibitem[{{Huber} {et~al.}(2016){Huber}, {Bryson}, {Haas}, {Barclay},
  {Barentsen}, {Howell}, {Sharma}, {Stello}, \& {Thompson}}]{huber_et_al2016}
{Huber}, D., {Bryson}, S.~T., {Haas}, M.~R., {et~al.} 2016, \apjs, 224, 2

\bibitem[{{Ikoma} {et~al.}(2000){Ikoma}, {Nakazawa}, \&
  {Emori}}]{ikoma_et_al2000}
{Ikoma}, M., {Nakazawa}, K., \& {Emori}, H. 2000, \apj, 537, 1013

\bibitem[{{Inamdar} \& {Schlichting}(2015)}]{inamdar+schlichting2015}
{Inamdar}, N.~K., \& {Schlichting}, H.~E. 2015, \mnras, 448, 1751

\bibitem[{{Inamdar} \& {Schlichting}(2016)}]{inamdar+schlichting2016}
---. 2016, \apjl, 817, L13

\bibitem[{{Ingalls} {et~al.}(2012){Ingalls}, {Krick}, {Carey}, {Laine},
  {Surace}, {Glaccum}, {Grillmair}, \& {Lowrance}}]{ingalls_et_al2012}
{Ingalls}, J.~G., {Krick}, J.~E., {Carey}, S.~J., {et~al.} 2012, in \procspie,
  Vol. 8442, Space Telescopes and Instrumentation 2012: Optical, Infrared, and
  Millimeter Wave, 84421Y

\bibitem[{{Kammer} {et~al.}(2015){Kammer}, {Knutson}, {Line}, {Fortney},
  {Deming}, {Burrows}, {Cowan}, {Triaud}, {Agol}, {Desert}, {Fulton}, {Howard},
  {Laughlin}, {Lewis}, {Morley}, {Moses}, {Showman}, \&
  {Todorov}}]{kammer_et_al2015}
{Kammer}, J.~A., {Knutson}, H.~A., {Line}, M.~R., {et~al.} 2015, \apj, 810, 118

\bibitem[{{Kempton} {et~al.}(2017){Kempton}, {Lupu}, {Owusu-Asare}, {Slough},
  \& {Cale}}]{kempton_et_al2017}
{Kempton}, E.~M.-R., {Lupu}, R., {Owusu-Asare}, A., {Slough}, P., \& {Cale}, B.
  2017, \pasp, 129, 044402

\bibitem[{{Kipping}(2013)}]{kipping2013}
{Kipping}, D.~M. 2013, ArXiv e-prints, arXiv:1311.1170

\bibitem[{{Kley} \& {Nelson}(2012)}]{kley+nelson2012}
{Kley}, W., \& {Nelson}, R.~P. 2012, \araa, 50, 211

\bibitem[{{Knutson} {et~al.}(2012){Knutson}, {Lewis}, {Fortney}, {Burrows},
  {Showman}, {Cowan}, {Agol}, {Aigrain}, {Charbonneau}, {Deming}, {D{\'e}sert},
  {Henry}, {Langton}, \& {Laughlin}}]{knutson_et_al2012}
{Knutson}, H.~A., {Lewis}, N., {Fortney}, J.~J., {et~al.} 2012, \apj, 754, 22

\bibitem[{{Kolbl} {et~al.}(2015){Kolbl}, {Marcy}, {Isaacson}, \&
  {Howard}}]{kolbl_et_al2015}
{Kolbl}, R., {Marcy}, G.~W., {Isaacson}, H., \& {Howard}, A.~W. 2015, \aj, 149,
  18

\bibitem[{{Kostov} {et~al.}(2014){Kostov}, {McCullough}, {Carter}, {Deleuil},
  {D{\'{\i}}az}, {Fabrycky}, {H{\'e}brard}, {Hinse}, {Mazeh}, {Orosz},
  {Tsvetanov}, \& {Welsh}}]{kostov_et_al2014}
{Kostov}, V.~B., {McCullough}, P.~R., {Carter}, J.~A., {et~al.} 2014, \apj,
  784, 14

\bibitem[{{Kramm} {et~al.}(2012){Kramm}, {Nettelmann}, {Fortney},
  {Neuh{\"a}user}, \& {Redmer}}]{kramm_et_al2012}
{Kramm}, U., {Nettelmann}, N., {Fortney}, J.~J., {Neuh{\"a}user}, R., \&
  {Redmer}, R. 2012, \aap, 538, A146

\bibitem[{{Kreidberg}(2015)}]{kreidberg2015}
{Kreidberg}, L. 2015, \pasp, 127, 1161

\bibitem[{{Kuchner} \& {Lecar}(2002)}]{kuchner+lecar2002}
{Kuchner}, M.~J., \& {Lecar}, M. 2002, \apjl, 574, L87

\bibitem[{{Lam} {et~al.}(2017){Lam}, {Faedi}, {Brown}, {Anderson}, {Delrez},
  {Gillon}, {H{\'e}brard}, {Lendl}, {Mancini}, {Southworth}, {Smalley},
  {Triaud}, {Turner}, {Hay}, {Armstrong}, {Barros}, {Bonomo}, {Bouchy},
  {Boumis}, {Collier Cameron}, {Doyle}, {Hellier}, {Henning}, {Jehin}, {King},
  {Kirk}, {Louden}, {Maxted}, {McCormac}, {Osborn}, {Palle}, {Pepe},
  {Pollacco}, {Prieto-Arranz}, {Queloz}, {Rey}, {S{\'e}gransan}, {Udry},
  {Walker}, {West}, \& {Wheatley}}]{lam_et_al2017}
{Lam}, K.~W.~F., {Faedi}, F., {Brown}, D.~J.~A., {et~al.} 2017, \aap, 599, A3

\bibitem[{{Lambrechts} \& {Johansen}(2012)}]{lambrechts+johansen2012}
{Lambrechts}, M., \& {Johansen}, A. 2012, \aap, 544, A32

\bibitem[{{Lanotte} {et~al.}(2014){Lanotte}, {Gillon}, {Demory}, {Fortney},
  {Astudillo}, {Bonfils}, {Magain}, {Delfosse}, {Forveille}, {Lovis}, {Mayor},
  {Neves}, {Pepe}, {Queloz}, {Santos}, \& {Udry}}]{lanotte_et_al2014}
{Lanotte}, A.~A., {Gillon}, M., {Demory}, B.-O., {et~al.} 2014, \aap, 572, A73

\bibitem[{{Lee} \& {Chiang}(2015)}]{lee+chiang2015}
{Lee}, E.~J., \& {Chiang}, E. 2015, \apj, 811, 41

\bibitem[{{Lee} \& {Chiang}(2016)}]{lee+chiang2016}
---. 2016, \apj, 817, 90

\bibitem[{{Lee} {et~al.}(2014){Lee}, {Chiang}, \& {Ormel}}]{lee_et_al2014}
{Lee}, E.~J., {Chiang}, E., \& {Ormel}, C.~W. 2014, \apj, 797, 95

\bibitem[{{Lega} \& {Lambrechts}(2016)}]{lega+lambrechts2016}
{Lega}, E., \& {Lambrechts}, M. 2016, in AAS/Division for Planetary Sciences
  Meeting Abstracts, Vol.~48, AAS/Division for Planetary Sciences Meeting
  Abstracts, 105.06

\bibitem[{{Leinhardt} \& {Stewart}(2012)}]{leinhardt+stewart2012}
{Leinhardt}, Z.~M., \& {Stewart}, S.~T. 2012, \apj, 745, 79

\bibitem[{{L{\'e}pine} \& {Gaidos}(2011)}]{lepine+gaidos2011}
{L{\'e}pine}, S., \& {Gaidos}, E. 2011, \aj, 142, 138

\bibitem[{{L{\'e}pine} \& {Shara}(2005)}]{lepine+shara2005}
{L{\'e}pine}, S., \& {Shara}, M.~M. 2005, \aj, 129, 1483

\bibitem[{{Lewis} {et~al.}(2013){Lewis}, {Knutson}, {Showman}, {Cowan},
  {Laughlin}, {Burrows}, {Deming}, {Crepp}, {Mighell}, {Agol}, {Bakos},
  {Charbonneau}, {D{\'e}sert}, {Fischer}, {Fortney}, {Hartman}, {Hinkley},
  {Howard}, {Johnson}, {Kao}, {Langton}, \& {Marcy}}]{lewis_et_al2013}
{Lewis}, N.~K., {Knutson}, H.~A., {Showman}, A.~P., {et~al.} 2013, \apj, 766,
  95

\bibitem[{{Liu} {et~al.}(2015){Liu}, {Hori}, {Lin}, \&
  {Asphaug}}]{liu_et_al2015}
{Liu}, S.-F., {Hori}, Y., {Lin}, D.~N.~C., \& {Asphaug}, E. 2015, \apj, 812,
  164

\bibitem[{{Lopez} \& {Fortney}(2013)}]{lopez+fortney2013}
{Lopez}, E.~D., \& {Fortney}, J.~J. 2013, \apj, 776, 2

\bibitem[{{Lopez} \& {Fortney}(2014)}]{lopez+fortney2014}
---. 2014, \apj, 792, 1

\bibitem[{{Mandel} \& {Agol}(2002)}]{mandel+agol2002}
{Mandel}, K., \& {Agol}, E. 2002, \apjl, 580, L171

\bibitem[{{Marcy} \& {Butler}(1992)}]{marcy+butler1992}
{Marcy}, G.~W., \& {Butler}, R.~P. 1992, \pasp, 104, 270

\bibitem[{{Mardling}(2007)}]{mardling2007}
{Mardling}, R.~A. 2007, \mnras, 382, 1768

\bibitem[{{Martinez} {et~al.}(2017){Martinez}, {Crossfield}, {Schlieder},
  {Dressing}, {Obermeier}, {Livingston}, {Ciceri}, {Peacock}, {Beichman},
  {L{\'e}pine}, {Aller}, {Chance}, {Petigura}, {Howard}, \&
  {Werner}}]{martinez_et_al2017}
{Martinez}, A.~O., {Crossfield}, I.~J.~M., {Schlieder}, J.~E., {et~al.} 2017,
  \apj, 837, 72

\bibitem[{{McLaughlin}(1924)}]{mclaughlin1924}
{McLaughlin}, D.~B. 1924, \apj, 60, doi:10.1086/142826

\bibitem[{{Mighell}(2005)}]{mighell2005}
{Mighell}, K.~J. 2005, \mnras, 361, 861

\bibitem[{{Mizuno}(1980)}]{mizuno1980}
{Mizuno}, H. 1980, Progress of Theoretical Physics, 64, 544

\bibitem[{{Mizuno} {et~al.}(1978){Mizuno}, {Nakazawa}, \&
  {Hayashi}}]{mizuno_et_al1978}
{Mizuno}, H., {Nakazawa}, K., \& {Hayashi}, C. 1978, Progress of Theoretical
  Physics, 60, 699

\bibitem[{{Mordasini} {et~al.}(2016){Mordasini}, {van Boekel}, {Molli{\`e}re},
  {Henning}, \& {Benneke}}]{mordasini_et_al2016}
{Mordasini}, C., {van Boekel}, R., {Molli{\`e}re}, P., {Henning}, T., \&
  {Benneke}, B. 2016, \apj, 832, 41

\bibitem[{{Morley} {et~al.}(2015){Morley}, {Fortney}, {Marley}, {Zahnle},
  {Line}, {Kempton}, {Lewis}, \& {Cahoy}}]{morley_et_al2015}
{Morley}, C.~V., {Fortney}, J.~J., {Marley}, M.~S., {et~al.} 2015, \apj, 815,
  110

\bibitem[{{Morley} {et~al.}(2017){Morley}, {Knutson}, {Line}, {Fortney},
  {Thorngren}, {Marley}, {Teal}, \& {Lupu}}]{morley_et_al2017}
{Morley}, C.~V., {Knutson}, H., {Line}, M., {et~al.} 2017, \aj, 153, 86

\bibitem[{{Morton}(2012)}]{morton2012}
{Morton}, T.~D. 2012, \apj, 761, 6

\bibitem[{{Morton}(2015)}]{morton2015}
---. 2015, {VESPA: False positive probabilities calculator}, Astrophysics
  Source Code Library, ascl:1503.011

\bibitem[{{Mui{\~n}os} \& {Evans}(2014)}]{muinos+evans2014}
{Mui{\~n}os}, J.~L., \& {Evans}, D.~W. 2014, Astronomische Nachrichten, 335,
  367

\bibitem[{{Mulders} {et~al.}(2015){Mulders}, {Pascucci}, \&
  {Apai}}]{mulders_et_al2015}
{Mulders}, G.~D., {Pascucci}, I., \& {Apai}, D. 2015, \apj, 798, 112

\bibitem[{{Mulders} {et~al.}(2016){Mulders}, {Pascucci}, {Apai}, {Frasca}, \&
  {Molenda-{\.Z}akowicz}}]{mulders_et_al2016}
{Mulders}, G.~D., {Pascucci}, I., {Apai}, D., {Frasca}, A., \&
  {Molenda-{\.Z}akowicz}, J. 2016, \aj, 152, 187

\bibitem[{{Nettelmann} {et~al.}(2011){Nettelmann}, {Fortney}, {Kramm}, \&
  {Redmer}}]{nettelmann_et_al2011}
{Nettelmann}, N., {Fortney}, J.~J., {Kramm}, U., \& {Redmer}, R. 2011, \apj,
  733, 2

\bibitem[{{{\"O}berg} {et~al.}(2011){{\"O}berg}, {Boogert}, {Pontoppidan}, {van
  den Broek}, {van Dishoeck}, {Bottinelli}, {Blake}, \&
  {Evans}}]{oberg_et_al2011}
{{\"O}berg}, K.~I., {Boogert}, A.~C.~A., {Pontoppidan}, K.~M., {et~al.} 2011,
  \apj, 740, 109

\bibitem[{{Pepper} {et~al.}(2017){Pepper}, {Rodriguez}, {Collins}, {Johnson},
  {Fulton}, {Howard}, {Beatty}, {Stassun}, {Isaacson}, {Col{\'o}n}, {Lund},
  {Kuhn}, {Siverd}, {Gaudi}, {Tan}, {Curtis}, {Stockdale}, {Mawet}, {Bottom},
  {James}, {Zhou}, {Bayliss}, {Cargile}, {Bieryla}, {Penev}, {Latham},
  {Labadie-Bartz}, {Kielkopf}, {Eastman}, {Oberst}, {Jensen}, {Nelson},
  {Sliski}, {Wittenmyer}, {McCrady}, {Wright}, {Relles}, {Stevens}, {Joner}, \&
  {Hintz}}]{pepper_et_al2017}
{Pepper}, J., {Rodriguez}, J.~E., {Collins}, K.~A., {et~al.} 2017, \aj, 153,
  215

\bibitem[{{Perri} \& {Cameron}(1974)}]{perri+cameron1974}
{Perri}, F., \& {Cameron}, A.~G.~W. 1974, \icarus, 22, 416

\bibitem[{{Petigura} {et~al.}(2013){Petigura}, {Howard}, \&
  {Marcy}}]{petigura_et_al2013b}
{Petigura}, E.~A., {Howard}, A.~W., \& {Marcy}, G.~W. 2013, Proceedings of the
  National Academy of Science, 110, 19273

\bibitem[{{Petigura} {et~al.}(2015){Petigura}, {Schlieder}, {Crossfield},
  {Howard}, {Deck}, {Ciardi}, {Sinukoff}, {Allers}, {Best}, {Liu}, {Beichman},
  {Isaacson}, {Hansen}, \& {L{\'e}pine}}]{petigura_et_al2015}
{Petigura}, E.~A., {Schlieder}, J.~E., {Crossfield}, I.~J.~M., {et~al.} 2015,
  \apj, 811, 102

\bibitem[{{Petigura} {et~al.}(2016){Petigura}, {Howard}, {Lopez}, {Deck},
  {Fulton}, {Crossfield}, {Ciardi}, {Chiang}, {Lee}, {Isaacson}, {Beichman},
  {Hansen}, {Schlieder}, \& {Sinukoff}}]{petigura_et_al2016}
{Petigura}, E.~A., {Howard}, A.~W., {Lopez}, E.~D., {et~al.} 2016, \apj, 818,
  36

\bibitem[{{Petigura} {et~al.}(2017){Petigura}, {Sinukoff}, {Lopez},
  {Crossfield}, {Howard}, {Brewer}, {Fulton}, {Isaacson}, {Ciardi}, {Howell},
  {Everett}, {Horch}, {Hirsch}, {Weiss}, \& {Schlieder}}]{petigura_et_al2017}
{Petigura}, E.~A., {Sinukoff}, E., {Lopez}, E.~D., {et~al.} 2017, \aj, 153, 142

\bibitem[{{Piso} \& {Youdin}(2014)}]{piso+youdin2014}
{Piso}, A.-M.~A., \& {Youdin}, A.~N. 2014, \apj, 786, 21

\bibitem[{{Pollack} {et~al.}(1996){Pollack}, {Hubickyj}, {Bodenheimer},
  {Lissauer}, {Podolak}, \& {Greenzweig}}]{pollack_et_al1996}
{Pollack}, J.~B., {Hubickyj}, O., {Bodenheimer}, P., {et~al.} 1996, \icarus,
  124, 62

\bibitem[{{Putnam} \& {Wiemer}(2014)}]{putnam+wiemer2014}
{Putnam}, D., \& {Wiemer}, D. 2014, Journal of the Astronautical Sciences, AAS,
  14

\bibitem[{{Rafikov}(2006)}]{rafikov2006}
{Rafikov}, R.~R. 2006, \apj, 648, 666

\bibitem[{{Rafikov}(2011)}]{rafikov2011}
---. 2011, \apj, 727, 86

\bibitem[{{Rauer} {et~al.}(2014){Rauer}, {Catala}, {Aerts}, {Appourchaux},
  {Benz}, {Brandeker}, {Christensen-Dalsgaard}, {Deleuil}, {Gizon}, {Goupil},
  {G{\"u}del}, {Janot-Pacheco}, {Mas-Hesse}, {Pagano}, {Piotto}, {Pollacco},
  {Santos}, {Smith}, {Su{\'a}rez}, {Szab{\'o}}, {Udry}, {Adibekyan}, {Alibert},
  {Almenara}, {Amaro-Seoane}, {Eiff}, {Asplund}, {Antonello}, {Barnes},
  {Baudin}, {Belkacem}, {Bergemann}, {Bihain}, {Birch}, {Bonfils}, {Boisse},
  {Bonomo}, {Borsa}, {Brand{\~a}o}, {Brocato}, {Brun}, {Burleigh}, {Burston},
  {Cabrera}, {Cassisi}, {Chaplin}, {Charpinet}, {Chiappini}, {Church},
  {Csizmadia}, {Cunha}, {Damasso}, {Davies}, {Deeg}, {D{\'{\i}}az}, {Dreizler},
  {Dreyer}, {Eggenberger}, {Ehrenreich}, {Eigm{\"u}ller}, {Erikson}, {Farmer},
  {Feltzing}, {de Oliveira Fialho}, {Figueira}, {Forveille}, {Fridlund},
  {Garc{\'{\i}}a}, {Giommi}, {Giuffrida}, {Godolt}, {Gomes da Silva},
  {Granzer}, {Grenfell}, {Grotsch-Noels}, {G{\"u}nther}, {Haswell}, {Hatzes},
  {H{\'e}brard}, {Hekker}, {Helled}, {Heng}, {Jenkins}, {Johansen},
  {Khodachenko}, {Kislyakova}, {Kley}, {Kolb}, {Krivova}, {Kupka}, {Lammer},
  {Lanza}, {Lebreton}, {Magrin}, {Marcos-Arenal}, {Marrese}, {Marques},
  {Martins}, {Mathis}, {Mathur}, {Messina}, {Miglio}, {Montalban}, {Montalto},
  {Monteiro}, {Moradi}, {Moravveji}, {Mordasini}, {Morel}, {Mortier},
  {Nascimbeni}, {Nelson}, {Nielsen}, {Noack}, {Norton}, {Ofir}, {Oshagh},
  {Ouazzani}, {P{\'a}pics}, {Parro}, {Petit}, {Plez}, {Poretti}, {Quirrenbach},
  {Ragazzoni}, {Raimondo}, {Rainer}, {Reese}, {Redmer}, {Reffert},
  {Rojas-Ayala}, {Roxburgh}, {Salmon}, {Santerne}, {Schneider}, {Schou},
  {Schuh}, {Schunker}, {Silva-Valio}, {Silvotti}, {Skillen}, {Snellen}, {Sohl},
  {Sousa}, {Sozzetti}, {Stello}, {Strassmeier}, {{\v S}vanda}, {Szab{\'o}},
  {Tkachenko}, {Valencia}, {Van Grootel}, {Vauclair}, {Ventura}, {Wagner},
  {Walton}, {Weingrill}, {Werner}, {Wheatley}, \& {Zwintz}}]{rauer_et_al2014}
{Rauer}, H., {Catala}, C., {Aerts}, C., {et~al.} 2014, Experimental Astronomy,
  38, 249

\bibitem[{{Ricker} {et~al.}(2014){Ricker}, {Winn}, {Vanderspek}, {Latham},
  {Bakos}, \& {et al.}}]{ricker_et_al2014}
{Ricker}, G.~R., {Winn}, J.~N., {Vanderspek}, R., {et~al.} 2014, Journal of
  Astronomical Telescopes, Instruments, and Systems, 1, 014003

\bibitem[{{Rogers} {et~al.}(2011){Rogers}, {Bodenheimer}, {Lissauer}, \&
  {Seager}}]{rogers_et_al2011}
{Rogers}, L.~A., {Bodenheimer}, P., {Lissauer}, J.~J., \& {Seager}, S. 2011,
  \apj, 738, 59

\bibitem[{{Rossiter}(1924)}]{rossiter1924}
{Rossiter}, R.~A. 1924, \apj, 60, doi:10.1086/142825

\bibitem[{{Santos} {et~al.}(2004){Santos}, {Israelian}, \&
  {Mayor}}]{santos_et_al2004}
{Santos}, N.~C., {Israelian}, G., \& {Mayor}, M. 2004, \aap, 415, 1153

\bibitem[{{Schlaufman}(2015)}]{schlaufman2015}
{Schlaufman}, K.~C. 2015, \apjl, 799, L26

\bibitem[{{Schlichting} {et~al.}(2015){Schlichting}, {Sari}, \&
  {Yalinewich}}]{schlichting_et_al2015}
{Schlichting}, H.~E., {Sari}, R., \& {Yalinewich}, A. 2015, \icarus, 247, 81

\bibitem[{{Schmitt} {et~al.}(2016){Schmitt}, {Tokovinin}, {Wang}, {Fischer},
  {Kristiansen}, {LaCourse}, {Gagliano}, {Tan}, {Schwengeler}, {Omohundro},
  {Venner}, {Terentev}, {Schmitt}, {Jacobs}, {Winarski}, {Sejpka}, {Jek},
  {Boyajian}, {Brewer}, {Ishikawa}, {Lintott}, {Lynn}, {Schawinski}, {Schwamb},
  \& {Weiksnar}}]{schmitt_et_al2016}
{Schmitt}, J.~R., {Tokovinin}, A., {Wang}, J., {et~al.} 2016, \aj, 151, 159

\bibitem[{Schwarz(1978)}]{schwarz1978}
Schwarz, G. 1978, The annals of statistics, 6, 461

\bibitem[{{Seager} \& {Mall{\'e}n-Ornelas}(2003)}]{seager+mallen-ornelas2003}
{Seager}, S., \& {Mall{\'e}n-Ornelas}, G. 2003, \apj, 585, 1038

\bibitem[{{Sinukoff} {et~al.}(2017{\natexlab{a}}){Sinukoff}, {Howard},
  {Petigura}, {Fulton}, {Crossfield}, {Isaacson}, {Gonzales}, {Crepp},
  {Brewer}, {Hirsch}, {Weiss}, {Ciardi}, {Schlieder}, {Benneke},
  {Christiansen}, {Dressing}, {Hansen}, {Knutson}, {Kosiarek}, {Livingston},
  {Greene}, {Rogers}, \& {L{\'e}pine}}]{sinukoff_et_al2017b}
{Sinukoff}, E., {Howard}, A.~W., {Petigura}, E.~A., {et~al.}
  2017{\natexlab{a}}, \aj, 153, 271

\bibitem[{{Sinukoff} {et~al.}(2017{\natexlab{b}}){Sinukoff}, {Howard},
  {Petigura}, {Fulton}, {Isaacson}, {Weiss}, {Brewer}, {Hansen}, {Hirsch},
  {Christiansen}, {Crepp}, {Crossfield}, {Schlieder}, {Ciardi}, {Beichman},
  {Knutson}, {Benneke}, {Dressing}, {Livingston}, {Deck}, {L{\'e}pine}, \&
  {Rogers}}]{sinukoff_et_al2017a}
---. 2017{\natexlab{b}}, \aj, 153, 70

\bibitem[{{Sozzetti} {et~al.}(2007){Sozzetti}, {Torres}, {Charbonneau},
  {Latham}, {Holman}, {Winn}, {Laird}, \& {O'Donovan}}]{sozzetti_et_al2007}
{Sozzetti}, A., {Torres}, G., {Charbonneau}, D., {et~al.} 2007, \apj, 664, 1190

\bibitem[{{Spalding} \& {Batygin}(2016)}]{spalding+batygin2016}
{Spalding}, C., \& {Batygin}, K. 2016, \apj, 830, 5

\bibitem[{{Stevenson}(1982)}]{stevenson1982}
{Stevenson}, D.~J. 1982, \planss, 30, 755

\bibitem[{{Tanaka} {et~al.}(2002){Tanaka}, {Takeuchi}, \&
  {Ward}}]{tanaka_et_al2002}
{Tanaka}, H., {Takeuchi}, T., \& {Ward}, W.~R. 2002, \apj, 565, 1257

\bibitem[{{Thorngren} {et~al.}(2016){Thorngren}, {Fortney}, {Murray-Clay}, \&
  {Lopez}}]{thorngren_et_al2016}
{Thorngren}, D.~P., {Fortney}, J.~J., {Murray-Clay}, R.~A., \& {Lopez}, E.~D.
  2016, \apj, 831, 64

\bibitem[{{Todorov} {et~al.}(2013){Todorov}, {Deming}, {Knutson}, {Burrows},
  {Fortney}, {Lewis}, {Cowan}, {Agol}, {Desert}, {Sada}, {Charbonneau},
  {Laughlin}, {Langton}, \& {Showman}}]{todorov_et_al2013}
{Todorov}, K.~O., {Deming}, D., {Knutson}, H.~A., {et~al.} 2013, \apj, 770, 102

\bibitem[{{Torres} {et~al.}(2008){Torres}, {Winn}, \&
  {Holman}}]{torres_et_al2008}
{Torres}, G., {Winn}, J.~N., \& {Holman}, M.~J. 2008, \apj, 677, 1324

\bibitem[{{Valenti} {et~al.}(1995){Valenti}, {Butler}, \&
  {Marcy}}]{valenti_et_al1995}
{Valenti}, J.~A., {Butler}, R.~P., \& {Marcy}, G.~W. 1995, \pasp, 107, 966

\bibitem[{{Van Eylen} {et~al.}(2016){Van Eylen}, {Nowak}, {Albrecht}, {Palle},
  {Ribas}, {Bruntt}, {Perger}, {Gandolfi}, {Hirano}, {Sanchis-Ojeda},
  {Kiilerich}, {Prieto-Arranz}, {Badenas}, {Dai}, {Deeg}, {Guenther},
  {Monta{\~n}{\'e}s-Rodr{\'{\i}}guez}, {Narita}, {B{\'e}jar}, {Shrotriya},
  {Winn}, \& {Sebastian}}]{vaneylen_et_al2016}
{Van Eylen}, V., {Nowak}, G., {Albrecht}, S., {et~al.} 2016, \apj, 820, 56

\bibitem[{{Vanderburg} \& {Johnson}(2014)}]{vanderburg+johnson2014}
{Vanderburg}, A., \& {Johnson}, J.~A. 2014, \pasp, 126, 948

\bibitem[{{Vanderburg} {et~al.}(2016){Vanderburg}, {Latham}, {Buchhave},
  {Bieryla}, {Berlind}, {Calkins}, {Esquerdo}, {Welsh}, \&
  {Johnson}}]{vanderburg_et_al2016}
{Vanderburg}, A., {Latham}, D.~W., {Buchhave}, L.~A., {et~al.} 2016, \apjs,
  222, 14

\bibitem[{{Vogt} {et~al.}(1994){Vogt}, {Allen}, {Bigelow}, {Bresee}, {Brown},
  {Cantrall}, {Conrad}, {Couture}, {Delaney}, {Epps}, {Hilyard}, {Hilyard},
  {Horn}, {Jern}, {Kanto}, {Keane}, {Kibrick}, {Lewis}, {Osborne},
  {Pardeilhan}, {Pfister}, {Ricketts}, {Robinson}, {Stover}, {Tucker}, {Ward},
  \& {Wei}}]{vogt_et_al1994}
{Vogt}, S.~S., {Allen}, S.~L., {Bigelow}, B.~C., {et~al.} 1994, in \procspie,
  Vol. 2198, Instrumentation in Astronomy VIII, ed. D.~L. {Crawford} \& E.~R.
  {Craine}, 362

\bibitem[{{von Braun} {et~al.}(2012){von Braun}, {Boyajian}, {Kane}, {Hebb},
  {van Belle}, {Farrington}, {Ciardi}, {Knutson}, {ten Brummelaar},
  {L{\'o}pez-Morales}, {McAlister}, {Schaefer}, {Ridgway}, {Collier Cameron},
  {Goldfinger}, {Turner}, {Sturmann}, \& {Sturmann}}]{vonbraun_et_al2012}
{von Braun}, K., {Boyajian}, T.~S., {Kane}, S.~R., {et~al.} 2012, \apj, 753,
  171

\bibitem[{{Wallace} {et~al.}(2017){Wallace}, {Tremaine}, \&
  {Chambers}}]{wallace_et_al2017}
{Wallace}, J., {Tremaine}, S., \& {Chambers}, J. 2017, \aj, 154, 175

\bibitem[{{Wang} \& {Fischer}(2015)}]{wang+fischer2015}
{Wang}, J., \& {Fischer}, D.~A. 2015, \aj, 149, 14

\bibitem[{{Ward}(1997)}]{ward1997}
{Ward}, W.~R. 1997, \icarus, 126, 261

\bibitem[{{Williams} \& {Cieza}(2011)}]{williams+cieza2011}
{Williams}, J.~P., \& {Cieza}, L.~A. 2011, \araa, 49, 67

\bibitem[{{Wolfgang} \& {Lopez}(2015)}]{wolfgang+lopez2015}
{Wolfgang}, A., \& {Lopez}, E. 2015, \apj, 806, 183

\bibitem[{{Youdin}(2011)}]{youdin2011}
{Youdin}, A.~N. 2011, \apj, 742, 38

\bibitem[{{Zacharias} {et~al.}(2013){Zacharias}, {Finch}, {Girard}, {Henden},
  {Bartlett}, {Monet}, \& {Zacharias}}]{zacharias_et_al2013}
{Zacharias}, N., {Finch}, C.~T., {Girard}, T.~M., {et~al.} 2013, \aj, 145, 44

\bibitem[{{Zhang} \& {Hamilton}(2008)}]{zhang+hamilton2008}
{Zhang}, K., \& {Hamilton}, D.~P. 2008, \icarus, 193, 267

\end{thebibliography}

\appendix
\section{The Inclusion \& Exclusion of K2-55 in K2 Guest Observer Programs}
\label{sec:appendix}
Although K2-55 is a dwarf star, it was not included in any of the approved K2 Guest Observer programs focused on dwarfs. In this section, we explore why K2-55 was proposed as part of a program focused on giant stars and excluded from programs studying dwarf stars. We first review the selection criteria for K2GO3051 (the program that nominated K2-55) and then consider three large programs focused on cool dwarfs.

K2GO3051 (PI: Dennis Stello) is a galactic archeology program designed to probe the chemical evolution of the Milky Way via asteroseismology of red giants.  Stello and collaborators selected their targets using a color-magnitude cut. They first restricted the sample to all stars redder than $J - Ks = 0.5$ and then ranked stars in order of decreasing brightness. While 90\% of the selected stars are expected to be giants, the proposers noted that their sample also includes red M and K dwarfs. The inclusion of \mbox{K2-55} in the K2GO3051 target list is therefore unsurprising, but its absence in any of the large Campaign~3 proposals targeting cool dwarfs (GO3069, GO3106, GO3107) is more noteworthy.\footnote{Note that there is no requirement that \emph{K2} target lists cannot overlap. On the contrary, many selected \emph{K2} targets were proposed by multiple teams.} 

K2-55 met the proper motion requirement of $>5$ mas/yr, the color cut of $0.7 < J-K < 1.1$, and the brightness requirement of $r < 17$ in the Carlsberg Meridian Catalogue \citep{muinos+evans2014} required by B.~Montet and collaborators for inclusion in GO3069, but the target failed the second color cut of $r-J > 2.0$. The $r-J$ color of K2-55 is $r- J = 1.799$. 

K2-55 was likely excluded from GO3106 and GO3107 because of its modest proper motion: -14.9 mas/yr in RA, -22.1 mas/yr in Dec \citep[UCAC4,][]{zacharias_et_al2013}. For GO3106, C. Beichman and collaborators crossmatched the 2MASS and WISE catalogs and selected targets based on both colors and reduced proper motions. Beichman et al. supplemented their target list by adding additional bright cool dwarfs from SIMBAD and SDSS. Finally, I. Crossfield, J. Schlieder, and S. Lepine proposed 4545 small stars for GO3107 by selecting targets from the SUPERBLINK proper motion survey \citep{lepine+shara2005, lepine+gaidos2011} and prioritizing them by planet detectability.

\clearpage
\enddocument